\documentclass[aps,%preprint,
showpacs,preprintnumbers,amsmath,amssymb,nofootinbib,longbibliography]{revtex4-2}
\usepackage[colorlinks=true, pdfstartview=FitV, linkcolor=red, citecolor=blue, urlcolor=black, pdftitle={},pdfauthor={},pdfsubject={}, pdfkeywords={}]{hyperref}
\usepackage{graphicx}
\usepackage{amsmath,amssymb,bm}
\usepackage{mathtools}
\usepackage{color}
\usepackage{physics}
\usepackage{comment}

\usepackage{colortbl}

\newcommand{\ri}{\mathrm{i}}

\newcommand{\vp}{\varphi}

\newcommand{\cd}{\! \cdot \!}

\newcommand{\Ra}{\rangle}
\newcommand{\La}{\langle}

\newcommand{\mf}[1]{\mathfrak{ #1}}
\newcommand{\cl}[1]{\mathcal{ #1}}
\newcommand{\tx}[1]{\mathrm{ #1}}
\newcommand{\red}[1]{{\color{red}#1}}

\newcommand{\eqn}[1]{\begin{equation}\begin{split}#1\end{split}\end{equation}}
\newcommand{\aln}[1]{\begin{align}#1\end{align}}
\newcommand{\mtx}[1]{\begin{matrix}#1\end{matrix}}
\newcommand{\pmtx}[1]{\begin{pmatrix}#1\end{pmatrix}}

\definecolor{lightgray}{RGB}{230,230,230}

\makeatletter

    \def\CT@@do@color{%
      \global\let\CT@do@color\relax
            \@tempdima\wd\z@
            \advance\@tempdima\@tempdimb
            \advance\@tempdima\@tempdimc
    \advance\@tempdimb\tabcolsep
    \advance\@tempdimc\tabcolsep
    \advance\@tempdima2\tabcolsep
            \kern-\@tempdimb
            \leaders\vrule
                    \hskip\@tempdima\@plus  1fill
            \kern-\@tempdimc
            \hskip-\wd\z@ \@plus -1fill }
    \makeatother

\begin{document}
%\begin{comment}
\preprint{KEK-TH-2453}
\preprint{YITP-22-136}
\author{Yoshimasa Hidaka}
\email{hidaka@post.kek.jp}
\affiliation{
KEK Theory Center, Tsukuba 305-0801, Japan
}
\affiliation{
Graduate University for Advanced Studies (Sokendai), Tsukuba 305-0801, Japan
}
\affiliation{
  Department of Physics, Faculty of Science, University of Tokyo, 7-3-1 Hongo Bunkyo-ku Tokyo 113-0033, Japan
}
\affiliation{
RIKEN iTHEMS, RIKEN, Wako 351-0198, Japan
}
\author{Satoshi Iso}
\email{satoshi.iso@kek.jp}
\affiliation{
KEK Theory Center, Tsukuba 305-0801, Japan
}
\affiliation{
Graduate University for Advanced Studies (Sokendai), Tsukuba 305-0801, Japan
}
\affiliation{International Center for Quantum-field Measurement Systems for Studies of the Universe and Particles (QUP), KEK, Oho 1-1, Tsukuba, Ibaraki 305-0801, Japan}
\author{Kengo Shimada}
\email{kengo.shimada@yukawa.kyoto-u.ac.jp}
\affiliation{
Center for Gravitational Physics and Quantum Information, Yukawa Institute for Theoretical Physics, Kyoto University, Kitashirakawa-Oiwakecho, Sakyo-Ku, Kyoto 606-8502, Japan
}

%\date{\today}
%\pacs{}

\title{Entanglement Generation  and Decoherence\\ in a Two-Qubit System Mediated by Relativistic Quantum Field}
\begin{abstract}
Motivated by the Bose {\it et al}.-Matletto-Vedral (BMV) proposal for detecting quantum superposition of
spacetime geometries, we study a toy model of a quantum entanglement generation between two
spins (qubits) mediated by a relativistic free scalar field. 
After time evolution, spin correlation is generated through the interactions with the field. Because of the associated particle creation into an open system, the quantum state of spins is partially decohered.
In this paper, we give a comprehensive study of the model based on the closed-time path formalism, focussing on relativistic causality and quantum mechanical complementarity. We calculate various quantities such as spin correlations, entanglement entropies, mutual information and negativity, and study their behaviors in various limiting situations. In particular, we calculate the mutual information of the two spins and compare it with spin correlation functions. 
We also discuss why no quantum entanglement can be generated unless both spins are causally affected by one another
while spin correlations are generated. 
\end{abstract}
\maketitle
%\end{comment}

%%%%%%%%%%%%%%%%%%%%%%%%%%%%%%%%%%%%%%%%%%%%%%%%%%%%
\section{Introduction} 
Relativistic quantum field theories (QFT) can successfully describe our Universe in a  compatible way between the principle of quantum mechanics and relativistic causality, and provides a nontrivial concept of quantum vacuum with vacuum fluctuations. 
Formulated on the curved spacetime, it can incorporate nontrivial effects of classical gravity.
Thus, we can say that QFT serves as a foundation for the success and further progress of modern descriptions of not only matter fields but also the spacetime itself in which matter lives.
Though it is natural to think that the gravitational field should also be quantized and described within the framework of the QFT paradigm,
it is not yet experimentally confirmed as no experiments have been done to observe the quantum nature of gravity.

Towards this direction, 
the so-called BMV tabletop experiment is proposed by Bose et al. \cite{Bose:2017nin} and Matletto, Vedral \cite{Marletto:2017kzi} to detect quantum entanglement between two massive objects generated by the Newtonian gravitational interaction. The generation of entanglement can be interpreted as being induced by a quantum superposition of different spacetime geometries. The experiment may become feasible in the foreseeable future by virtue of the experimental progress in the ground state cooling \cite{Delic:2020ndp,Tebbenjohanns2021}, and macroscopic superpositions \cite{Kovachy2015,Fein2019}.
The original analysis in the BMV proposal is very simple and intuitive based on the Newtonian approximation.
However, there have been discussions on its physical interpretation, especially about whether the detection of the gravity-mediated entanglement can be an experimental test of the quantum nature of gravity \cite{Mari:2015qva,Anastopoulos:2018drh,Belenchia:2018szb,Christodoulou:2018cmk,Danielson:2021egj,Christodoulou:2022vte,Fragkos:2022tbm,Christodoulou:2022knr,Chen:2022wro}.
Also, more detailed analyses based on field-theoretical models are given in the Hamiltonian formulation \cite{Chen:2022wro,Matsumura:2021tlu,Sugiyama:2022wcd,Sugiyama:2022ixw} and in the path-integral formulation \cite{Christodoulou:2022vte,Hidaka:2022tzk} to understand its possible impacts of the relativistic causality and the vacuum fluctuations of the gravitational field.
See Refs.~\cite{Carney:2018ofe,Huggett:2022uui} for reviews.

The purpose of the present work is to give a comprehensive description of the field-mediated quantum/classical correlations between two objects based on a simple toy model for the BMV-type setup.
Time evolution of the state containing two objects (qubits, or equivalently, spins) as well as the quantum field can be fully obtained by using the technique of closed-time path integral and given in terms of two types of Green's functions. 
One is the retarded/advanced Green's function and the other is the Keldysh Green's function. 
It is briefly reviewed in \cite{Hidaka:2022tzk} and in the Appendix \ref{app.Reduced density matrix}. 
The relativistic causality reflected in the retarded/advanced Green's functions controls the generation of the correlations via causal influence from one to the other.
On the other hand, the vacuum fluctuations manifested by the Keldysh Green's function have two important effects:  generation of correlations and decoherence of each object's quantum state by particle creations associated with nonadiabaticity of the setup. 
We discuss how these effects are related to each other in various cases, employing a sort of trade-off relation between entropies and mutual information. In particular, we compare the mutual information of two spins with the spin correlation functions to check the inequality between them. 
We also obtain the necessary conditions, in terms of the causality and the decoherence, for the quantum entanglement to be generated in our setup.
We show that the quantum entanglement can not be generated unless two spins are causally connected in both directions.

This paper is organized as follows; we use figures in each section to explain it. 
Our field-theoretical model is introduced in Sec.~\ref{sec.Setup}. A schematic picture of the model is given in Fig.~\ref{fig:setup}. 
We then discuss its Newtonian approximation in Sec.~\ref{sec.Newtonian} to describe entanglement generation due to the nonlocal interaction.  Fig.~\ref{fig:S_N} shows the evolution of entanglement as a function of the interaction time-interval.
In the Newtonian approximation, there is no decoherence due to the particle creations and the entanglement entropy simply oscillates. 
The notions of visibility, distinguishability and the separability condition by negativity are also introduced in this section. 
All the necessary calculations are given in this section. 

In Sec.~\ref{sec.Relativistic}, we introduce a dynamical field coupled to spin systems and solve the model exactly.  We obtain the reduced density matrix tracing out the field variables in which the effects of the causality and the vacuum fluctuations are automatically taken into account. 
Fig.~\ref{fig:sec4} shows the contents of this section, in particular various tools we introduce for investigating various properties. 
Due to the particle creation associated with the nonadiabaticity of the time evolution, various quantities such as spin correlations or quantum entanglement are suppressed 
by the adiabaticity parameters $\gamma_\tx{A}$ and $\gamma_\tx{B}$, given in Eq.~(\ref{gamma_A,B}),
written in terms of the Keldysh Green's function.

Using these results,
in Sec.~\ref{sec.limiting cases}, we consider four limiting cases to see how the causality and the vacuum fluctuations affect correlations between the objects. 
Figs.~\ref{fig:Adiabaticity} and \ref{fig:Causality} show the situations of these four cases, specified by their adiabaticity and causal relations between Alice and Bob. 
An adiabatic limit in Sec. \ref{sec.Adiabatic limit} corresponds to $\gamma_\tx{A}=1.$ There is no particle creation from Alice. 
Fig.~\ref{fig:Negativity} shows the entanglement negativity of Alice and Bob spins. Its nonvanishing property expresses quantum entanglement between the spins. 
We also depict the mutual information of Alice and Bob in Fig.~\ref{fig:I-adiabatic} and that of Alice and field in Fig.~\ref{fig:Isp-adiabatic}.
On the other hand, in the nonadiabatic limit of $\gamma_\tx{B}=0$ studied in Sec. \ref{sec.Non-adiabatic limit}, Bob is completely decohered. Thus the mutual information of Alice and Bob, depicted in Fig.~\ref{fig:I-nonadiabatic}, cannot be larger than the half value of the maximal one. 
The nonadiabatic limit can be considered as an analog of the Colella-Overhauser-Werner (COW) experiment in which Bob's spin is replaced by the earth and interacts with Alice's spin via gravitational interaction. 
The third and fourth cases are focussing relativistic causality. In Sec.~\ref{sec.Spacelike-separated}, we study spacelike separated case where both retarded Green's functions between Alice and Bob vanish as in Fig.~\ref{fig:setup-Spacelike}.
In this case, while they do not have the causal interaction, spin correlations appear from the entanglement of the vacuum state of the field.
Mutual information and spin correlations are depicted in Fig.~\ref{fig:IAB-spacelike}.
In Sec.~\ref{sec.One-way},
these quantities are evaluated in the case where only one of the retarded Green's functions is vanishing and depicted in Fig.~\ref{fig:IAB-Wald}.
In both of those cases, as proved in Sec.~\ref{sec.separaility-proof}, the reduced density matrix of the two-spin system is separable and there is no quantum entanglement between Alice's and Bob's spins. 
In Sec.~\ref{sec.short summary}, we have a short summary of the roles played by various Green's functions.
Finally, we summarize the paper in Sec.~\ref{sec.Summary}.

In Appendix \ref{app.Reduced density matrix}, we give a review of the closed time path integral formalism to calculate the reduced density matrix in our setup. In Appendix \ref{app.Correlation between field and spin}, we calculate correlations between the scalar field and the spin variables. In Appendix \ref{app.Particle creation and Keldysh function}, we prove that the strength  of decoherence due to the nonadiabaticity, $\gamma_\tx{A}$ and $\gamma_\tx{B}$,  are directly related to the number of particles created by the nonadiabatic change of the spin-field couplings. In Appendix \ref{app:Numerical evaluation}, we numerically evaluate integrations of various Green's functions. In Appendix \ref{app.Consistency condition}, we check inequality relations between Green functions as a consistency for the nonnegativity of the density matrix.

%%%%%%%%%%%%%%%%%%%%%%%%%%%%%%%%%%%%%%%%%%%%%%%%%%%%
\section{Setup \label{sec.Setup}}
%%%%%%%FIG%%%%%%%
\begin{figure}
\begin{center}
 \includegraphics[width=8cm]{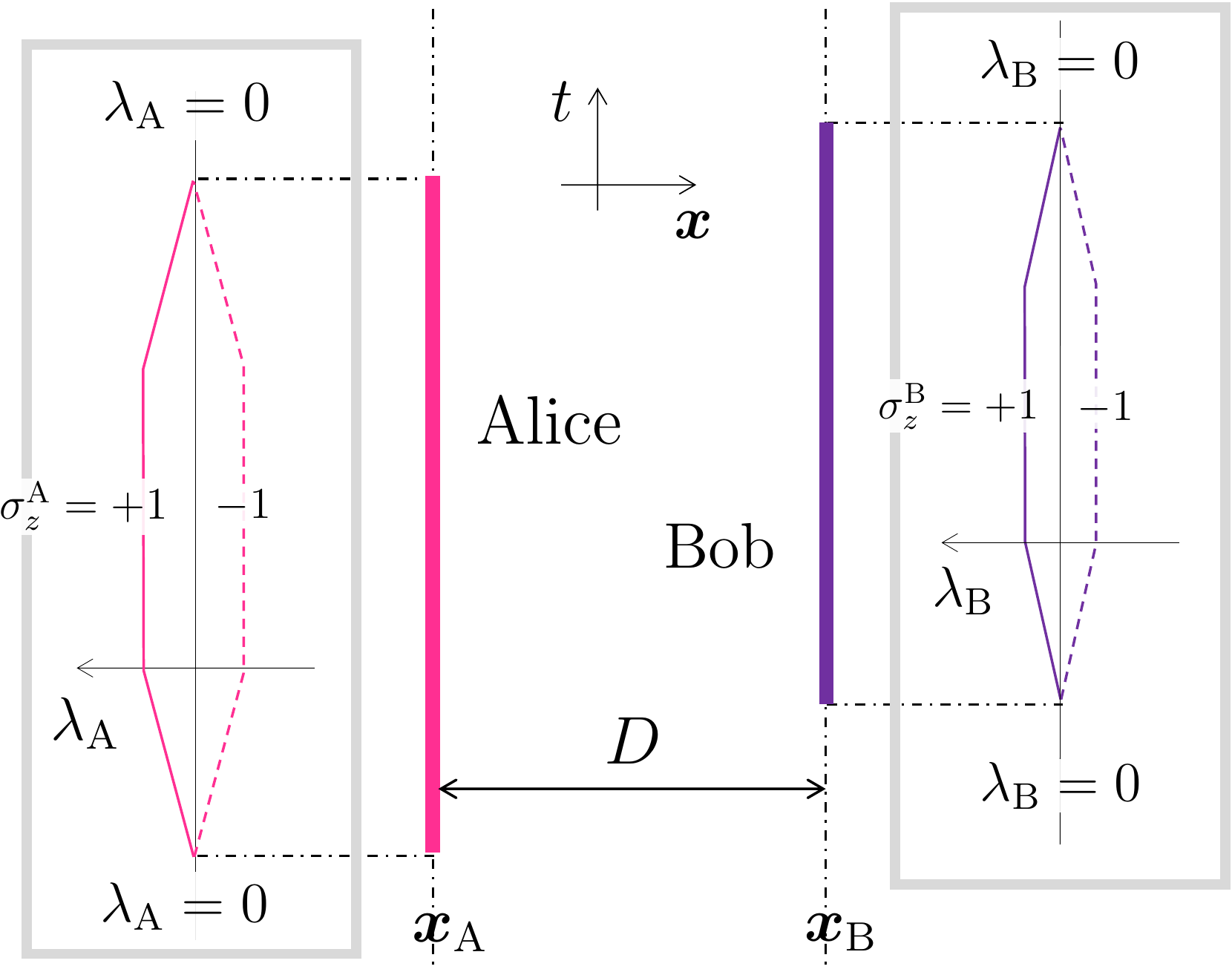}
\caption{A schematic view of the setup. Alice's spin and Bob's spin are located at $\vb*{x}= \vb*{x}_\tx{A}$ and $\vb*{x}= \vb*{x}_\tx{B}$, respectively, and interact with the field on the world lines colored pink and purple.
In the boxes, time-dependences of the strengths of the spin-field couplings $\lambda_\tx{A}$ and $\lambda_\tx{B}$ are depicted. Each coupling is nonvanishing only during a finite interval. The signs of the couplings are correlated with the signs of the $z$-direction spins $\hat{\sigma}_z^{\tx{A},\tx{B}  }$.}
\label{fig:setup}
\end{center}
\end{figure}
%%%%%%%%%%%%%%%%%%
We consider a system consisting of Alice's spin $\sigma^\tx{A}$, Bob's spin $\sigma^\tx{B}$ and a scalar field $\phi$ in $(3+1)$ dimensional spacetime. The Hamiltonian is given by
$\hat{H}=\hat{H}_\phi+ \hat{H}_\tx{A}+ \hat{H}_\tx{B}$ with
\eqn{
\hat{H}_\phi &=\int\dd[3]{x} \frac{1}{2} \qty[ \hat{\pi}^2(\vb*{x}) + (\vb*{\nabla}\hat{\phi}(\vb*{x}))^2 + m^2\hat{\phi}^2(\vb*{x}) ] \,,\\
\hat{H}_\tx{A} &= - \hat{\sigma}_z^\tx{A} \lambda_\tx{A}(t)\hat{\phi}(\vb*{x}_\tx{A}) ~,~~ \hat{H}_\tx{B}= - \hat{\sigma}_z^\tx{B} \lambda_\tx{B}(t)\hat{\phi}(\vb*{x}_\tx{B})~,\label{Hamiltonian}
}
where $m$ is the mass of the field and $\hat{\pi}(\vb*{x})$ is the canonical conjugate of $\hat{\phi}(\vb*{x})$. 
The strengths of the couplings, $\lambda_\tx{A}(t)$ and $\lambda_\tx{B}(t)$, are controlled by Alice and Bob; they are nonzero  in the finite time intervals as depicted in Fig.~\ref{fig:setup}. 
Without loss of generality, we suppose that they take positive values; $\lambda_\tx{A}(t) \geq 0$ and $\lambda_\tx{B}(t) \geq 0$.
The system is symmetric under an interchange of Alice and Bob, and
all the discussions below are interchangeable under  (A,B) $\to$ (B,A).

At the initial time $t=t_\tx{i}$ when Alice and Bob have not yet turned on the spin-field interactions,
it is assumed that the total system is in a pure and separable state:
\eqn{ 
\ket{\Psi_\tx{i}} = \ket{\Psi_\tx{i}}_\tx{AB}  \ket{\Omega}_\phi ~, \label{initial-state}
}
where $\ket{\Omega}_\phi$ is field's ground state with $\lambda_\tx{A} = \lambda_\tx{B}=0$, and
\eqn{
\ket{\Psi_\tx{i}}_\tx{AB}& \coloneqq  %\ket{+_x}_\tx{A} \ket{+_x}_\tx{B} \\=&
\frac{1}{2}(\ket{+}_\tx{A}+\ket{-}_\tx{A})
(\ket{+}_\tx{B}+\ket{-}_\tx{B})\\
&=\frac{1}{2} \qty( \ket{++} + \ket{+-} + \ket{-+} + \ket{--} ).
\label{spin-initial-state} }
Here $\ket{\pm}$ being the eigenstate of $\hat{\sigma}_z$ with the eigenvalue $\pm 1$.
We also introduced the notation
\eqn{
\ket{\sigma  \sigma' } \coloneqq \ket{\sigma}_\tx{A} \ket{\sigma' }_\tx{B} ~~\tx{with}~~ \sigma, \sigma' = \pm .~ \label{z-spin eigenstate}
}
Acting the time evolution operator $\hat{U}(t_\tx{f},t_\tx{i}) =  \tx{T} \exp \{ -\ri \int_{t_\tx{i}}^{t_\tx{f}} \dd t \hat{H}  \} $ on the initial state (\ref{initial-state}), we obtain a state at $t= t_\tx{f}$ by which time Alice and Bob have turned off the interactions:
\eqn{
\ket{\Psi_\tx{f}} = \hat{U}(t_\tx{f} ,t_\tx{i})   \ket{\Psi_\tx{i}} ~, \label{final-state}
}
which is no longer separable.
In the following sections, we clarify the nature of this final state
by solving the system exactly and evaluating various quantities such as correlation functions, entanglement entropies, entanglement negativity, and mutual informations.

%%%%%%%%%%%%%%%%%%%%%%%%%%%%%%%%%%%%%%%%%%%%%%%%%%%%%%%%%%%%%%%%%%%
\section{Entanglement generation in Newtonian approximation \label{sec.Newtonian}}
Before solving the system exactly, let us see how the entanglement between Alice's spin and Bob's spin develops in an effective theory where the interaction between two spins is replaced by a nonlocal Newtonian interaction.
The tabletop experiments \cite{Bose:2017nin, Marletto:2017kzi} to detect the quantum superposition of spacetime geometries are based on the analyses in this approximation.
Here, the entanglement between the two spins is induced by the Newtonian potential to generate spin correlations and affect quantum interferences, as shown in Fig.~\ref{fig:S_N}.
The entanglement entropy represents the amplitude of the correlations properly in the Newtonian picture.

\

The corresponding Hamiltonian is the ferromagnetic one:
\eqn{
\hat{H}_\tx{AB} =- \bar{J}  \hat{\sigma}_z^\tx{A} \hat{\sigma}_z^\tx{B} \times \theta(t_\tx{off} - t) \theta(t-t_\tx{on}) ~, \label{ferromagnetic-interaction}
}
where $\theta(t)$ is the Heaviside step function.
The distance between Alice and Bob $D = |\vb*{x}_\tx{A} - \vb*{x}_\tx{B}|$, the mass of the field $m$ and the interaction strengths $\lambda_\tx{A,B}$ are all contained in the coefficient,
\eqn{
\bar{J} \coloneqq \bar{\lambda}_\tx{A} \bar{\lambda}_\tx{B} \frac{\exp (-m D)}{4 \pi D}   ~.
}
Here we have assumed a simple time dependence, $\lambda_\tx{A}(t)\lambda_\tx{B}(t)= \bar{\lambda}_\tx{A} \bar{\lambda}_\tx{B} \theta(t_\tx{off} - t) \theta(t-t_\tx{on})$, where $\bar{\lambda}_\tx{A}$ and $\bar{\lambda}_\tx{B}$ are positive constants, $t_\tx{on} (> t_\tx{i})$ and $t_\tx{off}$ are the times at which the spin-spin interaction is turned on and off, respectively.

%%%%%%%%%%%%%%%%%%%%%%%%
\subsection{Generating entanglement and correlations}
Acting the time evolution operator with the Hamiltonian (\ref{ferromagnetic-interaction}) on the initial state (\ref{spin-initial-state}), we find
\eqn{
\ket{\Psi_\tx{f}}_\tx{AB} &= \frac{e^{\ri \Theta}}{2} \qty(\ket{++} + \ket{--}) + \frac{e^{-\ri \Theta}}{2} \qty(\ket{+-} + \ket{-+}) 
\eqqcolon \ket{\Theta}_\tx{AB} \label{Psi_AB-Newtonian}
}
for $t = t_\tx{f} > t_\tx{off}$, where 
$\Theta  \coloneqq \bar{J}\times (t_\tx{off} - t_\tx{on}) \label{Theta}$
is the dimensionless time interval for which the spin-spin interaction is effective.
It is convenient to write the density matrix in the
Bloch representation as
\eqn{
\hat{\rho}_\tx{AB} &= \ket{\Psi_\tx{f}}_\tx{AB}\!\bra{\Psi_\tx{f}}  \\
&= \frac{1}{4} \qty{ \hat{1}^\tx{A} \hat{1}^\tx{B} + \hat{\sigma}_x^\tx{A} \hat{\sigma}_x^\tx{B} +  \cos (2 \Theta) \qty[ \hat{\sigma}_x^\tx{A} \hat{1}^\tx{B} +  \hat{1}^\tx{A}\hat{\sigma}_x^\tx{B}  ]  - \sin (2 \Theta) \qty[ \hat{\sigma}_y^\tx{A} \hat{\sigma}_z^\tx{B} + \hat{\sigma}_z^\tx{A} \hat{\sigma}_y^\tx{B}  ]  } ~, \label{rho_AB-Newtonian}
}
where $\hat{1}^\tx{A}$ and $\hat{1}^\tx{B}$ are the unit operators acting on the Hilbert space of Alice's and Bob's spin.
By taking the partial trace, we get the following reduced density matrices respectively,
\eqn{
\hat{\rho}_\tx{A} &= \tr_\tx{B} \{ \hat{\rho}_\tx{AB} \} = \frac{1}{2} \qty{\hat{1}^\tx{A} +  \cos (2 \Theta) \hat{\sigma}_x^\tx{A}  }~, \\
\hat{\rho}_\tx{B} &= \tr_\tx{A} \{ \hat{\rho}_\tx{AB} \} = \frac{1}{2} \qty{\hat{1}^\tx{B} +  \cos (2 \Theta) \hat{\sigma}_x^\tx{B}  }~.
\label{rho_A(B)-Newtonian}}
Here, $\tr_\tx{A}$ and $\tr_\tx{B}$ represent the trace over Alice's and Bob's Hilbert space.

As is obvious from the above expressions (\ref{rho_AB-Newtonian}) and (\ref{rho_A(B)-Newtonian}), 
the expectation values of spins are given by
\eqn{
&\La  \hat{\sigma}_x^\tx{A} \hat{\sigma}_x^\tx{B} \Ra = 1 ~, \\
& \La \hat{\sigma}_y^\tx{A} \hat{\sigma}_z^\tx{B} \Ra = \La \hat{\sigma}_z^\tx{A} \hat{\sigma}_y^\tx{B} \Ra = - \sin (2 \Theta)   ~, \\
&\La \hat{\sigma}_x^\tx{A}  \Ra  =  \La  \hat{\sigma}_x^\tx{B}  \Ra  = \cos (2 \Theta)   ~, \label{expectation-value-Newtonian}
}
and all the others vanish. Here the expectation values are given by $\La  \hat{\cl{O}} \Ra \coloneqq \tr_\tx{AB} \{ \hat{\rho}_\tx{AB} \hat{\cl{O}}\}$.
In order to see the correlations between two spins,
we introduce $\delta \hat{\sigma}_w^\tx{A} \coloneqq \hat{\sigma}_w^\tx{A} - \La \hat{\sigma}_w^\tx{A} \Ra$ with $w=x,y,z$.
Nontrivial correlations are given by
\eqn{
&\La \delta \hat{\sigma}_x^\tx{A} \delta \hat{\sigma}_x^\tx{B} \Ra = \sin^2 (2 \Theta) ~, \\
&\La \delta \hat{\sigma}_y^\tx{A} \delta \hat{\sigma}_z^\tx{B} \Ra = \La \delta \hat{\sigma}_z^\tx{A} \delta \hat{\sigma}_y^\tx{B} \Ra = - \sin (2 \Theta) ~. \label{connected-part-Newtonian}
}
These correlations reflect the quantum entanglement between the two spins.
In the rest of this section, we see their relation by calculating entanglement entropy,  visibility,  distinguishability, and entanglement negativity.

%%%%%%%%%%%%%%%%%%%%%%%%
\subsection{Entanglement entropy}
\begin{figure}
\begin{center}
 \includegraphics[width=7.5cm]{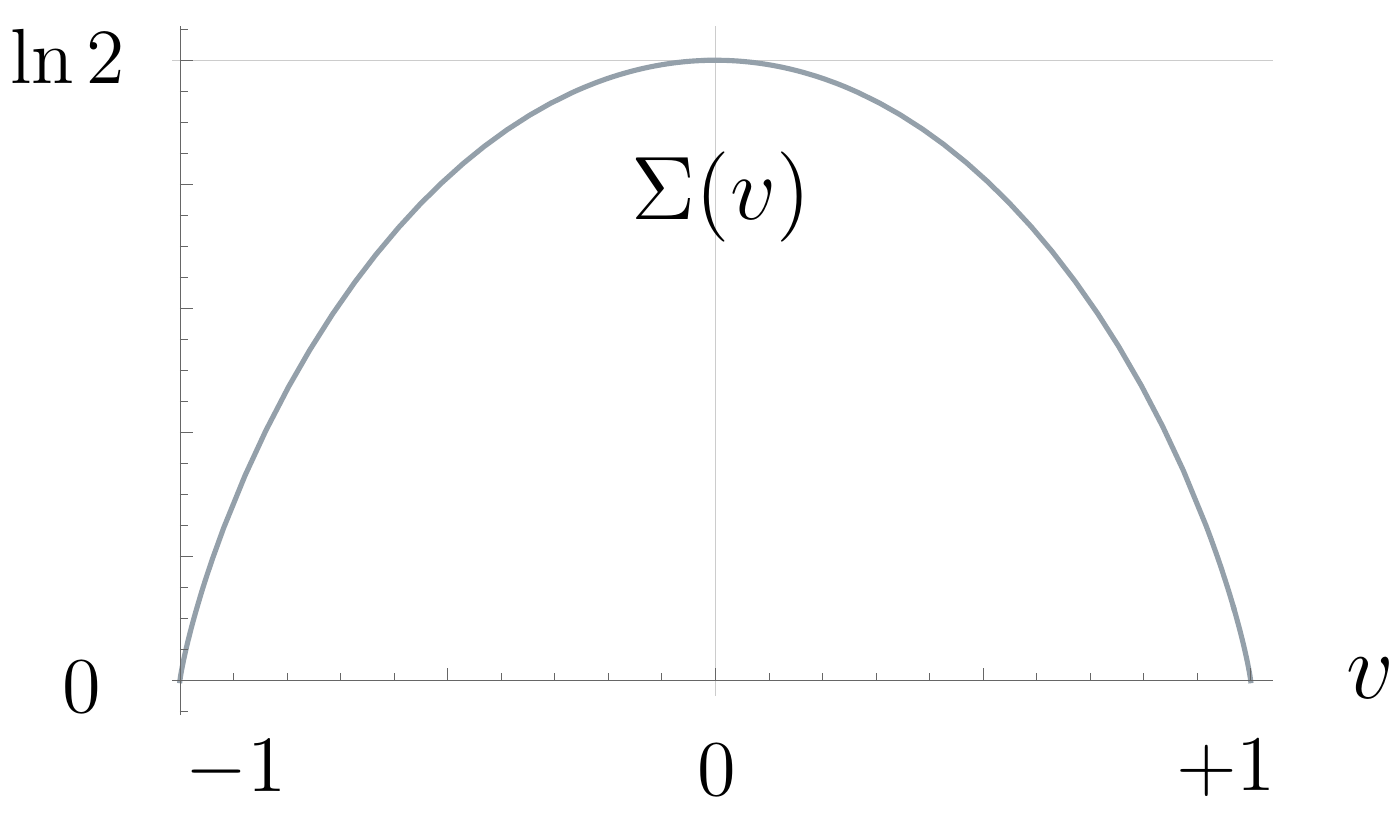}
\caption{$\Sigma (v)$ defined in (\ref{Sigma_v}) is the concave, even function of $v \in [-1,+1]$. It vanishes with $v=\pm 1$ and reaches the maximum value $\ln 2$ at $v=0$.}
\label{fig:Sigma}
\end{center}
\end{figure}

\begin{figure}
\begin{center}
 \includegraphics[width=9cm]{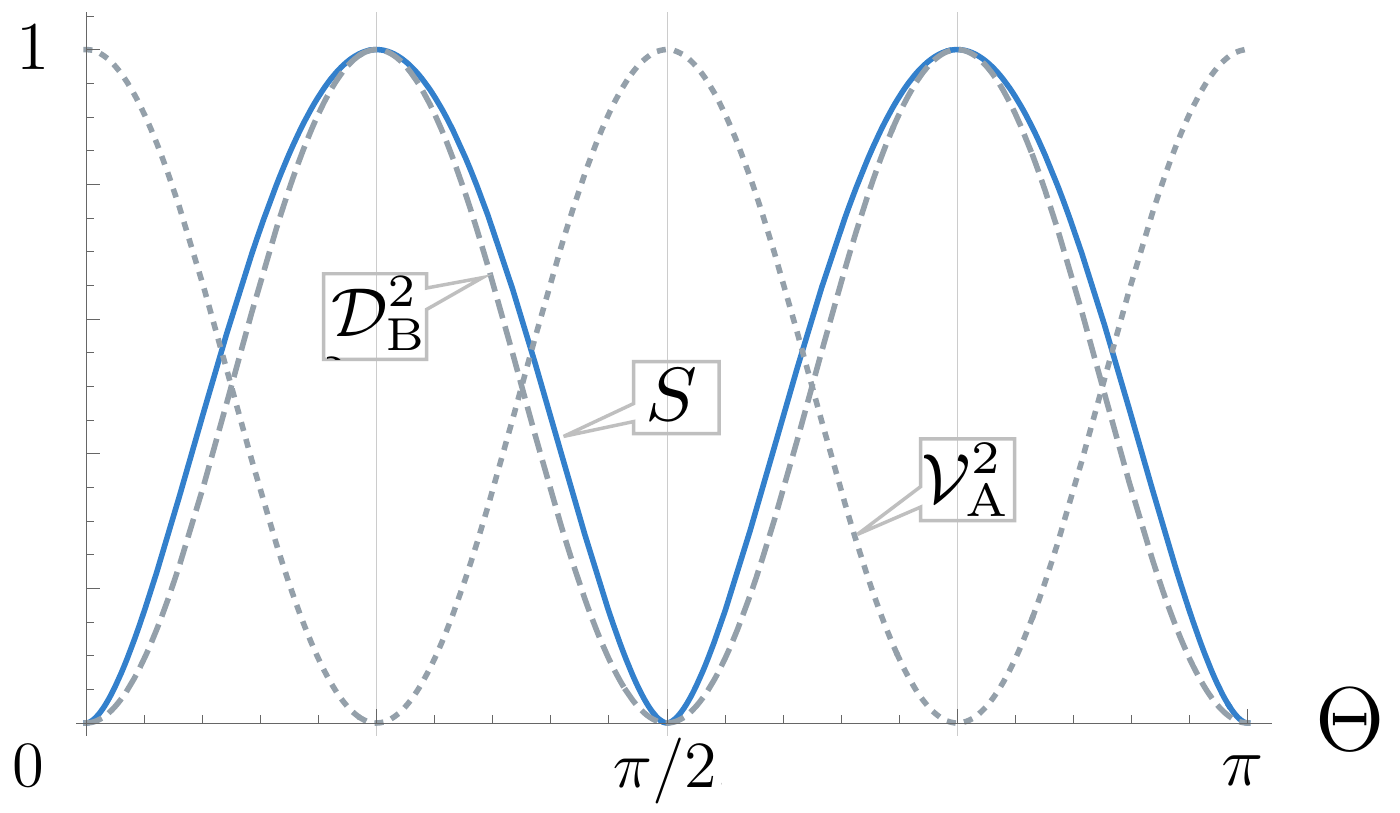}
\caption{Horizontal axis is $\Theta = \bar{J} (t_\tx{off} - t_\tx{on})$, and the dimensionless time interval for which the spin-spin interaction is effective. The solid line depicts the entanglement entropy $S(\hat{\rho}_\tx{A})/\ln 2$, which is positively correlated with the distinguishability $\cl{D}^2_\tx{B}(\Theta)= \La \delta \hat{\sigma}_y^\tx{A} \delta  \hat{\sigma}_z^\tx{B} \Ra^2 = \La \delta  \hat{\sigma}_z^\tx{A} \delta  \hat{\sigma}_y^\tx{B} \Ra^2 = \La \delta \hat{\sigma}_x^\tx{A} \delta  \hat{\sigma}_x^\tx{B} \Ra $ depicted with the dashed line.
On the other hand, the visibility $\cl{V}^2_\tx{A}(\Theta) =  \La \hat{\sigma}_x^\tx{A}  \Ra^2  =  \La  \hat{\sigma}_x^\tx{B}  \Ra^2$ depicted with the dotted-dashed line is negatively correlated with the entropy. }
\label{fig:S_N}
\end{center}
\end{figure}

The entanglement entropy is defined as the von Neumann entropy of a reduced state.
Since the eigenvalues of the reduced density matrix $\hat{\rho}_\tx{A}$ in (\ref{rho_A(B)-Newtonian}) is given by 
\eqn{
\mu_\tx{A}^s  = \frac{1+s \cos (2 \Theta)}{2}  ~~\tx{with}~~s=\pm 1 ~, \label{mu_N}
}
the entanglement entropy is computed as
\eqn{
 S(\hat{\rho}_\tx{A})  &= -\sum_{s=\pm} \mu_\tx{A}^s \ln \mu_\tx{A}^s =\Sigma (\cos (2 \Theta) )~,
\label{S_N}}
where the function
\eqn{
\Sigma (v) \coloneqq  - \frac{1 + v}{2} \ln \frac{1 + v}{2}  - \frac{1 - v}{2} \ln \frac{1 - v}{2}    \label{Sigma_v}
}
monotonically decreases from $\ln 2$ to $0$ when $|v|$ increases from $0$ to $1$, see Fig.~\ref{fig:Sigma}.
%%%%%%%FIG%%%%%%%
%%%%%%%%%%%%%%%%%%
In the Newtonian approximation, the entanglement entropy associated with Bob's spin takes the same value: $S(\hat{\rho}_\tx{B})  =  S(\hat{\rho}_\tx{A})=\Sigma (\cos (2 \Theta) )$.
As depicted in Fig.~\ref{fig:S_N}, the entanglement entropy oscillates with the dimensionless time interval $\Theta$ and it reaches the maximal value $\ln 2$
at $\Theta = \pi/4$ mod $\pi/2$.
This can be understood directly from the expression (\ref{Psi_AB-Newtonian}) at $\Theta = \pi/4$,
\eqn{
\ket{\Psi_\tx{f}}_\tx{AB}|_{\Theta = \pi/4} 
=\frac{1}{2\sqrt{2}}(\ket{+}_\tx{A}+\ket{-}_\tx{A})
(\ket{+}_\tx{B}+\ket{-}_\tx{B})
+\frac{\ri}{2\sqrt{2}}(\ket{+}_\tx{A}-\ket{-}_\tx{A})
(\ket{+}_\tx{B}-\ket{-}_\tx{B})
~, \label{Maximally-entangled}
}
as it is a maximally entangled state.
From the $\Theta$ dependence of the spin expectation values in (\ref{connected-part-Newtonian}),
the entanglement entropy faithfully represents the amount of the spin correlations in the Newtonian approximation. It is because  the dynamical field is absent and no information is transferred away from the subsystem of Alice and Bob.
%%%%%%%FIG%%%%%%%
%%%%%%%%%%%%%%%%%%
%%%%%%%%%%%%%%%%%%%%%%%%%%%%%%%%%%%%%%%%%%%%%%%%%%%%%
\subsection{Visibility and distinguishability \label{sec.Visibility and distinguishability-Newtonian}}
One can also introduce two types of quantities, visibility $\cl{V}$ and distinguishability $\cl{D}$,
to describe the ``wave particle duality'' \cite{PhysRevLett.77.2154}.
Suppose that the interaction described by (\ref{ferromagnetic-interaction}) is turned on for Bob to be able to make a guess on which eigenstate Alice would find $\ket{+}_\tx{A}$ or $\ket{-}_\tx{A}$.
Then, the quantum interference between these two eigenstates should be partially lost and the expectation values of $\hat{\sigma}^\tx{A}_x$ and $\hat{\sigma}^\tx{A}_y$ become smaller than the values that could be if the spin-spin interaction were not turned on.
In this sense, there is a trade-off relation between Alice's visibility of the quantum interference effects and Bob's distinguishability of the state of Alice's spin \cite{Sugiyama:2022wcd}.

\

Alice's visibility of the interference fringe can be quantified by
\eqn{
\cl{V}_\tx{A} \coloneqq & \sqrt{\La \hat{\sigma}^\tx{A}_x \Ra^2 + \La \hat{\sigma}^\tx{A}_y \Ra^2 } = | \La  \hat{\sigma}^\tx{A}_x + \ri \hat{\sigma}^\tx{A}_y  \Ra | = 2| \bra{-}\hat{\rho}_\tx{A} \ket{+}_\tx{A} | ~. \label{define-visibility}
}
Within the Newtonian approximation here, we have the visibility from \eqref{rho_A(B)-Newtonian} as
\eqn{
\cl{V}_\tx{A} = |\cos (2\Theta)| ~, \label{V_N}
}
which is nothing but the absolute value of 
$\La \hat{\sigma}_x^\tx{A}  \Ra $ in (\ref{expectation-value-Newtonian}).
As seen in Fig.~\ref{fig:S_N}, it is negatively correlated with the entanglement entropy (\ref{S_N}), which is simply because the entanglement entropy is now written as a function of the visibility:
\eqn{
S(\hat{\rho}_\tx{A}) = \Sigma \qty(\cl{V}_\tx{A}(\Theta) ) ~, \label{S_N<->Visibility}
}
where $\Sigma (v)$ is the monotonically decreasing function of $v$ for $v \geq 0$ defined in (\ref{Sigma_v}).
That is, the more entangled the two spins are, the weaker the quantum interference effects are.\footnote{\label{asymmetric-interferometer}
For an ``asymmetric interferometer''\cite{PhysRevLett.77.2154} with the predictability $\cl{P}_\tx{A} \coloneqq |\bra{\Psi_\tx{i}} \hat{\sigma}^\tx{A}_z \ket{\Psi_\tx{i}}| \ne 0$,
the entanglement entropy is given by $S(\hat{\rho}_\tx{A}) = \Sigma ( \sqrt{\cl{V}_\tx{A}^2 + \cl{P}_\tx{A}^2 })$, and then, the entropy is again the monotonically decreasing function of the visibility $\cl{V}_\tx{A}$.}

\

In order to define the distinguishability, let us first introduce Bob's density matrix under the condition that Alice's spin is found to be in the eigenstate of $\hat{\sigma}_z^\tx{A}$ with the eigenvalue $\pm1$:
\eqn{
\hat{\rho}_\tx{B}^{\pm} = 2\bra{\pm}\hat{\rho}_\tx{AB} \ket{\pm}_\tx{A} ~. \label{define-rho^pm}
}
Then, for the initial state given by (\ref{spin-initial-state}) with the vanishing expectation value of $\hat{\sigma}^\tx{A}_z$,
Bob's  distinguishability of Alice's  $z$-spin can be quantified by the trace distance between $\hat{\rho}_\tx{B}^{+}$ and $\hat{\rho}_\tx{B}^{-}$:
\eqn{
\cl{D}_\tx{B} = \frac{1}{2} \tr_\tx{B} |\hat{\rho}_\tx{B}^{+} - \hat{\rho}_\tx{B}^{-}| ~, \label{define-distinguishability}
}
where $|\,\cdot\,|$ is defined as $|\hat{\cl{O}}|\coloneqq\sqrt{\hat{\cl{O}}^\dag \hat{\cl{O}}}$ for an operator $\hat{\cl{O}}$~\cite{10.5555/1972505}.
The trace distance is a quantity that represents how close two states are. If the distance is large, two states can be distinguished.
In the special case of one-qubit that is the case of our interest,
when $\hat{\rho}_\tx{B}^{\pm}$ are represented 
in the Bloch representation,
\begin{equation}
    \hat{\rho}_\tx{B}^{\pm} =\frac{1}{2}(\hat{1}^\tx{B}+\vb*{r}_\pm\cdot\hat{\vb*{\sigma}}^\tx{B}),
    \label{eq:Bloch_one_qubit}
\end{equation}
where $\hat{\vb*{\sigma}}^\tx{B}=(\hat{\sigma}^\tx{B}_x,\hat{\sigma}^\tx{B}_y,\hat{\sigma}^\tx{B}_z)$, the trace distance is equal to the half of the Euclidean distance 
\begin{equation}
    \cl{D}_\tx{B}=\frac{1}{2}|\vb*{r}_+-\vb*{r}_-|
    \label{eq:trace_distance_one_qubit}
\end{equation}
 on the Bloch sphere.

Within the Newtonian approximation, (\ref{define-rho^pm}) turns out to be
\eqn{
\hat{\rho}_\tx{B}^{\pm}  =
\frac{1}{2} \qty{\hat{1}^\tx{B}  +  \cos (2 \Theta)  \hat{\sigma}_x^\tx{B}   \mp \sin (2 \Theta) \hat{\sigma}_y^\tx{B}  } 
 ~,
}
from \eqref{rho_AB-Newtonian}. 
Thus, we have $\vb*{r}_\pm=(0,\cos(2\Theta),\mp\sin(2\Theta))$ and the distinguishability is
\eqn{
\cl{D}_\tx{B} =\frac{1}{2}|\vb*{r}_+-\vb*{r}_-|
= |\sin (2 \Theta)|  ~. \label{D_N}
}
This is equal to the absolute value of $\La \delta \hat{\sigma}_y^\tx{A} \delta  \hat{\sigma}_z^\tx{B} \Ra = \La \delta  \hat{\sigma}_z^\tx{A} \delta  \hat{\sigma}_y^\tx{B} \Ra$ and $\sqrt{\La \delta \hat{\sigma}_x^\tx{A} \delta  \hat{\sigma}_x^\tx{B} \Ra}$ in (\ref{connected-part-Newtonian}), and also, positively correlated with the entanglement entropy (\ref{S_N}), see Fig.~\ref{fig:S_N}, since (\ref{S_N}) can be written as
\eqn{
S(\hat{\rho}_\tx{A}) = \Sigma \qty([1-\cl{D}_\tx{B}^2(\Theta)]^{1/2}) ~.
\label{S_N<->Distinguishability}}
The entanglement between the two spins makes it possible for Bob to tell the direction of Alice's spin by observing Bob's own spins.

Note that the visibility and distinguishability satisfy the trade-off relation
\eqn{
\cl{V}_\tx{A}^2 + \cl{D}_\tx{B}^2  = 1, \label{wave-particle-duality-Newtonian}
}
which implies $\hat{\rho}_\tx{B}^{\pm}$ are pure states~\cite{PhysRevLett.77.2154}. In fact, they are expressed by
\eqn{
\hat{\rho}_\tx{B}^{\pm}
=\ket{\pm \Theta}_\tx{B}\! \bra{\pm \Theta} ~,
}
where the state is nothing but the one that has evolved with the Hamiltonian $\bra{\pm_z}\hat{H}_\tx{AB} \ket{\pm_z}_\tx{A}$:
\eqn{
\ket{\pm \Theta}_\tx{B} = e^{\pm \ri \Theta \hat{\sigma}_z^\tx{B}}\frac{1}{\sqrt{2}}( \ket{+}_\tx{B}+\ket{-}_\tx{B} )
= \frac{1}{\sqrt{2}}( e^{\ri\Theta}\ket{+}_\tx{B}+e^{-\ri\Theta}\ket{-}_\tx{B} )
~. \label{Theta_A(B)}
}

As seen in Section \ref{sec.Visibility and distinguishability}, $\hat{\rho}_\tx{B}^{\pm}$ are no longer pure states if the field that creates potential between Alice and Bob is dynamical, and 
\eqref{wave-particle-duality-Newtonian} is to be replaced by inequalities. 

%%%%%%%%%%%%%%%%%%%%%%%%%%%%%%%%%%%
\subsection{Separability condition and negativity \label{sec.Separability condition1}}
Whether quantum entanglement has been generated by the time evolution can be judged by whether the density matrix is separable.
When the density matrix, which can be in a mixed or pure state, is written as a mixture of product states as
\eqn{
\hat{\rho}_\tx{AB} = \sum_i p_i \hat{\rho}^{(i)}_\tx{A} \hat{\rho}^{(i)}_\tx{B} \label{def-separable}
}
with the probabilities $p_i > 0$ satisfying $\sum_i p_i = 1$, the state is said to be separable.
Such a state can be prepared by local operations and classical communication (LOCC) \cite{Horodecki:2009zz}; in this sense, spin correlations existing in a separable state are regarded as classical correlations.

For the two-qubit system, the necessary and sufficient condition for a density matrix to be separable is the positivity of the partial transposition \cite{PhysRevLett.77.1413,HORODECKI19961}.
As a measure of nonseparability or 
the quantum entanglement between the two spins,
we take the entanglement negativity, summation of the absolute values of negative ones in \eqref{PT-mu_AB_Newtonian}:
\eqn{
\cl{N} \coloneqq \sum_{n} \theta (-\tilde{\mu}_n) ~ |\tilde{\mu}_n| ~, \label{entanglement-negativity}
}
where $\tilde{\mu}_n$ are eigenvalues of the transposed density matrix.
Note that if $\hat{\rho}_\tx{AB}$ is a pure state, $\cl{N}\neq0$ is equivalent to $S(\hat{\rho}_\tx{A})\neq0$.

Let us evaluate the negativity in the Newtonian approximation.
Noting the transposition of Pauli matrices,
$(\hat{1}^T,\hat{\sigma}_x^T,\hat{\sigma}_y^T,\hat{\sigma}_z^T) = (\hat{1},\hat{\sigma}_x,-\hat{\sigma}_y,\hat{\sigma}_z)$,
the partial transposition (acting on Bob's spin) of the density matrix \eqref{rho_AB-Newtonian} is obtained as
\eqn{
\hat{{\rho}}^{T_\tx{B}}_\tx{AB} =& \ket{\Psi_\tx{f}}_\tx{AB}\!\bra{\Psi_\tx{f}}  \\
&= \frac{1}{4} \qty{ \hat{1}^\tx{A} \hat{1}^\tx{B} + \hat{\sigma}_x^\tx{A} \hat{\sigma}_x^\tx{B} +  \cos (2 \Theta) \qty[ \hat{\sigma}_x^\tx{A} \hat{1}^\tx{B} +  \hat{1}^\tx{A}\hat{\sigma}_x^\tx{B}  ]  - \sin (2 \Theta) \qty[ \hat{\sigma}_y^\tx{A} \hat{\sigma}_z^\tx{B} - \hat{\sigma}_z^\tx{A} \hat{\sigma}_y^\tx{B}  ]  } ~.
}
From this, we obtain the eigenvalues of the partial transposition:
\eqn{
\tilde{\mu}_\tx{AB}^{s_1 s_2}
&= \frac{1}{4} \biggl\{ 1+s_2 
+ s_1 \sqrt{2 + 2 s_2 \cos (4\Theta)} \biggr\}
~. \label{PT-mu_AB_Newtonian}
}
There is one negative eigenvalue, $\tilde{\mu}_\tx{AB}^{--}=-|\sin(2\Theta)|/2$ for $\Theta \neq 0 $ mod $\pi/2$,
and thus, the negativity is 
\begin{equation}
    \cl{N} =\frac{1}{2}|\sin(2\Theta)|.
    \label{eq:negativity_Newtonian}
\end{equation}
Therefore, the Hamiltonian~\eqref{ferromagnetic-interaction} in the Newtonian approximation generates the quantum entanglement between Bob and Alice, except $\Theta= 0 $ mod $\pi/2$, where the state is recursed to the initial one,
\eqn{\hat{\rho}_\tx{AB} = \frac{\hat{1}^\tx{A} + \hat{\sigma}_x^\tx{A} }{2} \frac{\hat{1}^\tx{B} + \hat{\sigma}_x^\tx{B} }{2} 
=\ket{\Psi_\tx{i}}\bra{\Psi_\tx{i}}
~.  \label{product-state0}}
As we will see below, when the field is dynamical, the negativity can vanish for various reasons discussed in Secs.~\ref{sec.Non-adiabatic limit}, \ref{sec.Spacelike-separated} and \ref{sec.One-way}, even though the entanglement entropy is nonvanishing.

%%%%%%%%%%%%%%%%%%%%%%%%%%%%%%%%%%%%%%%%%%%%%%%%%%%%%%%%%%%%%%
\section{Relativistic treatment with a dynamical field \label{sec.Relativistic}}
Hereafter, we go back to the original model described by the Hamiltonian (\ref{Hamiltonian}) with the initial state (\ref{initial-state}), and study its time evolution without any approximation.
With the dynamics of the field taken into account, the resultant final state turns out to be different from the one obtained in the previous section: it respects the causality expressed by the retarded Green's function of the field; and also describes particle creation and decoherence of the quantum state of the spins expressed by the Keldysh Green's function. 

%%%%%%%FIG%%%%%%%
\begin{figure}
\begin{center}
 \includegraphics[width=13cm]{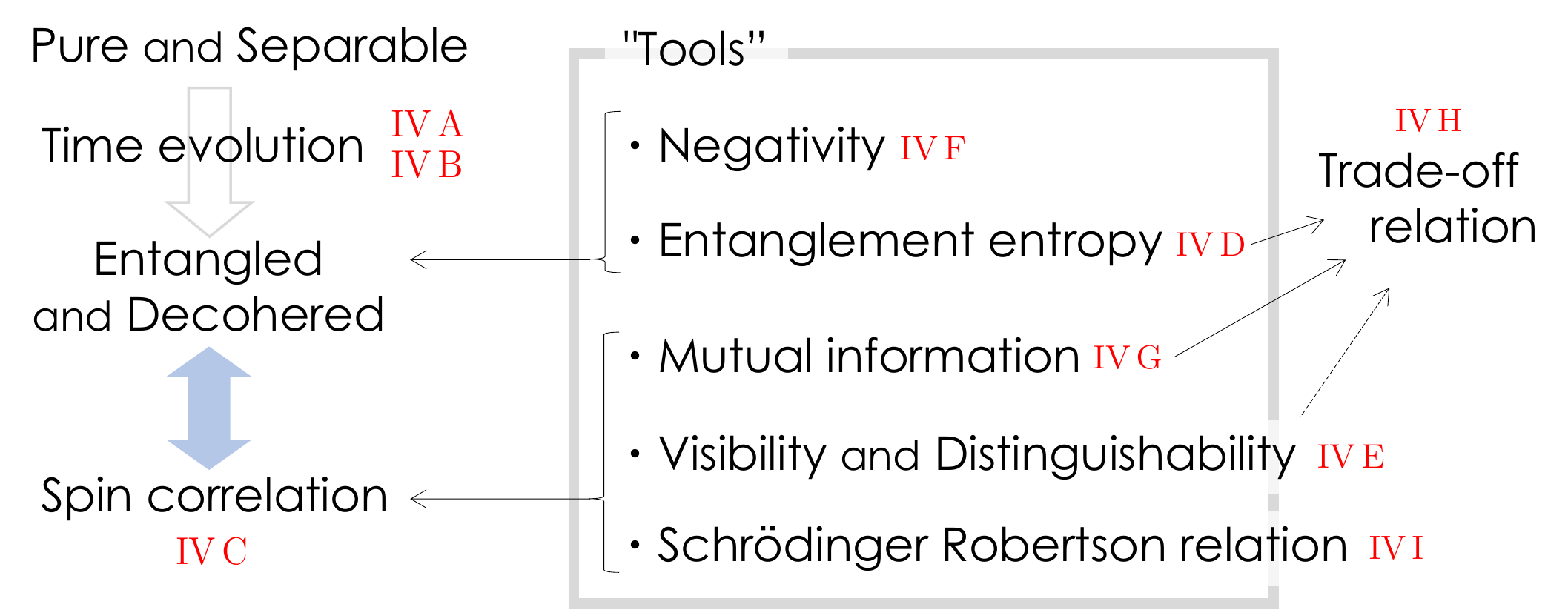}
\caption{In Sec.~\ref{sec.Relativistic}, we compute the reduced density matrix of the final state of the spins, and then, introduce various ``tools'' in each subsection to understand its entanglement structure as well as the induced spin-spin correlations. 
Also, we discuss trade-off relations to be satisfied with the spin-field correlations taken into account.
The red letters indicate the corresponding subsections.
Armed with these tools, we discuss the physical consequences of the spin-spin correlations in various limiting cases in Sec.~\ref{sec.limiting cases}.
}
\label{fig:sec4}
\end{center}
\end{figure}
%%%%%%%%%%%%%%%%%%

\

In this section, we expand general discussions including some ``tools'' to understand the quantum state after the time evolution, see Fig.~\ref{fig:sec4}; one may skip this section to Sec.~\ref{sec.limiting cases} to see the physical consequences of the dynamical field in limiting cases and come back to this section to check some definitions or notations.

In Secs.~\ref{sec.Reduced density matrix} and \ref{Results of the Schwinger-Keldish effective action}, we first obtain the density matrix of the reduced system, composed of the two spins, realized after the time evolution with the interaction between the spins and the dynamical field.
In Sec.~\ref{sec.Spin correlation}, we calculate the spin correlations to see how they are different from the ones obtained in the Newtonian limit due to the causality and the decoherence.
In Sec.~\ref{sec.Entanglement entropy}, we see some general behaviors of the entanglement entropies in our setup.
In Sec.~\ref{sec.Visibility and distinguishability}, the visibility and the distinguishability are obtained and discussed from the viewpoint of the relation with the entropies.
In Sec.~\ref{sec.Separability condition},  we discuss a sufficient condition for the density matrix of the reduced system of the two spins to be separable in our setup using the entanglement negativity.
In Sec.~\ref{sec.Mutual information}, various mutual informations are constructed from the entropies to quantify the amount of the spin-spin and spin-field correlations.
In Sec.~\ref{sec.Trade-off relation}, we discuss
trade-off relations to be satisfied by the mutual informations and their similarity to and difference from inequalities satisfied by the visibility and the distinguishability.

In addition, in order to approach various limiting cases considered in Sec.~\ref{sec.limiting cases}, we discuss one more thing in this section.
In Sec.~\ref{sec.RS inequalities}, inequalities to be satisfied by the propagators are derived from the Robertson-Schr\"{o}dinger uncertainty relation.

%%%%%%%%%%%%%%%%%%%%%%%%%%%%%%%%%%%%%%%
\subsection{Reduced density matrix of the final state \label{sec.Reduced density matrix}}
We do not observe the scalar field directly.
Then, it is sufficient to know the reduced density matrix obtained by tracing out the scalar field from the total density matrix constructed from (\ref{final-state}):
\eqn{
\hat{\rho}_\tx{AB} = \tr_\phi \{ \ket{\Psi_\tx{f}}\bra{\Psi_\tx{f}}  \} = \tr_\phi \{ \hat{U}(t_\tx{f} ,t_\tx{i})  \ket{\Psi_\tx{i}}\bra{\Psi_\tx{i}}  \hat{U}(t_\tx{i} ,t_\tx{f})  \} ~, \label{rho_AB-def}
}
which is a mixed state unlike (\ref{rho_AB-Newtonian}) in general.
For spins, the Hamiltonian only depends on $\hat{\sigma}_z^{\tx{A}}$ and $\hat{\sigma}_z^{\tx{B}}$, i.e., the time evolution operator can be expanded by the $z$-spin basis~(\ref{z-spin eigenstate}), in which each component of (\ref{rho_AB-def}) can be written as
\eqn{
\bra{\sigma_1^\tx{A} \sigma_1^\tx{B}} \hat{\rho}_\tx{AB} \ket{\sigma_2^\tx{A} \sigma_2^\tx{B}} = \frac{1}{4} \times Z[J_1,J_2] %~~\tx{with}~~ \sigma_1^\tx{A,B}, \sigma_2^\tx{A,B} = \pm 
~. \label{rho_AB-z-basis}
}
Here, $\sigma^\tx{A,B}_{1}, \sigma^\tx{A,B}_{2} = \pm 1$ and the prefactor comes from the initial state (\ref{spin-initial-state}) as
\eqn{
\bra{\sigma^\tx{A}_{1} \sigma^\tx{B}_{1}} \ket{\Psi_\tx{i}}_\tx{AB}\! \bra{\Psi_\tx{i}} \ket{\sigma^\tx{A}_{2} \sigma^\tx{B}_2}  = 1/4
}
and
\eqn{ Z[J_1,J_2] = \bra{\Omega}  [\hat{U}_\phi (t_\tx{f} , t_\tx{i} ; J_2)]^\dag  \hat{U}_\phi (t_\tx{f} ,t_\tx{i}; J_1)  \ket{\Omega}_\phi \label{Z}
}
with the time evolution operator 
\eqn{
\hat{U}_\phi (t_\tx{f} ,t_\tx{i}; J_i) : & = \bra{\sigma^\tx{A}_{i} \sigma^\tx{B}_{i}}  \hat{U}(t_\tx{f} ,t_\tx{i}) \ket{\sigma^\tx{A}_{i} \sigma^\tx{B}_{i}} \\
&= \tx{T} \exp \qty[ -\ri \int_{t_\tx{i}}^{t_\tx{f}} \dd t  \qty{  \hat{H}_\phi - \int \dd^3 x   J_i (t,\vb*{x}) \hat{\phi}(\vb*{x}) }  ] ~. \label{time-evolution-op}
}
The ``source'' for the scalar field 
\eqn{J_i (t,\vb*{x}) \coloneqq \sigma_{i}^\tx{A}  \lambda_\tx{A}(t)  \delta^{(3)} (\vb*{x} - \vb*{x}_\tx{A}) + \sigma_{i}^\tx{B}  \lambda_\tx{B}(t) \delta^{(3)} (\vb*{x} - \vb*{x}_\tx{B})  \label{source}
}
is different in 
$\hat{U}_\phi (t_\tx{f} , t_\tx{i} ; J_2)$ and $\hat{U}_\phi (t_\tx{f} , t_\tx{i} ; J_1)$
and therefore $Z[J_1,J_2]$ is not trivial.

\subsection{Results of the Schwinger-Keldysh effective action
\label{Results of the Schwinger-Keldish effective action}}
Since the field's configuration is traced out, it is convenient to use the path-integral formulation \cite{Hidaka:2022tzk} on the closed time path \cite{Keldysh:1964ud}, Schwinger-Keldysh formalism.
The Hamiltonian $\hat{H}_\phi$ is the free-field part in (\ref{Hamiltonian}), and hence, (\ref{Z}) is easily computed as a Gaussian functional integral.
Putting details of the computation in Appendix \ref{app.Reduced density matrix},
let us present the result;
\eqn{
Z[J_1 ,J_2] =&  \exp \biggl\{ \ri \qty( \sigma^\tx{B}_\tx{a} \mf{G}^\tx{BA}_\tx{R} \sigma^\tx{A}_\tx{r} + \sigma^\tx{A}_\tx{a} \mf{G}^\tx{AB}_\tx{R} \sigma^\tx{B}_\tx{r} ) \\
& ~~~~~~~ - \frac{1}{2} \qty( \sigma^\tx{A}_\tx{a} \mf{G}^\tx{AA}_\tx{K} \sigma^\tx{A}_\tx{a} + \sigma^\tx{B}_\tx{a} \mf{G}^\tx{BB}_\tx{K} \sigma^\tx{B}_\tx{a} +2 \sigma^\tx{B}_\tx{a} \mf{G}^\tx{BA}_\tx{K} \sigma^\tx{A}_\tx{a} )  \biggr\}  ~,
\label{CTPresults}
}
where $\sigma^\tx{A,B}_\tx{r} \coloneqq (\sigma^\tx{A,B}_\tx{1} + \sigma^\tx{A,B}_\tx{2})/2$ and $\sigma^\tx{A,B}_\tx{a} \coloneqq \sigma^\tx{A,B}_\tx{1} - \sigma^\tx{A,B}_\tx{2}$ have been introduced for convenience.

In the first line of (\ref{CTPresults}), we have real quantities,
\eqn{
  \mf{G}_\tx{R}^\tx{BA} \coloneqq& \int_{t_\tx{i}}^{t_\tx{f}} \dd{t}\int_{t_\tx{i}}^{t_\tx{f}} \dd{t'}\lambda_\tx{B}(t) G_\tx{R}(t-t',\vb*{x}_\tx{B}-\vb*{x}_\tx{A}) \lambda_\tx{A}(t') ~,  \\
  \mf{G}_\tx{R}^\tx{AB} \coloneqq& \int_{t_\tx{i}}^{t_\tx{f}} \dd{t}\int_{t_\tx{i}}^{t_\tx{f}}  \dd{t'}\lambda_\tx{A}(t) G_\tx{R}(t-t',\vb*{x}_\tx{A}-\vb*{x}_\tx{B}) \lambda_\tx{B}(t')  ~,  \label{mathfrak-G_R}
  }
which connect Alice's spin and Bob's spin in a causal way with the retarded/advanced Green's function of the source-free\footnote{\label{source-free}
Here, the Heisenberg operator $\hat{\phi}(t,\vb*{x}) \coloneqq \hat{U}(t_\tx{i}, t ; 0) \hat{\phi}(\vb*{x}) \hat{U}(t ,t_\tx{i} ; 0 )$ is introduced with the interaction with the spins turned off. Since the scalar field is originally free apart from the interaction with the spins, one may regard $\hat{\phi}(t,\vb*{x})$ as the interaction picture operator.
It satisfies free-field's equation of motion: $(\partial^2 -m^2) \hat{\phi}(t,\vb*{x}) = 0$.
Note that, because of this definition, the one-point function vanishes: $\bra{\Omega} \hat{\phi}(t,\vb*{x}) \ket{\Omega}_\phi = 0$.
}
real scalar field. The retarded Green's function
\eqn{G_\tx{R}(t-t',\vb*{x}-\vb*{x}')=G_\tx{A}(t'-t,\vb*{x}'-\vb*{x}) = \ri\theta(t-t')\bra{\Omega}[\hat{\phi}(t,\vb*{x}),\hat{\phi}(t',\vb*{x}')]\ket{\Omega}_\phi ~, \label{Retarded}
}
 satisfies 
\eqn{ (\partial^2 -m^2) G_\tx{R(A)}(t-t',\vb*{x}-\vb*{x}') = - \delta^{(4)}(x-x'). } 
 Hence, $\mf{G}_\tx{R}^\tx{BA}$ is nothing but the time integration of the retarded potential that Alice's spin induces at Bob's position.
Similarly, $\mf{G}_\tx{R}^\tx{AB}$ is the retarded potential that Bob's spin induces at Alice's position.

In the second line of (\ref{CTPresults}),
\eqn{
  \mf{G}_\tx{K}^\tx{AA} \coloneqq&  \int_{t_\tx{i}}^{t_\tx{f}} \dd{t}\int_{t_\tx{i}}^{t_\tx{f}} \dd{t'}\lambda_\tx{A}(t) G_\tx{K}(t-t',\vb*{0})\lambda_\tx{A}(t') ~, \\
  \mf{G}_\tx{K}^\tx{BB} \coloneqq& \int_{t_\tx{i}}^{t_\tx{f}} \dd{t}\int_{t_\tx{i}}^{t_\tx{f}} \dd{t'}\lambda_\tx{B}(t) G_\tx{K}(t-t',\vb*{0})\lambda_\tx{B}(t') ~, \\
\mf{G}_\tx{K}^\tx{AB} =   \mf{G}_\tx{K}^\tx{BA} \coloneqq& \int_{t_\tx{i}}^{t_\tx{f}} \dd{t}\int_{t_\tx{i}}^{t_\tx{f}} \dd{t'}\lambda_\tx{B}(t) G_\tx{K}(t-t',\vb*{x}_\tx{B}-\vb*{x}_\tx{A})\lambda_\tx{A}(t') \label{mathfrak-G_K}
}
are the contributions of the vacuum fluctuations described by the Keldysh Green's function,
\eqn{
 G_\tx{K}(t-t',\vb*{x}-\vb*{x}') \coloneqq\frac{1}{2}\bra{\Omega} \{\hat{\phi}(t,\vb*{x}),\hat{\phi}(t',\vb*{x}')\}
  \ket{\Omega}_\phi  ~ \label{Keldysh}
}
where $\{\cdot,\cdot\}$ represents the anti-commutator.
The behavior of the retarded and Keldysh Green's functions are depicted in Fig.~\ref{fig:Gs-Wald}.
Here, note that we have the following inequalities
\eqn{
& \mf{G}^\tx{AA}_\tx{K} =  \left\| \int \dd t \lambda_\tx{A}(t) \hat{\phi}(t,\vb*{x}_\tx{A}) \ket{\Omega}_{\phi} \right\|^2 \geq 0 ~ ,~~~ \\
& \mf{G}^\tx{BB}_\tx{K} =  \left\| \int \dd t \lambda_\tx{B}(t) \hat{\phi}(t,\vb*{x}_\tx{B}) \ket{\Omega}_{\phi} \right\|^2 \geq 0~ , \label{G_K-positivity-1}
}
\eqn{
\mf{G}^\tx{AA}_\tx{K} + \mf{G}^\tx{BB}_\tx{K} \pm 2  \mf{G}^\tx{BA}_\tx{K} 
=  \left\| \int \dd t \qty[ \lambda_\tx{A}(t) \hat{\phi}(t,\vb*{x}_\tx{A})\pm \lambda_\tx{B}(t) \hat{\phi}(t,\vb*{x}_\tx{B}) ] \ket{\Omega}_{\phi} \right\|^2 \geq 0 ~. \label{G_K-positivity-2}
}
As computed in Appendix \ref{app.Particle creation and Keldysh function}, in our model without nonlinear interactions,
these quantities are related to the number of particles $N$ created due to the nonadiabaticity of Alice's and Bob's protocols:
\eqn{
\bra{\Psi_\tx{f}} \hat{N} \ket{\Psi_\tx{f}} &=  \mf{G}_\tx{K}^\tx{AA} + \mf{G}_\tx{K}^\tx{BB} ~, \label{number-of-particles-no-projection} 
}
\eqn{
\bra{\Psi_\tx{f}} \! :\! \! \hat{N}^2 \! \! :\! \ket{\Psi_\tx{f}}-\bra{\Psi_\tx{f}} \hat{N}  \ket{\Psi_\tx{f}}^2 &= 4 (\mf{G}_\tx{K}^\tx{AB})^2 ~,
}
where $:\!\!\hat{N}\! \!:$ represents the normal ordering of $\hat{N}$.
Especially, (\ref{G_K-positivity-2}) corresponds to the number of particles created when Alice and Bob observe the $z$-spin eigenstates with their respective eigenvalues $\sigma^\tx{A}$ and $\sigma^\tx{B}$:
\eqn{
\bra{\Psi_\tx{f}}\qty{ \ket{\sigma^\tx{A} \sigma^\tx{B} } \hat{N} \bra{\sigma^\tx{A} \sigma^\tx{B}} }\ket{\Psi_\tx{f}} = \mf{G}^\tx{AA}_\tx{K} + \mf{G}^\tx{BB}_\tx{K} + \sigma^\tx{A} \sigma^\tx{B} 2  \mf{G}^\tx{BA}_\tx{K} ~. \label{number-of-particles-produced}
}

\subsection{Reduced density matrix and spin correlation \label{sec.Spin correlation}}
We can now calculate the reduced density matrix of Alice and Bob, and the spin correlation functions by using the results in the previous section. 
The density matrix (\ref{rho_AB-z-basis}) can be written in the Bloch representation as
\eqn{
\hat{\rho}_\tx{AB} =& \frac{1}{4} \Bigl\{ \hat{1}^\tx{A} \hat{1}^\tx{B}
+ C_{xx} \hat{\sigma}_x^\tx{A} \hat{\sigma}_x^\tx{B} 
+ C_{x0} \hat{\sigma}_x^\tx{A} \hat{1}^\tx{B} 
+ C_{0x}  \hat{1}^\tx{A}\hat{\sigma}_x^\tx{B}  \\
&~~~~~~~~~~~~~~~~~~~+ C_{yy} \hat{\sigma}_y^\tx{A} \hat{\sigma}_y^\tx{B}  
+ C_{yz} \hat{\sigma}_y^\tx{A} \hat{\sigma}_z^\tx{B} 
+ C_{zy} \hat{\sigma}_z^\tx{A} \hat{\sigma}_y^\tx{B}    \Bigr\} ~, \label{rho_AB}
}
and it is further reduced to
\eqn{
\hat{\rho}_\tx{A} = \tr_\tx{B}\{ \hat{\rho}_\tx{AB}  \} = \frac{1}{2} \qty{ \hat{1}^\tx{A}  + C_{x0} \hat{\sigma}_x^\tx{A}  } ~, \label{rho_A}
}
\eqn{
\hat{\rho}_\tx{B} = \tr_\tx{A}\{ \hat{\rho}_\tx{AB}  \} = \frac{1}{2} \qty{ \hat{1}^\tx{B}  + C_{0x} \hat{\sigma}_x^\tx{B}  } ~, \label{rho_B}
}
where the coefficients are obtained by computing the expectation values of the spin operators,
\aln{
C_{xx} &\coloneqq \La \hat{\sigma}^\tx{A}_x \hat{\sigma}^\tx{B}_x \Ra 
= \gamma_\tx{A} \gamma_\tx{B}  \cosh (4 \mf{G}^\tx{BA}_\tx{K}) ~,~~ 
&C_{yy} &\coloneqq \La \hat{\sigma}^\tx{A}_y \hat{\sigma}^\tx{B}_y \Ra 
=  \gamma_\tx{A} \gamma_\tx{B}  \sinh (4 \mf{G}^\tx{BA}_\tx{K}) ~, \notag \\
C_{x0} &\coloneqq \La \hat{\sigma}^\tx{A}_x \Ra 
=  \gamma_\tx{A}  \cos ( 2 \mf{G}^\tx{AB}_\tx{R} )~,~~
&C_{0x} &\coloneqq \La \hat{\sigma}^\tx{B}_x \Ra 
=  \gamma_\tx{B}  \cos ( 2 \mf{G}^\tx{BA}_\tx{R} ) ~,   \label{Cs} \\
C_{yz} &\coloneqq\La \hat{\sigma}^\tx{A}_y \hat{\sigma}^\tx{B}_z \Ra 
= - \gamma_\tx{A}   \sin ( 2 \mf{G}^\tx{AB}_\tx{R} ) ~,~~ 
&C_{zy}&\coloneqq\La \hat{\sigma}^\tx{A}_z \hat{\sigma}^\tx{B}_y \Ra 
= - \gamma_\tx{B}   \sin ( 2 \mf{G}^\tx{BA}_\tx{R} )~. \notag
}
Their overall amplitudes  are suppressed by two quantities
\begin{align}
\gamma_\tx{A} =   \exp (-2 \mf{G}^\tx{AA}_\tx{K} )  ~,~~ \gamma_\tx{B} =   \exp (-2 \mf{G}^\tx{BB}_\tx{K} )  \label{gamma_A,B}
\end{align}
representing the nonadiabaticity of Alice's and Bob's protocols. 
Observe that, because of $\mf{G}^\tx{AA}_\tx{K}, \mf{G}^\tx{BB}_\tx{K}, \mf{G}^\tx{BA}_\tx{K} \ne 0$ and $\mf{G}^\tx{BA}_\tx{R} \ne \mf{G}^\tx{AB}_\tx{R}$ in general, the reduced density matrices deviate from those in the Newtonian approximation in (\ref{rho_AB-Newtonian}) and (\ref{rho_A(B)-Newtonian}).
Note that the inequalities in (\ref{G_K-positivity-1}) guarantee $0 \leq \gamma_\tx{A}, \gamma_\tx{B} \leq 1$.
With (\ref{G_K-positivity-2}), it is also guaranteed that $0\leq C_{xx} \leq 1$ and $0\leq  |C_{yy} | \leq 1$.

From (\ref{Cs}), it is straightforward to see that the correlations between the two spins are obtained as
\eqn{
&\La \delta \hat{\sigma}_x^\tx{A} \delta \hat{\sigma}_x^\tx{B} \Ra = \gamma_\tx{A} \gamma_\tx{B} \qty[ \cosh (4 \mf{G}^\tx{BA}_\tx{K}) - \cos (2 \mf{G}^\tx{BA}_\tx{R}) \cos (2 \mf{G}^\tx{AB}_\tx{R})]  ~, \\
&\La \delta \hat{\sigma}_y^\tx{A} \delta \hat{\sigma}_y^\tx{B} \Ra =\gamma_\tx{A} \gamma_\tx{B} \sinh (4 \mf{G}^\tx{BA}_\tx{K})  ~,  \\
&\La \delta \hat{\sigma}_y^\tx{A} \delta \hat{\sigma}_z^\tx{B} \Ra= - \gamma_\tx{A}   \sin ( 2 \mf{G}^\tx{AB}_\tx{R} )  ~,  \\
&\La \delta \hat{\sigma}_z^\tx{A} \delta \hat{\sigma}_y^\tx{B} \Ra = - \gamma_\tx{B}   \sin ( 2 \mf{G}^\tx{BA}_\tx{R} ) ~,
\label{connected-part}}
and all the others vanish.
The situation here can be compared with (\ref{connected-part-Newtonian}) in the Newtonian approximation where all the nonvanishing correlation functions depend only on the single parameter $\Theta$.

%%%%%%%%%%%%%%%%%%%%%%%%%%%%%%%%

%%%%%%%%%%%%%%%%%%%%%%%%%%%%%%%%%%%%%
\subsection{Entanglement entropy \label{sec.Entanglement entropy}}
From the reduced density matrix and the spin correlations obtained in Sec. \ref{sec.Spin correlation}, we can calculate various quantities. 
In this section, we first calculate the entanglement entropy.
\subsubsection{$S(\hat{\rho}_\tx{A})$ and $S(\hat{\rho}_\tx{B})$}
The eigenvalues of Alice's reduced density matrix (\ref{rho_A}) and Bob's one (\ref{rho_B}) are given by
\eqn{
\mu_\tx{A}^s \coloneqq \frac{1+s C_{x0}}{2} ~~\tx{with}~~s=\pm 1 ~,  \label{mu_A}
}
\eqn{
\mu_\tx{B}^s \coloneqq \frac{1+s C_{0x}}{2} ~~\tx{with}~~s=\pm 1 ~,  \label{mu_B}
}
respectively. These are nonnegative since $|C_{x0}|, |C_{0x}| \leq 1$ as shown below (\ref{gamma_A,B}).
Then,
the entanglement entropies are computed as
\eqn{S(\hat{\rho}_\tx{A}) &= -\sum_{s=\pm} \mu_\tx{A}^s \ln \mu_\tx{A}^s = \Sigma (C_{x0}) = \Sigma \qty(\gamma_\tx{A} \cos (2 \mf{G}_\tx{R}^\tx{AB})) ~, \label{S_A}}
\eqn{S(\hat{\rho}_\tx{B}) &= -\sum_{s=\pm} \mu_\tx{B}^s \ln \mu_\tx{B}^s = \Sigma (C_{0x}) = \Sigma \qty(\gamma_\tx{B} \cos (2 \mf{G}_\tx{R}^\tx{BA})) ~. \label{S_B}}
Instead of $\cos (2\Theta)$ in the Newtonian picture (\ref{S_N}),  we have two different quantities $C_{x0} = \La \hat{\sigma}_x^\tx{A} \Ra$ and $C_{x0} = \La \hat{\sigma}_x^\tx{B} \Ra$ given in (\ref{Cs}) for Alice's spin and Bob's spin, respectively.
Remember that $\Sigma (v)$ monotonically decreases with $v$ increasing.
Take the reduced system of Alice's spin.
When $C_{x0}$ vanishes, $S(\hat{\rho}_\tx{A})$ takes the maximal\footnote{In our setup, the maximum value of $S(\hat{\rho}_\tx{A})$ reaches the maximal one $\ln 2$ because the predictability $\cl{P}_\tx{A}$ vanishes with the spin initial state (\ref{spin-initial-state}). See the footnote \ref{asymmetric-interferometer}.}
value.
Since  $\gamma_\tx{A}$ defined in (\ref{gamma_A,B}) does not exceed unity, the entanglement entropies are restricted to take their values in the region
\eqn{
0\leq  \Sigma (\gamma_\tx{A}) \leq \Sigma \qty(\gamma_\tx{A} \cos (2 \mf{G}_\tx{R}^\tx{AB})) = S(\hat{\rho}_\tx{A}) \leq \ln 2 ~. 
}

The minimal value $0$ 
is realized when the reduced system is in a pure state, that is, Alice's spin does not entangle with the rest of the total system composed of Bob's spin and the field.
The necessary and sufficient condition for the vanishing $S(\hat{\rho}_\tx{A})$ is that $\mf{G}_\tx{R}^\tx{AB} = 0$ mod $\pi/2$ and 
$\gamma_\tx{A}=1$, namely  an adiabatic limit of Alice; $\mf{G}_\tx{K}^\tx{AA} \to 0$.

On the other hand, $S(\hat{\rho}_\tx{A})$ can take the maximal value $\ln 2$ for two reasons.
One is the case of maximal entanglement between Alice's spin and Bob's spin via the field $\phi$ as the retarded potential. In this case,  $C_{x0}$ vanishes owing to $\mf{G}_\tx{R}^\tx{AB} = \pi/4$ mod $\pi/2$ even when $\gamma_\tx{A}=1$. The other case is realized by the entanglement between Alice's spin and the on-shell excitations of the field due to the nonadiabaticity of Alice's protocol.  That is,  if Alice abruptly turns on/off the spin-field interaction,  $\mf{G}_\tx{K}^\tx{AA} \to \infty$ and $\gamma_\tx{A} \to 0$ and the maximal value of the entropy is obtained.

In contrast to the Newtonian case where $S(\hat{\rho}_\tx{A}) = S(\hat{\rho}_\tx{B})$, the causal structure also plays important roles in  presence of dynamical field.
For instance, when Alice is not in Bob's causal future, the retarded Green's function from Bob to Alice disappears and so does $\mf{G}_\tx{R}^\tx{AB}$.
Then, $S(\hat{\rho}_\tx{A})$ takes its nonzero minimum value $\Sigma (\gamma_\tx{A})$, whereas $S(\hat{\rho}_\tx{B})$ is still possible to take the maximal value $\ln 2$.

\subsubsection{$S(\hat{\rho}_\tx{AB})=S(\hat{\rho}_\phi)$}
Since the total system is in a pure state, the entanglement entropy of the subsystem of Alice and Bob $S(\hat{\rho}_\tx{AB})$ is equal to that of the scalar field $S(\hat{\rho}_\phi)$. Let us calculate it. 

The eigenvalues of (\ref{rho_AB}) are computed as
\eqn{
\mu_\tx{AB}^{s_1 s_2} :&= \frac{1}{4} \qty{ 1+s_2 C_{xx} + s_1 \sqrt{ (C_{x0} +s_2 C_{0x})^2 + (C_{yz} +s_2 C_{zy})^2+ C_{yy}^2  } } \\
&= \frac{1}{4} \biggl\{ 1+s_2 \gamma_\tx{A} \gamma_\tx{B} \cosh (4 \mf{G}^\tx{BA}_\tx{K})  \\
&~~~~~~+ s_1 \sqrt{\gamma_\tx{A}^2 + \gamma_\tx{B}^2 + 2 s_2 \gamma_\tx{A}  \gamma_\tx{B} \cos (2 (\mf{G}^\tx{AB}_\tx{R}- \mf{G}^\tx{BA}_\tx{R}))  + \gamma_\tx{A}^2 \gamma_\tx{B}^2 \sinh^2 (4 \mf{G}^\tx{BA}_\tx{K})   } \biggr\}
\label{mu_AB}}
with $s_{1,2} = \pm 1$.
Thus, the entanglement entropy is obtained by
\eqn{
S(\hat{\rho}_\tx{AB}) &= -\sum_{s_1 , s_2 =\pm} \mu_\tx{AB}^{s_1 s_2} \ln \mu_\tx{AB}^{s_1 s_2} 
\label{S_AB}}
with the eigenvalues (\ref{mu_AB}).
Unlike $S(\hat{\rho}_\tx{A})$ and $S(\hat{\rho}_\tx{B})$ discussed above,
it depends on all the spin expectation values in (\ref{Cs}) and quantifies the entanglement between the field and the two spins.

The minimal value $0$ of $S(\hat{\rho}_\tx{AB})$ is found when the reduced system with $\hat{\rho}_\tx{AB}$ is in a pure state, or equivalently, when the largest eigenvalue $\mu_\tx{AB}^{++}$ is the only finite one to be unity.
The necessary and sufficient condition is that both Alice and Bob adiabatically change the spin-field coupling so that all the $\mf{G}_\tx{K}$'s vanish: $\mf{G}_\tx{K}^\tx{AA}, \mf{G}_\tx{K}^\tx{BB}, \mf{G}_\tx{K}^\tx{BA} \to 0$.

$S(\hat{\rho}_\tx{AB})$ takes the maximal value $2\ln 2$ if all the eigenvalues equate: $\mu_\tx{AB}^{s_1 s_2} = 1/4$, which is realized only in the nonadiabatic limit where $\mf{G}_\tx{K}^\tx{AA}, \mf{G}_\tx{K}^\tx{BB} \to \infty$ so that all the $C$'s vanish.\footnote{
As shown in Appendix \ref{app.Correlation between field and spin}, as far as $D= | \vb*{x}_\tx{A} - \vb*{x}_\tx{B}|$ is fixed to be finite, 
$\mf{G}_\tx{K}^\tx{AA}+\mf{G}_\tx{K}^\tx{BB} - 2 |\mf{G}_\tx{K}^\tx{BA}| \to \infty$ is guaranteed, and then, $C_{xx}, C_{yy} \to 0$.
}
Note that, when all the spin correlation functions in (\ref{connected-part}) vanish, or equivalently,
\eqn{C_{yy} =C_{yz}= C_{zy} = 0~, ~~ C_{xx} = C_{x0}C_{0x} ~, }
we have 
\eqn{
S(\hat{\rho}_\tx{AB}) = S(\hat{\rho}_\tx{A}) + S(\hat{\rho}_\tx{B}) ~.  \label{S_AB=S_A+S_B}}
It is because the density matrix (\ref{rho_AB}) can be written as the tensor product state:
\eqn{
\hat{\rho}_\tx{AB} = \frac{1}{4} \qty{ \hat{1}^\tx{A} \hat{1}^\tx{B} + C_{x0} C_{0x}\hat{\sigma}_x^\tx{A} \hat{\sigma}_x^\tx{B}   + C_{x0} \hat{\sigma}_x^\tx{A} \hat{1}^\tx{B} + C_{0x}  \hat{1}^\tx{A}\hat{\sigma}_x^\tx{B}  } =  \hat{\rho}_\tx{A}  \hat{\rho}_\tx{B} ~. \label{tensor-product-state}
}

%%%%%%%%%%%%%%%%%%%%%%%%%%%%%%%

\subsection{Visibility and distinguishability \label{sec.Visibility and distinguishability}}
The visibility and the distinguishability introduced in Sec.~\ref{sec.Visibility and distinguishability-Newtonian} are discussed in \cite{Sugiyama:2022wcd} in a specific situation. Here we discuss them in a bit more general situation. 
According to the definition (\ref{define-visibility}), Alice's and Bob's visibility of their interference fringes are computed from \eqref{rho_A} and \eqref{rho_B} as
\eqn{
\cl{V}_\tx{A} = |C_{x0}| =\gamma_\tx{A} |\cos (2 \mf{G}_\tx{R}^\tx{AB} )| ~, ~~~ \cl{V}_\tx{B} = |C_{0x}| =\gamma_\tx{B} |\cos (2 \mf{G}_\tx{R}^\tx{BA} )|
\label{V} ~,
}
which are nothing but the absolute values of $\La \hat{\sigma}^\tx{A}_x \Ra$ and $\La \hat{\sigma}^\tx{B}_x \Ra$, respectively, seen in (\ref{Cs}).
Although $\cl{V}_\tx{A} \ne \cl{V}_\tx{B}$, note that the relation with the entanglement entropy (\ref{S_N<->Visibility}) generally holds:
\eqn{
S(\hat{\rho}_\tx{A}) = \Sigma (\cl{V}_\tx{A})~,~~S(\hat{\rho}_\tx{B}) = \Sigma (\cl{V}_\tx{B}) ~, \label{S_A(B)<->Visibility}
}
as long as $\expval*{\hat{\sigma}^\tx{A}_z}=\expval*{\hat{\sigma}^\tx{B}_z}=0$.
This means that the entanglement entropy and the visibility embody the same kind of information.

In order to compute the distinguishability, we first obtain
\eqn{
\hat{\rho}_\tx{A}^{\pm}  &= \frac{1}{2} \Bigl\{ \hat{1}^\tx{A}
+ C_{x0} \hat{\sigma}_x^\tx{A}
\pm C_{yz} \hat{\sigma}_y^\tx{A}\Bigr\}, \\
\hat{\rho}_\tx{B}^{\pm}  &=
\frac{1}{2} \Bigl\{ \hat{1}^\tx{B}
+ C_{0x} \hat{\sigma}_x^\tx{B}
\pm C_{zy} \hat{\sigma}_y^\tx{B}\Bigr\}
~, \label{rho^pm}
}
from \eqref{rho_AB} according to the definition (\ref{define-rho^pm}).
Then, (\ref{eq:Bloch_one_qubit}) and \eqref{eq:trace_distance_one_qubit} tells us that Alice's distinguishability of Bob's spin and Bob's distinguishability of Alice's spin can be quantified by
\eqn{
\cl{D}_\tx{A}=|C_{yz}| = \gamma_\tx{A} |\sin (2 \mf{G}_\tx{R}^\tx{AB} )| ~, ~~~ \cl{D}_\tx{B}=|C_{zy}| = \gamma_\tx{B} |\sin (2 \mf{G}_\tx{R}^\tx{BA} )| ~, \label{D}
}
which are nothing but the absolute values of $\La \delta \hat{\sigma}^\tx{A}_y \delta \hat{\sigma}^\tx{B}_z \Ra$ and $\La \delta \hat{\sigma}^\tx{A}_z \delta \hat{\sigma}^\tx{B}_y \Ra$, respectively.
However, unlike in the Newtonian picture, these differ from $\sqrt{\La \delta \hat{\sigma}_x^\tx{A} \delta  \hat{\sigma}_x^\tx{B} \Ra}$.
And also, while the entanglement entropy can be regarded as a monotonically increasing function of the distinguishability as $S(\hat{\rho}_\tx{A}) = \Sigma (\cl{V}_\tx{A}) = \Sigma ( (\gamma_\tx{A}^2 - \cl{D}^2_\tx{A})^{1/2} )$,
the entanglement entropy and the distinguishability no longer have the same information due to the field dynamics that affects the value of $\gamma_\tx{A}$.

\

The presence of the dynamical field also spoils the equality (\ref{wave-particle-duality-Newtonian}).
It is known that 
\eqn{\cl{V}_\tx{A}^2 + \cl{D}_\tx{B}^2  \leq 1 ~, ~~ \cl{V}_\tx{B}^2 + \cl{D}_\tx{A}^2  \leq 1 ~ \label{wave-particle-duality}
}
hold to quantify the notion of wave particle duality \cite{PhysRevLett.77.2154}, see discussions in \cite{Sugiyama:2022wcd} for the applicability to systems with dynamical fields.
In our setup, these imply
\aln{
\gamma_\tx{A}^2 \cos^2 ( 2 \mf{G}^\tx{AB}_\tx{R} ) + \gamma_\tx{B}^2   \sin^2 ( 2 \mf{G}^\tx{BA}_\tx{R} ) \leq 1 ~, \label{wave-particle-duality-1}  \\
\gamma_\tx{B}^2  \cos^2 ( 2 \mf{G}^\tx{BA}_\tx{R} ) + \gamma_\tx{A}^2   \sin^2 ( 2 \mf{G}^\tx{AB}_\tx{R} ) \leq 1 ~, \label{wave-particle-duality-2}
}
respectively.

Remember that, in the Newtonian approximation, the two-spin system is a  pure state and 
 the equality (\ref{wave-particle-duality-Newtonian}) holds. In the presence of the dynamical field, the two-spin system is no longer in a pure state and the equality is replaced by an inequality. 
 In order to know how much the inequality deviates from the equality, 
 we need a different type of trade-off relation which can properly take the information of the dynamical field into account.
 In later sections, we study a trade-off relation of mutual information for this purpose.

%%%%%%%%%%%%%%%%%%%%%%%%%%%%%%%%%%%
\subsection{Separability condition and negativity \label{sec.Separability condition}}

The reduced density matrix $\hat{\rho}_\tx{AB}$ in (\ref{rho_AB}) describes a mixed state, and 
it can be a separable state. Then, Alice's spin and Bob's spin are not entangled and the spin correlations in (\ref{connected-part}) are  classical. The separability for the two-qubit system is quantified by the negativity.
If the negativity vanishes, the density matrix is separable.
The negativity~\eqref{entanglement-negativity} is evaluated by the eigenvalue of the partial transposition (acting on Bob's spin) of the reduced density matrix (\ref{rho_AB}),
\eqn{
\hat{{\rho}}^{T_\tx{B}}_\tx{AB} =& \frac{1}{4} \Bigl\{ \hat{1}^\tx{A} \hat{1}^\tx{B} +C_{xx} \hat{\sigma}_x^\tx{A} \hat{\sigma}_x^\tx{B} - C_{yy} \hat{\sigma}_y^\tx{A} \hat{\sigma}_y^\tx{B} \\
&~~~~~~ +C_{x0} \hat{\sigma}_x^\tx{A} \hat{1}^\tx{B} + C_{0x}  \hat{1}^\tx{A}\hat{\sigma}_x^\tx{B}  + C_{yz} \hat{\sigma}_y^\tx{A} \hat{\sigma}_z^\tx{B} - C_{zy} \hat{\sigma}_z^\tx{A} \hat{\sigma}_y^\tx{B}    \Bigr\} ~,  
}
which is nothing but (\ref{rho_AB}) with the following replacements,
\eqn{
C_{yy} \to - C_{yy} ~, ~~~ C_{zy} \to - C_{zy} ~.
}
Then, making these replacements in (\ref{mu_AB}), we get the eigenvalues of the partial transposition:
\eqn{
\tilde{\mu}_\tx{AB}^{s_1 s_2} &= \frac{1}{4} \qty{ 1+s_2 C_{xx} + s_1 \sqrt{ (C_{x0} +s_2 C_{0x})^2 + (C_{yz} - s_2 C_{zy})^2+ C_{yy}^2  } }\\
&= \frac{1}{4} \biggl\{ 1+s_2 \gamma_\tx{A} \gamma_\tx{B} \cosh (4 \mf{G}^\tx{BA}_\tx{K})  \\
&~~~~~~+ s_1 \sqrt{\gamma_\tx{A}^2 + \gamma_\tx{B}^2 + 2 s_2 \gamma_\tx{A}  \gamma_\tx{B} \cos (2 (\mf{G}^\tx{AB}_\tx{R}+ \mf{G}^\tx{BA}_\tx{R}))  + \gamma_\tx{A}^2 \gamma_\tx{B}^2 \sinh^2 (4 \mf{G}^\tx{BA}_\tx{K})   } \biggr\}
~. \label{PT-mu_AB}
}
The entanglement negativity is the summation of the absolute values of negative eigenvalues:
\eqn{
\cl{N} \coloneqq \sum_{s_1,s_2 = \pm} \theta (-\tilde{\mu}_\tx{AB}^{s_1 s_2}) ~ |\tilde{\mu}_\tx{AB}^{s_1 s_2}| ~.
\label{negativity-definition}
}
Note that, when $C_{yz} \times C_{zy} =0$, we have $\tilde{\mu}_\tx{AB}^{s_1 s_2} = \mu_\tx{AB}^{s_1 s_2} $ which can not be negative for consistency,
and thus, $\hat{\rho}_\tx{AB}$ is separable.
In other words, 
it is a necessary condition to be nonsparable that both Alice and Bob have finite distinguishabilities, $\cl{D}_\tx{A}\neq0$ and $\cl{D}_\tx{B}\neq0$ in our setup.

%%%%%%%%%%%%%%%%%%%%%%%%%%%%%%%%%%%%%%%%%%%%%%%%%%%%%
\subsection{Mutual information\label{sec.Mutual information}}
\subsubsection{Definition of mutual information}
The mutual information $I_\tx{XY}$ quantifies the amount of correlations between two subsystems X and Y.
In our setup, each of X and Y is identified as either Alice's spin, Bob's spin, the field $\phi$, or compositions of them.
In terms of the mutual information,
correlations between two subsystems are related to correlations between another choice of two subsystems as seen below.

Consider three more reduced density matrices besides (\ref{rho_AB})-(\ref{rho_B}) as
\eqn{
\hat{\rho}_{\tx{B}\phi} \coloneqq \tr_{\tx{A}} \{ \hat{\rho}_{\tx{AB}\phi} \}   ~,~~ \hat{\rho}_{\tx{A}\phi} \coloneqq \tr_{\tx{B}}  \{\hat{\rho}_{\tx{AB}\phi} \} ~, ~~\hat{\rho}_{\phi} \coloneqq \tr_\tx{AB}  \{ \hat{\rho}_{\tx{AB}\phi}  \} ~,
}
where $\hat{\rho}_{\tx{AB}\phi} : = \ket{\Psi_\tx{f}}\bra{\Psi_\tx{f}}$.
Since the total system is in the pure state, the entanglement entropies associated with these density matrices are computed as
\aln{S(\hat{\rho}_{\tx{B}\phi})= S(\hat{\rho}_\tx{A})  ~, ~~
S(\hat{\rho}_{\tx{A}\phi})= S(\hat{\rho}_\tx{B}) ~,~~
S(\hat{\rho}_{\phi})= S(\hat{\rho}_\tx{AB})  ~. 
}
From these three, we can compose three types of the mutual informations as 
\aln{
I_\tx{AB} &\coloneqq 
S(\hat{\rho}_\tx{A}) + S(\hat{\rho}_\tx{B}) - S(\hat{\rho}_\tx{AB}) ~, \label{I_AB}\\
I_{\tx{A}\phi} &\coloneqq 
 S(\hat{\rho}_\tx{A}) + S(\hat{\rho}_\phi)  - S(\hat{\rho}_{\tx{A}\phi}) 
=  S(\hat{\rho}_\tx{A}) + S(\hat{\rho}_\tx{AB}) - S(\hat{\rho}_\tx{B}) ~, \label{I_Aphi} \\
I_{\tx{B}\phi } &\coloneqq 
S(\hat{\rho}_\tx{B}) + S(\hat{\rho}_\phi) - S(\hat{\rho}_{\tx{B} \phi})
= S(\hat{\rho}_\tx{B}) + S(\hat{\rho}_\tx{AB}) - S(\hat{\rho}_\tx{A}) ~,  \label{I_Bphi}
}
which are guaranteed to be nonnegative by the subadditivity for von Neumann entropy.\footnote{
Those can be understood as the strong subadditivity inequalities
\[
S(\hat{\rho}_{\tx{B}\phi}) + S(\hat{\rho}_{\tx{A}\phi}) \geq  S(\hat{\rho}_{\phi}) + S(\hat{\rho}_{\tx{AB}\phi}) ~,
\]
and ones that are obtained by permuting A, B and $\phi$. Here, the total state $\hat{\rho}_{\tx{AB}\phi} \coloneqq \ket{\Psi_\tx{f}}\bra{\Psi_\tx{f}}$ is pure, and hence, $S(\hat{\rho}_{\tx{AB}\phi}) = 0$.
}
Then, from the above expressions, it is easy to see that they are bounded  from above,
\eqn{0 \leq I_\tx{AB}, I_{\tx{A}\phi}, I_{\tx{B}\phi} \leq 2 \ln 2 ~. }
% Note also that the entanglement entropies, (\ref{S_A}), (\ref{S_B}) and (\ref{S_AB}), themselves are also regarded as the mutual informations in our setup:
% \eqn{
% 2 S(\hat{\rho}_\tx{A}) &= I_{\tx{A}(\tx{B}\phi)} \coloneqq S(\hat{\rho}_\tx{A}) + S(\hat{\rho}_{\tx{B}\phi}) - S(\hat{\rho}_{\tx{AB}\phi})  ~,\\
% 2 S(\hat{\rho}_\tx{B}) &= I_{\tx{B}(\tx{A}\phi)} \coloneqq S(\hat{\rho}_\tx{B}) + S(\hat{\rho}_{\tx{A}\phi}) - S(\hat{\rho}_{\tx{AB}\phi}) ~,\\
% 2 S(\hat{\rho}_\tx{AB}) &= I_{\phi(\tx{AB})} \coloneqq S(\hat{\rho}_\phi) + S(\hat{\rho}_\tx{AB}) - S(\hat{\rho}_{\tx{AB}\phi})~.
% }
% Since the total system is in the pure state, those quantify the amount of correlations that are all originated from the quantum entanglement.

\subsubsection{Mutual information and correlation functions}
In general, the mutual information vanishes if and only if there is no correlation between variables from one reduced system and variables from the other.
Especially, if all the spin correlation functions vanish, then $I_\tx{AB}$ also vanishes because the reduced density matrix is tensor product state, $\hat{\rho}_\tx{AB} = \hat{\rho}_\tx{A} \hat{\rho}_\tx{B}$, as explicitly seen in (\ref{tensor-product-state}).
If $I_\tx{AB}$ vanishes, all the correlation functions vanish because the mutual information gives an upper bound on the correlation functions~\cite{PhysRevLett.100.070502}. In our case, the inequality is given by
\eqn{
I_\tx{AB} \geq \frac{ \La \delta \hat{\sigma}^\tx{A}_{w} \delta \hat{\sigma}^\tx{B}_{w'} \Ra ^2 }{2 \|\hat{\sigma}^\tx{A}_{w}\|^2 \|\hat{\sigma}^\tx{B}_{w'}\|^2} =  \frac{ \La \delta \hat{\sigma}^\tx{A}_{w} \delta \hat{\sigma}^\tx{B}_{w'} \Ra ^2 }{2} \label{WVHC}
}
for any pair of spin variables,  $w,w'=x,y,z$. Here, $\|\cdot\|$ represents the operator norm.
Note that, even though the spin correlations are built up by the interaction mediated by the quantum field, they are not necessarily attributed to the quantum entanglement between Alice's and Bob's spins.
The mutual information does not care if the correlations are quantum or classical, and hence, it can not be a measure of the entanglement in general.

%%%%%%%%%%%%%%%%%%%%%%%%%%%%%%%%%%%%%%%%%%%%%%%%
\subsection{Trade-off relation in terms of entropy \label{sec.Trade-off relation}}
Let us first observe from the definitions \eqref{I_AB}-\eqref{I_Bphi} that, when the total system is in a pure state, the mutual informations satisfy\footnote{When the total system is in a mixed state, inequalities $I_\tx{AB} +  I_{\tx{A}\phi} \leq  2 S(\hat{\rho}_\tx{A})$ and $I_\tx{AB} +  I_{\tx{B}\phi} \leq  2 S(\hat{\rho}_\tx{B})$ hold in general \cite{10.5555/1972505}.
}
\eqn{I_\tx{AB} +  I_{\tx{A}\phi} =  2 S(\hat{\rho}_\tx{A}) ~,~~~ I_\tx{AB} +  I_{\tx{B}\phi} =  2 S(\hat{\rho}_\tx{B}) ~. \label{Information-trade-off}
}
When Alice's spin is not entangled with the rest of the system and $S(\hat{\rho}_\tx{A})=0$, there is neither the correlation $I_\tx{AB}$
between Alice's spin and Bob's spin  nor the correlation $I_{\tx{A}\phi}$ between Alice's spin and the field.
One can also regard these as trade-off relations: With the right-hand side $S(\hat{\rho}_\tx{A})$ known and fixed, the more correlation $I_{\tx{A}\phi}$ 
 between the field and Alice's spin, the less correlation $I_\tx{AB}$ between Bob's spin and Alice's spin.
Since the mutual informations are nonnegative because of the subadditivity as mentioned above,
we have an  inequalities
\eqn{
I_\tx{AB}  \leq   2~ \tx{min} \qty{ S(\hat{\rho}_\tx{A}), S(\hat{\rho}_\tx{B}) } ~.  \label{inequalities}
}
by dropping  $I_{\tx{A}\phi}$ and $I_{\tx{B}\phi}$ from the equalities. 
Thus both spins must be
 entangled with the rest of the system in order for the spin correlations to exist.

We have already seen that,  in (\ref{S_A(B)<->Visibility}), the visibility and the entropy are in one-to-one correspondence.
Let us here introduce an entropic counterpart of the visibility\footnote{For an asymmetric interferometer mentioned in the footnote \ref{asymmetric-interferometer}, one may define it by $P_\tx{A} \coloneqq 2 \qty[ \Sigma(\cl{P}_\tx{A} ) - S(\hat{\rho}_\tx{A}) ]$ so that it vanishes when $\cl{V}_\tx{A}=0$.}
as
\eqn{
P_\tx{A} \coloneqq& 2 \qty[ \ln 2 - S(\hat{\rho}_\tx{A}) ] \\
 =& (1+ \cl{V}_\tx{A})\ln (1+ \cl{V}_\tx{A}) + (1 - \cl{V}_\tx{A})\ln (1 - \cl{V}_\tx{A}) ~, \label{P} 
}
which monotonically increases from $0$ to $2\ln 2$ as $\cl{V}_\tx{A}$ increases from $0$ to $1$.
For Bob's spin, $P_\tx{B}$ is defined in the same manner.
Then, the inequalities in (\ref{inequalities}) turn out to be
\eqn{
I_\tx{AB} \leq   {2 \ln 2}  - P_\tx{A}    ~, ~~~  
I_\tx{AB} \leq   {2 \ln 2}  - P_\tx{B} ~.
\label{Information-trade-off-inequality}}
These relations can be regarded as analogs of the 
 relations between the visibility and the distinguishability in (\ref{wave-particle-duality}),
\eqn{
  \cl{D}_\tx{B}^2  \leq  1 - \cl{V}_\tx{A}^2   ~, ~~  \cl{D}_\tx{A}^2   \leq  1 -   \cl{V}_\tx{B}^2   ~. \label{bound-on-D}
}
The mutual information $I_\tx{AB}$ plays the role of the distinguishability. 
Unlike the distinguishabilities in (\ref{D}),
it is neither ``Alice's'' mutual information nor ``Bob's'' mutual information, but the mutual information of Alice and Bob. 
It is because $I_\tx{AB}$ cares about all the spin correlations including both $\La \delta \hat{\sigma}^\tx{A}_y \delta \hat{\sigma}^\tx{B}_z \Ra$ and $\La \delta \hat{\sigma}^\tx{A}_z \delta \hat{\sigma}^\tx{B}_y \Ra$, as will be discussed in Sec.~\ref{sec.Spacelike-separated}.
In this sense, the mutual information $I_\tx{AB}$ rather corresponds to the averaged distinguishability of Bob and Alice, $(\mathcal{D}_\tx{A}^2+\mathcal{D}_\tx{B}^2)/2$.
The averaged distinguishability satisfies the following inequality:
\begin{equation}
    \frac{\mathcal{D}_\tx{A}^2+\mathcal{D}_\tx{B}^2}{2}\leq 1-\frac{\mathcal{V}_\tx{A}^2+\mathcal{V}_\tx{B}^2}{2},
    \label{eq:averaged_trade_off_relation}
\end{equation}
whose counter part of mutual information is given by
\begin{equation}
    I_\tx{AB} \leq 2\ln2 - \frac{P_\tx{A}+P_\tx{B}}{2}.
    \label{eq:modified_trade_off_relation}
\end{equation}
We will see in Secs.~\ref{sec.Adiabatic limit} and \ref{sec.Non-adiabatic limit} that the mutual information and the average distinguishability behave similarly.
Note that the mutual information contains information of not only distinguishabilities but also other correlations.
Therefore, the mutual information can be nonzero even if both Alice's and Bob's distinguishability.
An advantage of using \eqref{Information-trade-off-inequality} or \eqref{eq:modified_trade_off_relation} instead of \eqref{bound-on-D} or \eqref{eq:averaged_trade_off_relation} is that
we know what is missing to saturate the equalities. The inequalities \eqref{Information-trade-off-inequality} can 
be made to become equalities  (\ref{Information-trade-off})
by taking into account the mutual informations $I_{\tx{A}\phi}$ and $I_{\tx{B}\phi}$: 
\eqn{
I_\tx{AB} = {2 \ln 2}   - {P_\tx{A}} - {I_{\tx{A}\phi}}   ~, ~~~  
I_\tx{AB} = {2 \ln 2}   - {P_\tx{B}} - {I_{\tx{B}\phi}} ~,
\label{tradeof-equality-MI}
}
and (\ref{eq:modified_trade_off_relation}) becomes
\begin{equation}
    I_\tx{AB} = {2 \ln 2}   - \frac{{P_\tx{A}}+{P_\tx{B}}}{2} - S(\hat{\rho}_\tx{AB}),
\end{equation}
where $({{I_{\tx{A}\phi}}+{I_{\tx{B}\phi}}})/{2}=S(\hat{\rho}_\tx{AB})$ is used.

%%%%%%%%%%%%%%%%%%%%%%%%%%%%%%%%%%%%%%%%%%%%%%%%%%%%

\subsection{Robertson-Schr\"{o}dinger inequalities \label{sec.RS inequalities}}
In the reduced density matrix (\ref{rho_AB}), we get $\mf{G}$'s, the various time integrations of the propagators given in (\ref{mathfrak-G_R}) and (\ref{mathfrak-G_K}).
Those are not independent from each other; there are inequalities to be satisfied among those quantities, due to the quantum uncertainty discussed in \cite{Sugiyama:2022wcd}.

Given a density matrix $\hat{\rho}$, for two Hermitian operators $\delta \hat{\cl{O}}_{1,2}$ with  $\La \delta \hat{\cl{O}}_{1,2} \Ra \coloneqq \tr \{ \hat{\rho} \delta \hat{\cl{O}}_{1,2}  \} = 0$, their vanishing expectation value under $\rho$, the Robertson-Schr\"{o}dinger inequality
\eqn{
\La \delta \hat{\cl{O}}_1^2 \Ra \La  \delta \hat{\cl{O}}_2^2 \Ra \geq \frac{1}{4} \qty(   |\La \{ \delta \hat{\cl{O}}_1, \delta \hat{\cl{O}}_2 \} \Ra |^2  + |\La [\delta \hat{\cl{O}}_1, \delta \hat{\cl{O}}_2] \Ra |^2 )
\label{Robertson-Schrodinger}}
holds, which follows from the Cauchy-Schwarz inequality.
If two operators are chosen as 
\eqn{
\delta \hat{\cl{O}}_1 = \int_{t_\tx{i}}^{t_\tx{f}} \dd t  \lambda_\tx{A}(t) \hat{\phi}(t,\vb*{x}_\tx{A})  ~, ~~ \delta \hat{\cl{O}}_2 = \int_{t_\tx{i}}^{t_\tx{f}} \dd t  \lambda_\tx{B}(t) \hat{\phi}(t,\vb*{x}_\tx{B}) 
\label{Operators}}
(see footnote \ref{source-free} for the definition of the Heisenberg operator here), and the expectation value is taken under the vacuum state $\ket{\Omega}_\phi\!\bra{\Omega}$,
we get
$\La \delta\hat{\cl{O}}_1^2 \Ra=\mf{G}^\tx{AA}_\tx{K} $,
$\La \delta\hat{\cl{O}}_2^2 \Ra=\mf{G}^\tx{BB}_\tx{K} $,
$\La \{ \delta \hat{\cl{O}}_1, \delta \hat{\cl{O}}_2 \}\Ra=2\mf{G}^\tx{AB}_\tx{K}$, and   
$\La [\delta \hat{\cl{O}}_1, \delta \hat{\cl{O}}_2] \Ra
=\mf{G}^\tx{BA}_\tx{R}-\mf{G}^\tx{AB}_\tx{R}$.
Therefore, we have
\eqn{
 (\mf{G}^\tx{BA}_\tx{R}-\mf{G}^\tx{AB}_\tx{R})^2 \leq 4 \qty( \mf{G}^\tx{AA}_\tx{K} \mf{G}^\tx{BB}_\tx{K} - (\mf{G}^\tx{BA}_\tx{K})^2 )  \label{RS-inequality}
}
with, for consistency,
\eqn{
  \mf{G}^\tx{AA}_\tx{K} \mf{G}^\tx{BB}_\tx{K}  \geq (\mf{G}^\tx{BA}_\tx{K})^2   ~,   \label{RS-inequality-K}
}
which is stronger than (\ref{G_K-positivity-2}).
Similar inequalities can be obtained as consistency conditions for the eigenvalues of the density matrix (\ref{rho_AB}) to be nonnegative in a certain limit, see Appendix \ref{app.Consistency condition}.
These inequalities are to be used in Secs.~\ref{sec.Adiabatic limit}, \ref{sec.Spacelike-separated} and \ref{sec.One-way} to take adiabatic limit properly.

%%%%%%%%%%%%%%%%%%%%%%%%%%%%%%%%%%%%%%%%
%%%%%%%%%%%%%%%%%%%%%%%%%%%%%%%%%%%%%%
\section{Four limiting cases \label{sec.limiting cases}}
\begin{figure}
\begin{center}
 \includegraphics[width=7cm]{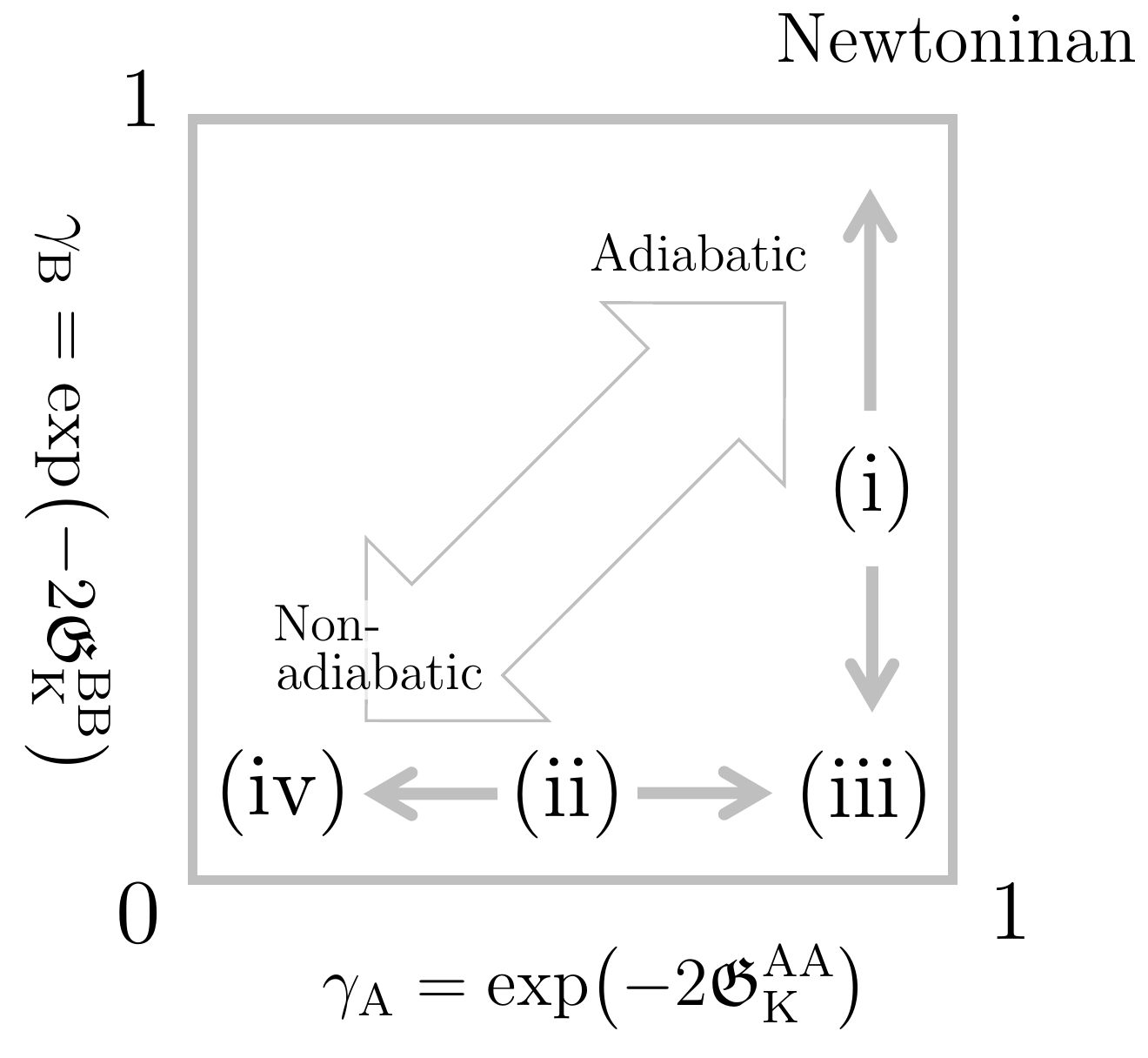}
\caption{
Various limiting cases in terms of Alice's (horizontal) and Bob's (vertical) adiabaticities.
The Upper right corner corresponds to the Newtonian approximation discussed in Section \ref{sec.Newtonian} which is realized as the adiabatic limit and a part of the limit (i). 
The lower right corner (iii) is included both in the limit (i) and (ii).
The lower left corner (iv) corresponds to the nonadiabatic limit where both Alice and Bob turn on or off their spin-field interactions abruptly which is a part of the limit (ii).
}
\label{fig:Adiabaticity}
\end{center}
\end{figure}

\begin{figure}
\begin{center}
 \includegraphics[width=9cm]{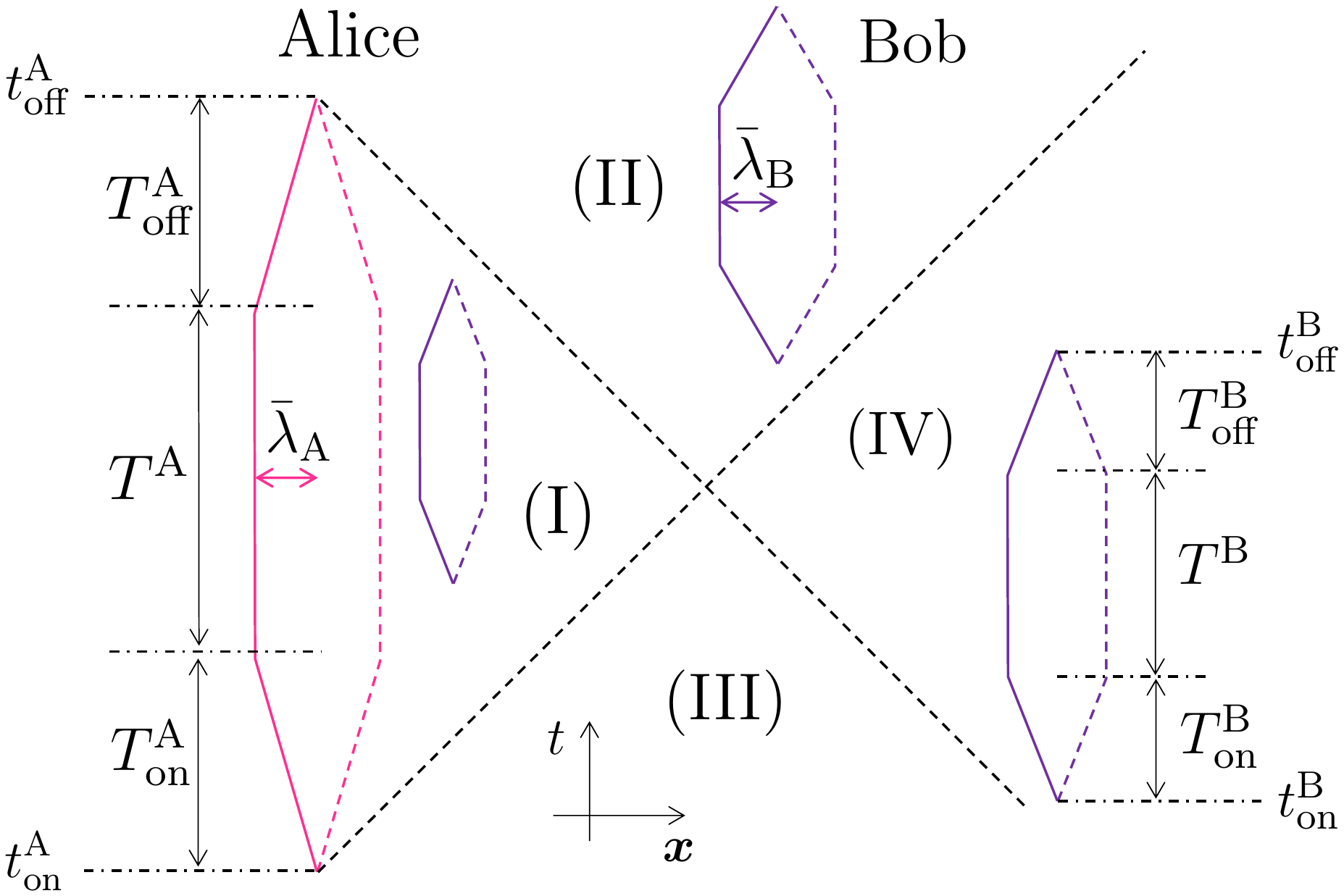}
\caption{
The four distinctive spacetime locations of Bob's protocol in terms of causality.
The dashed lines are lightlike curves in the flat spacetime under consideration.
$T^\tx{A}_\tx{on (off)}$ is the length of time Alice takes to turn on (off) the spin-field interaction. The coupling takes a constant value $\lambda_\tx{A}(t)= \bar{\lambda}_\tx{A}$ for the period $T^\tx{A}$. See (\ref{time-dependence}) in Appendix~\ref{app:Numerical evaluation} for its explicit time dependence. Note that the pink and purple curves are just to show the strengths of the couplings changing in time at the same spatial points $\vb*{x}_\tx{A}$ and $\vb*{x}_\tx{B}$. The distance between the two spatial points denoted by $D$ does not change in time as in Fig.~\ref{fig:setup}.}
\label{fig:Causality}
\end{center}
\end{figure}

In the previous sections, we have exactly solved the model and obtained various quantities such as the spin correlations and negativity.
In this section, we study their properties in some limiting cases. Compared to the Newtonian approximation, decoherence due to particle creations 
and relativistic causality are two important new ingredients in relativistic theories. Two figures~\ref{fig:Adiabaticity} and \ref{fig:Causality} show various situations.
In Fig.~\ref{fig:Adiabaticity}, one can consider various situations depending on how much the particle creations occur, which is described by the Keldysh Green's functions.
Figure~\ref{fig:Causality} shows the causal structure described by the retarded Green's functions connecting Alice's spin and Bob's spin.  
These relativistic effects of particle creations and causality 
are not totally independent due to the inequalities in Sec.~\ref{sec.RS inequalities}.

%%%%%%%%FIG%%%%%%
%%%%%%%%%%%%%%%%%%%%

In Sec.~\ref{sec.Adiabatic limit}, we take the adiabatic limit of Alice's protocol so that $\gamma_\tx{A} = 1$; the region (i) in Fig.~\ref{fig:Adiabaticity}.
On the other hand, in Sec.~\ref{sec.Non-adiabatic limit}, we take the nonadiabatic limit of Bob's protocol so that $\gamma_\tx{B} = 0$; the region (ii).
In terms of the causal structure, case (i) necessarily falls within the region (I) in Fig.~\ref{fig:Causality} as mentioned later, whereas case (ii) can be anywhere.
Then, in Sec.~\ref{sec.Spacelike-separated}, we assume Alice's and Bob's protocols are spacelike separated from each other; the region (IV) in Fig.~\ref{fig:Causality}.
In Sec.~\ref{sec.One-way}, we discuss the case where Alice can influence Bob's spin in a causal way while Bob can not influence Alice's spin; the region (II).
In Sec.~\ref{sec.short summary}, we summarize which propagator is responsible for the spin-spin correlation in each limiting case, mentioning the necessary condition for the quantum entanglement between the two spins to be generated.

%%%%%%%%FIG%%%%%%
%%%%%%%%%%%%%%%%%

%%%%%%%%%%%%%%%%%%%%%%%%%%%%%%%%%%%%%%
\subsection{Adiabatic limit\label{sec.Adiabatic limit}}
Let us first consider the region (i) in Fig.~\ref{fig:Adiabaticity} characterized by $\gamma_\tx{A} = 1$.
It can be realized by taking the adiabatic limit of $T^\tx{A}_\tx{on, off} \to \infty$ with the strength of the coupling $\bar{\lambda}_\tx{A}$ kept finite, 
and we have
\eqn{
\mf{G}^\tx{AA}_\tx{K} = 0 ~. \label{G_K^AA=0}
}
Then Bob's protocol must be in the region (I) in Fig.~\ref{fig:Causality}. It is necessary 
because of the Robertson-Schr\"{o}dinger inequality (\ref{RS-inequality}). Indeed the condition (\ref{G_K^AA=0}) requires
\eqn{\mf{G}^\tx{BA} _\tx{K} = 0 ~, ~~ \mf{G}^\tx{AB}_\tx{R} = \mf{G}^\tx{BA}_\tx{R} =:  \Theta  ~, }
where
$\Theta = \bar{J} \ \bar{T}^\tx{B}$
is a product of  the effective duration of the interaction between Bob's spin and the field\footnote{
With an explicit form of the time dependence of $\lambda_\tx{B}$ given by (\ref{time-dependence}),
we have $\bar{T}^\tx{B}= T^\tx{B} + (T^\tx{B}_\tx{off} + T^\tx{B}_\tx{on})/2$.
}
\eqn{
\bar{T}^\tx{B} \coloneqq \frac{1}{\bar{\lambda}_\tx{B}}\int_{-\infty}^{\infty} \dd{t} \lambda_\tx{B}(t)
}
and the effective coupling between two spins
\eqn{
\bar{J} \coloneqq&  \bar{\lambda}_\tx{B} \int_{-\infty}^{\infty} \dd{t}' G_\tx{R}(t-t',\vb*{x}_\tx{B}-\vb*{x}_\tx{A})  \lambda_\tx{A}(t') \\
=& \bar{\lambda}_\tx{B} \int_{-\infty}^{\infty} \dd{t}' G_\tx{A}(t-t',\vb*{x}_\tx{B}-\vb*{x}_\tx{A})  \lambda_\tx{A}(t')  = \bar{\lambda}_\tx{A}\bar{\lambda}_\tx{B} \times  \frac{\exp (-m D)}{4 \pi D} ~ .
}
It is nothing but the coefficient $\bar{J}$ in (\ref{ferromagnetic-interaction}).
We have taken $t_\tx{i} \to -\infty$ and $t_\tx{f} \to \infty$  to realize the limit of $T^\tx{A}_\tx{on, off} \to \infty$.

\subsubsection{Nonseparability and quantum entanglement}
In this case, Alice and Bob are quantum entangled as a nonseparable state. It can be seen from the nonvanishing negativity.
Let us first look at the reduced density matrix of Alice and Bob.
The spin correlations in (\ref{Cs}) are given by 
\aln{
C_{xx} &= \La \hat{\sigma}^\tx{A}_x \hat{\sigma}^\tx{B}_x \Ra = \gamma_\tx{B} ~,~~ &
C_{yy} &= \La \hat{\sigma}^\tx{A}_y \hat{\sigma}^\tx{B}_y \Ra = 0  ~, \notag \\
C_{x0} &= \La \hat{\sigma}^\tx{A}_x \Ra =  \cos ( 2 \Theta )~,~~&
C_{0x} &= \La \hat{\sigma}^\tx{B}_x \Ra = \gamma_\tx{B} \cos ( 2 \Theta ) ~,  \label{Cs-adiabatic}  \\
C_{yz} &=\La \hat{\sigma}^\tx{A}_y \hat{\sigma}^\tx{B}_z \Ra = -   \sin ( 2 \Theta ) ~,~~ &
C_{zy}&=\La \hat{\sigma}^\tx{A}_z \hat{\sigma}^\tx{B}_y \Ra = -  \gamma_\tx{B} \sin ( 2 \Theta ) ~. \notag
\label{eq:spin_correlation_addiabatic}
}
From these spin correlations, 
we can obtain the density matrix,
\eqn{
\hat{\rho}_\tx{AB} = \frac{1+ \gamma_\tx{B}}{2} \ket{\Theta}_\tx{AB}\!\bra{\Theta} + \frac{1- \gamma_\tx{B}}{2} \hat{\sigma}^\tx{B}_z \ket{\Theta}_\tx{AB}\!\bra{\Theta} \hat{\sigma}^\tx{B}_z ~, \label{rho_AB-adiabatic}
}
where $\ket{\Theta}_\tx{AB}$ is defined in (\ref{Psi_AB-Newtonian}).
Its simple form, written as a statistical sum of the two states $\ket{\Theta}_\tx{AB}\!\bra{\Theta}$ and $\hat{\sigma}^\tx{B}_z \ket{\Theta}_\tx{AB}\!\bra{\Theta} \hat{\sigma}^\tx{B}_z$, can be derived from the eigenvalues (\ref{mu_AB}) with the spin correlations in (\ref{Cs-adiabatic}),
\eqn{
\mu^{-s_2}_\tx{AB} =0  ~, ~~\mu^{+s_2}_\tx{AB} = \frac{1 + s_2 \gamma_\tx{B}}{2} \label{mu_AB-adiabatic} ~.
}

\begin{figure}
\begin{center}
 \includegraphics[width=9cm]{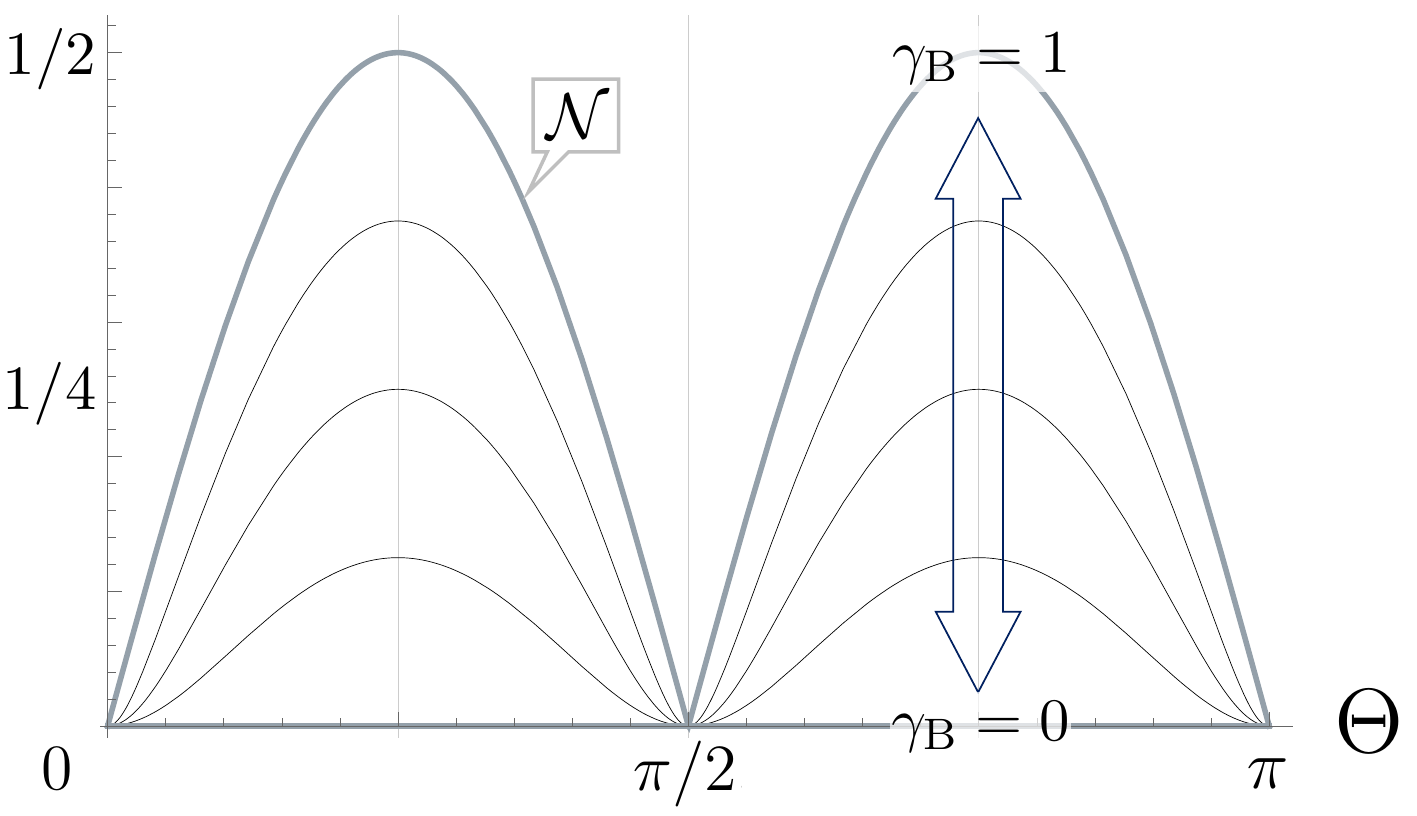}
\caption{The entanglement negativity $\cl{N} = |\tilde{\mu}_\tx{AB}^{--}|$ as the function of $\Theta = \bar{J} \times \bar{T}^\tx{B}$ given by (\ref{PT-mu_AB^--})
with $\gamma_\tx{B}=0$, $0.25$, $0.5$, $0.75$ and $1$.
Regardless of $\gamma_\tx{B}$, it takes maximum values with $\Theta = \pi /4$ mod $\pi /2$.
In the limit of $\gamma_\tx{B} \to 1$, it gives the negativity of the pure state consisting of the two spins discussed in Section \ref{sec.Newtonian} and behaves consistently with the entanglement entropy $\Sigma \qty(\cos ( 2 \Theta ))$ depicted in Fig.~\ref{fig:S_N}.
In $\gamma_\tx{B} \to 0$, it vanishes regardless of the values of $\Theta$ due to the total decoherence of Bob's spin. }
\label{fig:Negativity}
\end{center}
\end{figure}

The density matrix (\ref{rho_AB-adiabatic}) is nonseparable since one of the eigenvalues of the partial transposition (\ref{PT-mu_AB}) with $s_1 = s_2 = -1$,
\eqn{
\tilde{\mu}_\tx{AB}^{--} = \frac{1}{4} \qty{ 1- \gamma_\tx{B} - \sqrt{ 1+ \gamma_\tx{B}^2 -  2 \gamma_\tx{B} \cos (4 \Theta) } }  \label{PT-mu_AB^--}
}
becomes negative.
It shows that the two spins are quantum entangled as long as $\gamma_\tx{B} >0$.
In this case, the negativity (\ref{entanglement-negativity}) is merely the absolute value of $\tilde{\mu}_\tx{AB}^{--}$ which depicted in Fig.~\ref{fig:Negativity} as the function of $\Theta$ with different values of $\gamma_\tx{B}$.
In the adiabatic limit of Bob, $\gamma_\tx{B}\to1$, the negativity reduces to the Newtonian one~\eqref{eq:negativity_Newtonian}.
The negativity takes the maximum value with $\Theta = \pi /4$ mod $\pi /2$ irrespective of $\gamma_\tx{B}$. 
It is consistent with the fact that the entanglement entropy (\ref{S_N}) in the Newtonian picture depicted in Fig.~\ref{fig:S_N} takes the maximal value with $\Theta = \pi /4$.
At $\Theta = 0$ mod $\pi /2$, the negativity vanishes as
 the reduced state (\ref{rho_AB-adiabatic}) is disentangled to be
\eqn{\hat{\rho}_\tx{AB}|_{\Theta = 0} = \frac{\hat{1}^\tx{A} + \hat{\sigma}_x^\tx{A} }{2} \frac{\hat{1}^\tx{B} +\gamma_\tx{B} \hat{\sigma}_x^\tx{B} }{2} ~.  \label{product-state}}
In the nonadiabatic limit, $\gamma_\tx{B}=0$ with a general value of $\Theta$, the negativity vanishes and the state becomes separable, which is the special case of Sec.~\ref{sec.Non-adiabatic limit}, and 
the explicit form of the density matrix is given in \eqref{rho_AB-non-adiabatic}.
%%%%%%Fig%%%%%%
%%%%%%%%%%%%%%%

\subsubsection{Spin-spin correlations}
\begin{figure}
\begin{center}
 \includegraphics[width=9cm]{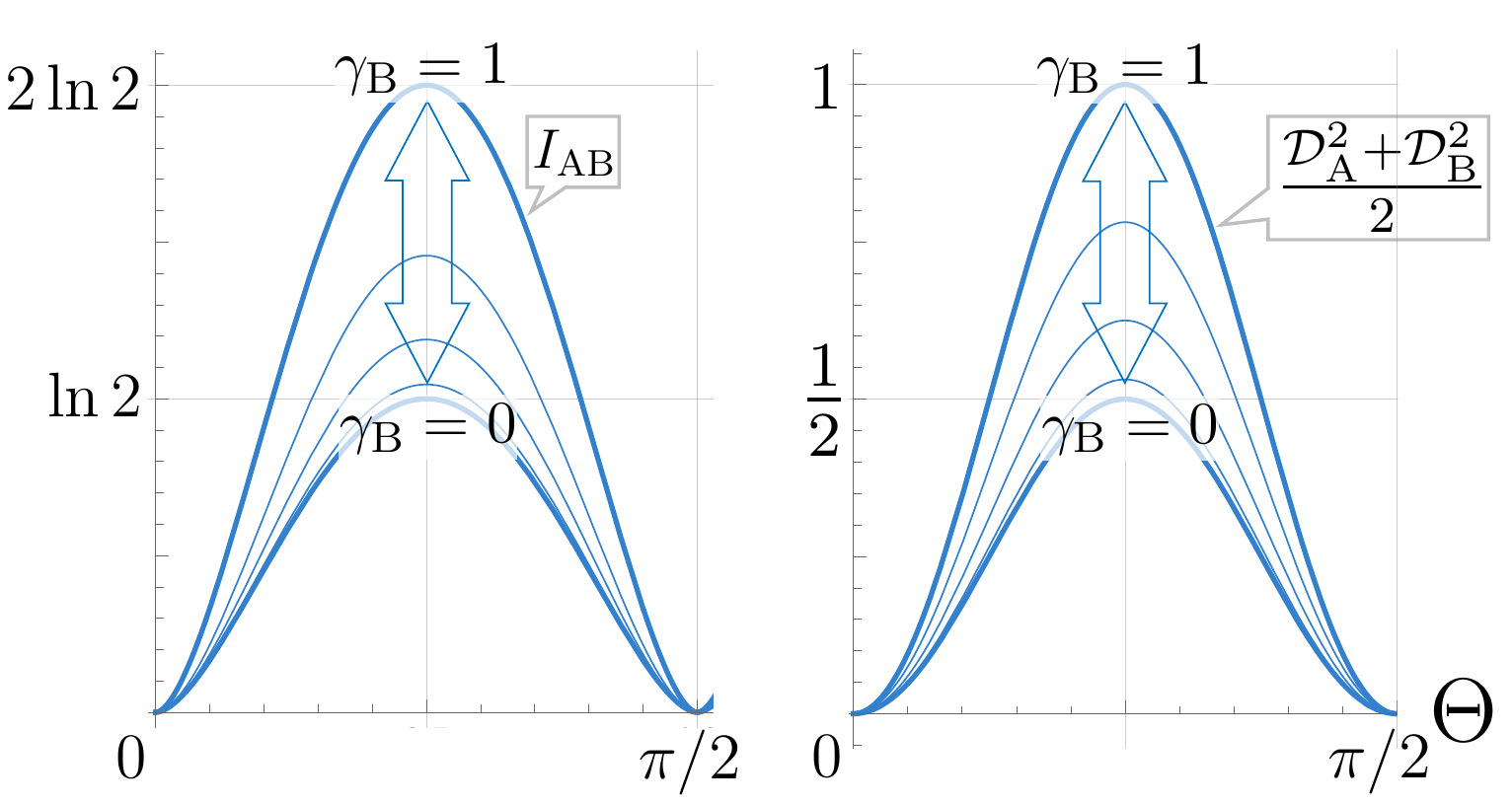}
\caption{Left panel: The mutual information $I_\tx{AB}$ between the two spins in the adiabatic limit of Alice  $\gamma_\tx{A}=1$.
$\Theta = \bar{J} \times \bar{T}^\tx{B}$ is given by (\ref{I_AB-adiabatic}) and each line corresponds to 
Bob's nonadiabaticity, $\gamma_\tx{B}=0$, $0.25$, $0.5$, $0.75$ and $1$.
It is positively correlated with the negativity $\cl{N} = |\tilde{\mu}_\tx{AB}^{--}|$ depicted in Fig.~\ref{fig:Negativity}.
In the limit of $\gamma_\tx{B} \to 1$, $I_\tx{AB}/2 $  coincides with the entanglement entropy $\Sigma \qty(\cos ( 2 \Theta ))$ in the Newtonian approximation (\ref{S_N}). $I_\tx{AB}$ decreases as Bob's nonadiabaticity is increased. 
Even in the limit $\gamma_\tx{B} \to 0$, the mutual information $I_\tx{AB}$ is nonvanishing unlike the negativity.
Right panel: The averaged distinguishability $(\cl{D}_\tx{A}^2 + \cl{D}_\tx{B}^2 )/2$
behaves in a similar way, and it takes the maximum value $(1+\gamma_\tx{B}^2 )/2$ with $\Theta = \pi /4$ mod $\pi /2$.
}
\label{fig:I-adiabatic}
\end{center}
\end{figure}

The adiabatic limit can be compared to the Newtonian approximation as we vary Bob's nonadiabaticity $\gamma_\tx{B}$ from $1$ to $0$.
Let us evaluate the mutual information $I_\tx{AB}$  and compare it to the Newtonian case of $\gamma_\tx{B}=1$.
From (\ref{Cs-adiabatic}), we obtain
\eqn{
&\La \delta \hat{\sigma}_x^\tx{A} \delta \hat{\sigma}_x^\tx{B} \Ra = \gamma_\tx{B} \sin^2 (2 \Theta ) ~, ~~~ \La \delta \hat{\sigma}_y^\tx{A} \delta \hat{\sigma}_y^\tx{B} \Ra = 0  ~,  \\
&\La \delta \hat{\sigma}_y^\tx{A} \delta \hat{\sigma}_z^\tx{B} \Ra= -   \sin ( 2 \Theta )  ~,  ~~~ \La \delta \hat{\sigma}_z^\tx{A} \delta \hat{\sigma}_y^\tx{B} \Ra = - \gamma_\tx{B}   \sin ( 2 \Theta ) ~.
\label{connected-part-adiabatic}}
From the eigenvalues (\ref{mu_AB-adiabatic}) of $\hat{\rho}_\tx{AB}$, the entropy (\ref{S_AB}) quantifying the entanglement between the field and the two spins is obtained as
\eqn{
S(\hat{\rho}_\tx{AB}) &= \Sigma (\gamma_\tx{B}) ~, \label{S_AB-adiabatic}
}
which depends only on $\gamma_\tx{B}$, see (\ref{Sigma_v}) for the definition of the function $\Sigma$.
In the adiabatic limit $\gamma_\tx{B} \to 1$, it vanishes so that the reduced system of the two spins becomes pure and (\ref{rho_AB-adiabatic}) reproduces the Newtonian one (\ref{rho_AB-Newtonian}) as expected.
The eigenvalues of $\hat{\rho}_\tx{A}$ given by (\ref{mu_A}) are the same as (\ref{mu_N}) in the Newtonian picture.
On the other hand,
the eigenvalues of $\hat{\rho}_\tx{B}$ given by (\ref{mu_B}) deviates from them by $\gamma_\tx{B}\ne 1$.
The entanglement entropies of Alice and Bob are given respectively by
\aln{
S(\hat{\rho}_\tx{A}) &=  \Sigma \qty(\cos ( 2 \Theta ) )~, \label{S_A-adiabatic}\\
S(\hat{\rho}_\tx{B}) &= \Sigma \qty(\gamma_\tx{B} \cos ( 2 \Theta ) ) ~. \label{S_B-adiabatic} 
}
From (\ref{S_AB-adiabatic}), (\ref{S_A-adiabatic}) and (\ref{S_B-adiabatic}), we find the mutual information between Alice's and Bob's spin in (\ref{I_AB}) as
\eqn{
I_\tx{AB} &= \Sigma \qty(\cos ( 2 \Theta ) ) + \Sigma \qty(\gamma_\tx{B} \cos ( 2 \Theta ) )  -  \Sigma (\gamma_\tx{B}) ~.  \label{I_AB-adiabatic}
}
It takes the maximum value $2 \ln 2 - \Sigma (\gamma_\tx{B})$ at $\Theta = \pi /4$ mod $\pi /2$,
as seen in Fig.~\ref{fig:I-adiabatic}.
On the other hand, it vanishes at $\Theta = 0$ when $\hat{\rho}_\tx{AB}$ becomes a tensor product state (\ref{product-state}).
The mutual information is comparable to the averaged distinguishability of Alice and Bob,
\begin{align}
\frac{\mathcal{D}_\tx{A}^2+\mathcal{D}_\tx{B}^2}{2}&=
\frac{|C_{yz}|^2+|C_{zy}|^2}{2}
=\frac{1+\gamma_\tx{B}^2}{2}\sin^2(2\Theta) ~.
\end{align}
Figure~\ref{fig:I-adiabatic} shows that behaviors of the mutual information and the averaged distinguishability agree with each other.
%%%%%%%%FIG%%%%%%%%
%%%%%%%%%%%%%%%%%%%

\subsubsection{Spin-field correlations}
The mutual information of Alice and Bob  $I_\tx{AB}$ decreases as the nonadiabaticity of Bob is increased from the Newtonian limit $\gamma_\tx{B}=1$
to the nonadiabatic limit of Bob with $\gamma_\tx{B}=0$.
This behavior can be understood as a consequence of one of the trade-off relations in (\ref{Information-trade-off}) written as
\eqn{  I_\tx{AB}  = 2 S(\hat{\rho}_\tx{A})   - I_{\tx{A}\phi} = 2 \Sigma \qty(\cos ( 2 \Theta ))- I_{\tx{A}\phi}     ~.  \label{Information-trade-off-(i)A}}
The first term is the value expected  in the Newtonian approximation.
$I_\tx{AB}$ is smaller than this value by the second term representing correlations between the field and Alice's spin. 
We can also get another trade-off relation for $I_\tx{AB}$ focussing on the correlation of field and Bob spin; $I_\tx{AB} =  2 S(\hat{\rho}_\tx{B})   - I_{\tx{B}\phi}= 2 \Sigma \qty(\gamma_\tx{B} \cos ( 2 \Theta ))- I_{\tx{B}\phi} $.
These trade-off relations reflect that a part of information of the subsystem of Alice and Bob is carried out by the emission of radiation.

\begin{figure}
\begin{center}
 \includegraphics[width=9cm]{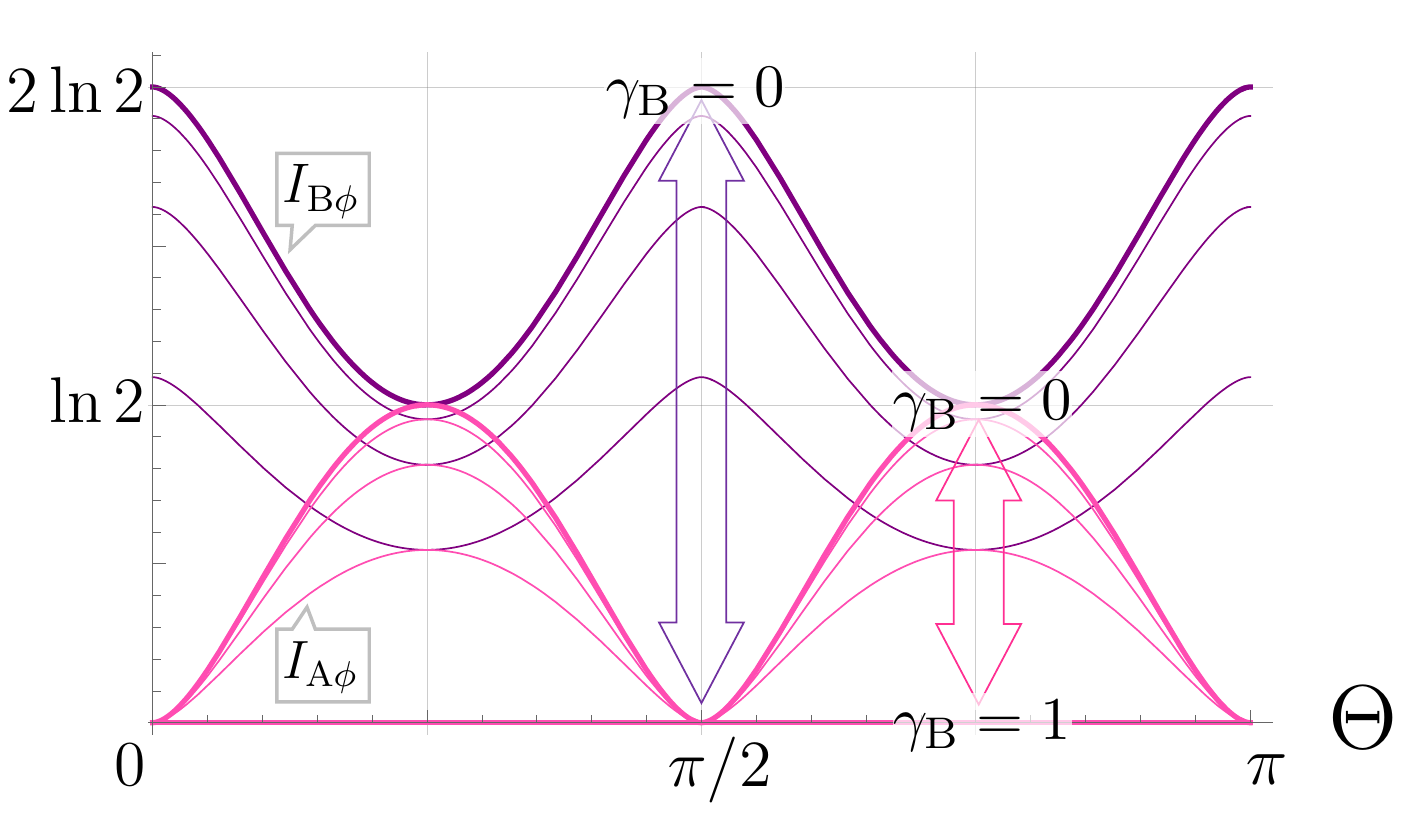}
\caption{The mutual informations between the field and one of the spins, $I_{\tx{A}\phi}$ in pink and $I_{\tx{B}\phi}$ in purple, in the adiabatic limit of Alice $\gamma_\tx{A}=1$ as functions of $\Theta = \bar{J} \times \bar{T}^\tx{B}$ given in (\ref{I_A(B)phi-adiabatic}) with $\gamma_\tx{B}=0$, $0.25$, $0.5$, $0.75$ and $1$.
In the limit of $\gamma_\tx{B}\to 1$, both vanish since the field decouples with the spins. 
$I_{\tx{B}\phi}$ does not vanish with $\gamma_\tx{B} < 1$.
On the other hand,
with $\Theta$ changed, $I_{\tx{A}\phi}$ vanishes with $\Theta = 0$ mod $\pi /2$ and positively correlated with $I_\tx{AB}$ depicted in Fig.~\ref{fig:I-adiabatic}, which is because the field is correlated with Alice's spin only via the correlation with Bob's spin in the current limit of $\gamma_\tx{A} = 1$.
}
\label{fig:Isp-adiabatic}
\end{center}
\end{figure}

The carried-out information,  $I_{\tx{A}\phi}$ and $I_{\tx{B}\phi}$,  are given by (\ref{I_Aphi}) and (\ref{I_Bphi}),
\eqn{
I_{\tx{A}\phi} &=   \Sigma \qty(\cos ( 2 \Theta ) ) + \Sigma (\gamma_\tx{B})  - \Sigma \qty(\gamma_\tx{B} \cos ( 2 \Theta ) )  ~, \\
I_{\tx{B}\phi} &=  \Sigma \qty(\gamma_\tx{B} \cos ( 2 \Theta ) )   + \Sigma (\gamma_\tx{B})  -\Sigma \qty(\cos ( 2 \Theta ) )  ~, \label{I_A(B)phi-adiabatic}
}
and  depicted in Fig.~\ref{fig:Isp-adiabatic}. 
The mutual information of Alice and Bob $I_\tx{AB}$, a counterpart of distinguishability, 
satisfies the trade-off relation (\ref{tradeof-equality-MI}) with the visibility represented by $S(\hat{\rho}_\tx{A})$  (or $P_\tx{A}$)
and the carried-out information $I_{\tx{A}\phi}$.
%%%%%%%%FIG%%%%%%%%
%%%%%%%%%%%%%%%%%%%

One can explicitly see that the spin-field correlation functions behave correspondingly.
For instance, the correlation with Alice's spin,
\eqn{
\bra{\Psi_\tx{f}}  \delta \hat{\sigma}^\tx{A}_{y} \delta \hat{\phi}(\vb*{x}) \ket{\Psi_\tx{f}} &=  \La \delta \hat{\sigma}^\tx{A}_y \delta \hat{\sigma}^\tx{B}_z \Ra  \int_{t_\tx{i}} ^{t_\tx{f}} dt G_\tx{R}(t_\tx{f}-t,\vb*{x}-\vb*{x}_\tx{B}) \lambda_\tx{B}(t) ~, 
}
can not vanish at $^\forall  \vb*{x}$ on the final time slice with $\gamma_\tx{B}  = \exp (-2 \mf{G}_\tx{K}^\tx{BB} ) < 1$ except with $\Theta = 0$ mod $\pi /2$. Here, $\delta \hat{\phi}(\vb*{x})=\hat{\phi}(\vb*{x})-\expval*{\hat{\phi}(\vb*{x})}$.
Note that Alice's spin is correlated with the field only through the correlation with Bob's spin, $\La \delta \hat{\sigma}^\tx{A}_y \delta \hat{\sigma}^\tx{B}_z \Ra = -   \sin ( 2 \Theta )$.
On the other hand, the correlation with Bob's spin,
\eqn{
\bra{\Psi_\tx{f}}  \delta \hat{\sigma}^\tx{B}_{z} \delta \hat{\phi} (\vb*{x}) \ket{\Psi_\tx{f}} &=  \int_{t_\tx{i}} ^{t_\tx{f}}\dd t G_\tx{R}(t_\tx{f}-t, \vb*{x}- \vb*{x}_\tx{B})  \lambda_\tx{B}(t)  ~,
}
can not vanish at $^\forall  \vb*{x}$ on the final time slice with $\gamma_\tx{B} < 1$.

%%%%%%%%%%%%%%%%%%%%%%%%%%%%%%%%%%%%%%%%%%%%%%%%%%%%%%%%%%%
\subsection{Nonadiabatic limit\label{sec.Non-adiabatic limit}}
Let us consider the region (ii) in Fig.~\ref{fig:Adiabaticity} characterized by $\gamma_\tx{B} = 0$, where Bob turns on and/or turns off the spin-field interaction abruptly so that
\begin{equation}
\mf{G}^\tx{BB}_\tx{K} \to \infty  ~,
\end{equation} 
while $\mf{G}^\tx{AA}_\tx{K}$ and $\mf{G}^\tx{BA}_\tx{K}$ remain finite.
As seen shortly, the quantum entanglement between Alice's spin and Bob's spin is not generated in this case because of the total decoherence of Bob's spin.
The COW experiment falls within this limiting case in the sense mentioned below.

\subsubsection{Separability of spins}
The expectation values in (\ref{Cs}) become
\aln{
C_{xx} &= \La \hat{\sigma}^\tx{A}_x \hat{\sigma}^\tx{B}_x \Ra = 0 ~,~~ &
C_{yy} &= \La \hat{\sigma}^\tx{A}_y \hat{\sigma}^\tx{B}_y \Ra =  0 ~, \notag \\
C_{x0} &= \La \hat{\sigma}^\tx{A}_x \Ra = \gamma_\tx{A}  \cos ( 2 \mf{G}^\tx{AB}_\tx{R} ) ~,~~&
C_{0x} &= \La \hat{\sigma}^\tx{B}_x \Ra =  0 ~, \label{C-non-adiabatic}  \\
C_{yz} & =\La \hat{\sigma}^\tx{A}_y \hat{\sigma}^\tx{B}_z \Ra =- \gamma_\tx{A}   \sin ( 2 \mf{G}^\tx{AB}_\tx{R} )  ~,~~ &
C_{zy}& =\La \hat{\sigma}^\tx{A}_z \hat{\sigma}^\tx{B}_y \Ra = 0 ~. \notag
}
In this case, Bob's distinguishability of Alice's spin $\cl{D}_\tx{B}=|C_{zy}|=0$ vanishes due to the nonadiabaticity. From the expression in Sec.~\ref{sec.Separability condition}, we can see that 
the entanglement negativity vanishes since  
\eqn{
\mu_\tx{AB}^{s_1 +} = \mu_\tx{AB}^{s_1 -} = \frac{1 + s_1 \gamma_\tx{A}}{4} ~. \label{mu_AB-non-adiabatic}
}
Accordingly the density matrix (\ref{rho_AB}) is written in a separable form:
\eqn{
\hat{\rho}_\tx{AB} &= \frac{1}{2} \qty( \hat{1}^\tx{A} \hat{1}^\tx{B} + C_{x0} \hat{\sigma}^\tx{A}_x \hat{1}^\tx{B} + C_{yz} \hat{\sigma}^\tx{A}_y \hat{\sigma}^\tx{B}_z ) \\
&= \hat{\rho}_\tx{A}^{+} \frac{\ket{+}_\tx{B}\!\bra{+}}{2}  + \hat{\rho}_\tx{A}^{-} \frac{\ket{-}_\tx{B}\!\bra{-}}{2}  ~, \label{rho_AB-non-adiabatic}
}
where $\hat{\rho}_\tx{A}^{+}$ and $\hat{\rho}_\tx{A}^{-}$ are defined in (\ref{rho^pm}), and
written explicitly as
\eqn{
 \hat{\rho}_\tx{A}^{\pm}  &= \frac{1 + \gamma_\tx{A}}{2} \ket{\pm \Theta}_\tx{A}\! \bra{\pm \Theta} +  \frac{1 - \gamma_\tx{A}}{2} \hat{\sigma}_z^\tx{A}\ket{\pm \Theta}_\tx{A}\! \bra{\pm \Theta} \hat{\sigma}_z^\tx{A} ~~~\tx{with}~~~ \Theta = \mf{G}_\tx{R}^\tx{AB} ~, \label{rho_A^pm}
 }
where $\ket{\pm \Theta}_\tx{A}$ are defined as (\ref{Theta_A(B)}).
The separability in the nonadiabatic case $\gamma_\tx{B}=0$  is  because Bob's spin is totally decohered due to the violent particle creation.\footnote{
In principle, we can take a time slice in the course of the protocol where $\lambda_\tx{B}(t)$ is still nonzero to compute the reduced density matrix.
It is effectively the same as the sudden turn-off, and hence, we get the same result as (\ref{rho_AB-non-adiabatic}).
However, in this case, the disappearance of the interference fringe is not because of the particle creation, but rather because of the ``false decoherence''\cite{2011arXiv1110.2199U} due to the orthogonality between two states with different Newtonian potentials.
}

Therefore, spin correlations discussed below are classical unlike in the adiabatic limit discussed in the previous section.

%%%%%%%%Fig%%%%%%%
\begin{figure}
\begin{center}
 \includegraphics[width=9cm]{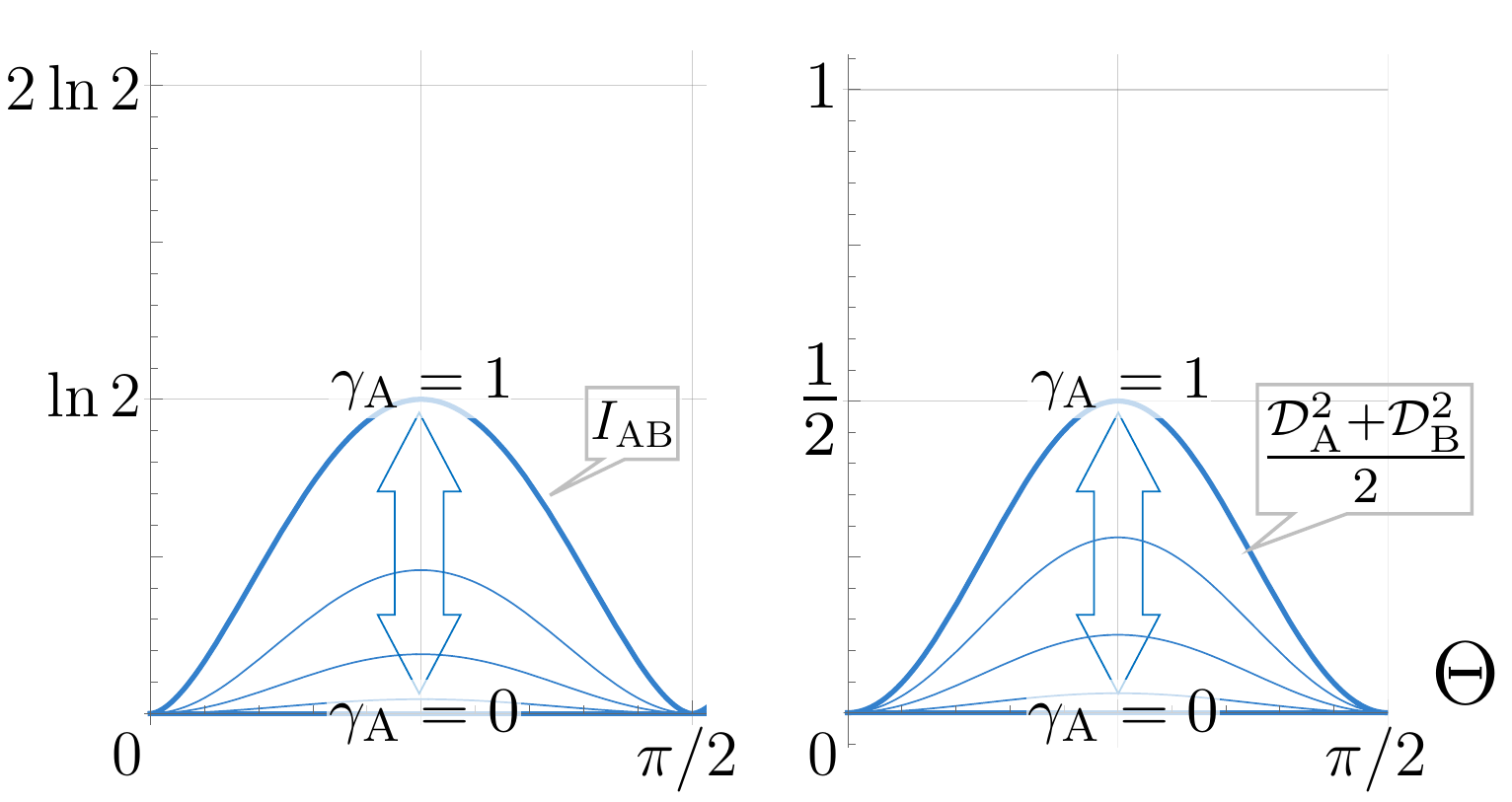}
\caption{Left panel: The mutual information $I_\tx{AB}$ in the nonadiabatic limit of Bob $\gamma_\tx{B}=0$ as the function of $\Theta = \mf{G}_\tx{R}^\tx{AB}$ given by (\ref{I_AB-nonadiabatic})  with $\gamma_\tx{A}=0$, $0.25$, $0.5$, $0.75$ and $1$.
It takes maximum value with $\Theta = \pi /4$ mod $\pi /2$.
In the limit of $\gamma_\tx{A} \to 1$ toward the region (iii) in Fig.~\ref{fig:Adiabaticity}, it corresponds to the one obtained in the previous section with $\gamma_\tx{A} = 1$ and $\gamma_\tx{B} \to 0$, see Fig.~\ref{fig:I-adiabatic}.
On the other hand, $\gamma_\tx{A} \to 0$ leads to the region (iv) in Fig.~\ref{fig:Adiabaticity}, and then, $I_\tx{AB}$ vanishes.
Right panel: The averaged distinguishability $(\cl{D}_\tx{A}^2 + \cl{D}_\tx{B}^2 )/2$
behaves in a similar way, and it takes the maximum value $\gamma_\tx{A}^2/2$ with $\Theta = \pi /4$ mod $\pi /2$.
}
\label{fig:I-nonadiabatic}
\end{center}
\end{figure}
%%%%%%%%%%%%%%%%%%%

\subsubsection{Spin-spin correlations}
From (\ref{C-non-adiabatic}),  most of the correlation functions vanish and the only nonvanishing one is 
\eqn{
\La \delta \hat{\sigma}_y^\tx{A} \delta \hat{\sigma}_z^\tx{B} \Ra = - \gamma_\tx{A}   \sin ( 2 \Theta )  ~,
\label{connected-part-nonadiabatic}}
whose absolute value is nothing but Alice's distinguishability of Bob's spin, 
$\cl{D}_\tx{A}=\gamma_\tx{A} | \sin ( 2 \Theta )|$ given in (\ref{D}).
Thus, the mutual information $I_\tx{AB}$  and $\cl{D}_\tx{A}$ embody the same information in this case.

Let us look at it more explicitly. In the nonadiabatic limit of Bob, the entanglement entropies of Alice and Bob in  (\ref{S_A}) and (\ref{S_B}) are given by
\aln{
S(\hat{\rho}_\tx{A}) &= \Sigma \qty( \gamma_\tx{A}  \cos ( 2 \Theta ) ) ~, \label{S_A-nonadiabatic} \\
S(\hat{\rho}_\tx{B}) &=  \ln 2 ~. \label{S_B-nonadiabatic}
}
From (\ref{S_B-nonadiabatic}), we find $P_\tx{B}=2(\ln 2-S(\hat{\rho}_\tx{B}))=0$, and equivalently, the visibility of Bob vanishes, ${\cal V}_\tx{B}=0$.
It is because the large nonadiabaticity completely destroys the interference between $\hat{\sigma}_z^\tx{B}=\pm 1$ states.
Using  (\ref{mu_AB-non-adiabatic}), the entanglement entropy (\ref{S_AB}) of the field is given by 
\eqn{
S(\hat{\rho}_\tx{AB})
=\ln 2+\Sigma \qty( \gamma_\tx{A}  )   ~,
}
and the mutual information is obtained as
\eqn{
I_\tx{AB} = \Sigma \qty( \gamma_\tx{A}  \cos ( 2 \Theta ) ) - \Sigma \qty( \gamma_\tx{A}  ) ~. 
\label{I_AB-nonadiabatic}
}
It behaves in a similar way as the averaged distinguishability does, as seen in Fig.~\ref{fig:I-nonadiabatic}.

In the nonadiabatic limit where the coherence of Bob's spin is totally lost, we can get a stronger upper bound on the mutual information $I_\tx{AB}$:
When the reduced density matrix $\hat{\rho}_\tx{AB}$ is given by the form\footnote{In a more general setup with the asymmetric interferometer with $\cl{P}_\tx{B} \ne 0$ mentioned in Footnote \ref{asymmetric-interferometer}, one finds
$\hat{\rho}_\tx{AB} = \sum_{k=\pm} p_{k} \hat{\rho}^{k}_\tx{A} \ket{k}_\tx{B}\!\bra{k}$
with $p_\pm$ such that $|p_+ -p_-| = \cl{P}_\tx{B}$ and $p_+ + p_- = 1$,
and then, $S(\hat{\rho}_\tx{AB}) = S(\hat{\rho}_\tx{B}) +  \sum_{k=\pm} p_{k} S(\hat{\rho}^k_\tx{A})$ where $S(\hat{\rho}_\tx{B}) = \Sigma( \cl{P}_\tx{B} )$.
Therefore, $S(\hat{\rho}_\tx{AB}) \geq S(\hat{\rho}_\tx{B})$.
Based on the fact that projective measurements never decrease entropy, $S(\hat{\rho}_\tx{AB}) \geq S(\hat{\rho}_\tx{A})$ can also be shown \cite{10.5555/1972505}.}
of (\ref{rho_AB-non-adiabatic}), one can show that
\eqn{
S(\hat{\rho}_\tx{AB}) \geq \tx{max} \qty{ S( \hat{\rho}_\tx{A} ) , S( \hat{\rho}_\tx{B} )}
\label{conditional entropy}
}
which corresponds to the inequality, always satisfied in classical information theory, to guarantee nonnegativity of conditional entropy.
Then, the mutual information defined by (\ref{I_AB}) satisfies
\eqn{
I_\tx{AB} \leq \tx{min} \qty{ S( \hat{\rho}_\tx{A} ) , S( \hat{\rho}_\tx{B} )} ~. \label{stronger bound on I_AB}
}
Since $S( \hat{\rho}_\tx{A} ), S( \hat{\rho}_\tx{B} ) \leq \ln 2$, we necessarily have $I_\tx{AB} \leq \ln 2$. 
This can be considered as an entropic counterpart of 
the upper bound of the averaged distinguishability $({\cal D}_\tx{A}^2+{\cal D}_\tx{B}^2)/2 \le 1/2$,
which obeys from the vanishing $\mathcal{D}_\tx{B}$ and $\mathcal{D}_\tx{A}\leq1$.

\subsubsection{Spin-field correlations}
\begin{figure}
\begin{center}
 \includegraphics[width=9cm]{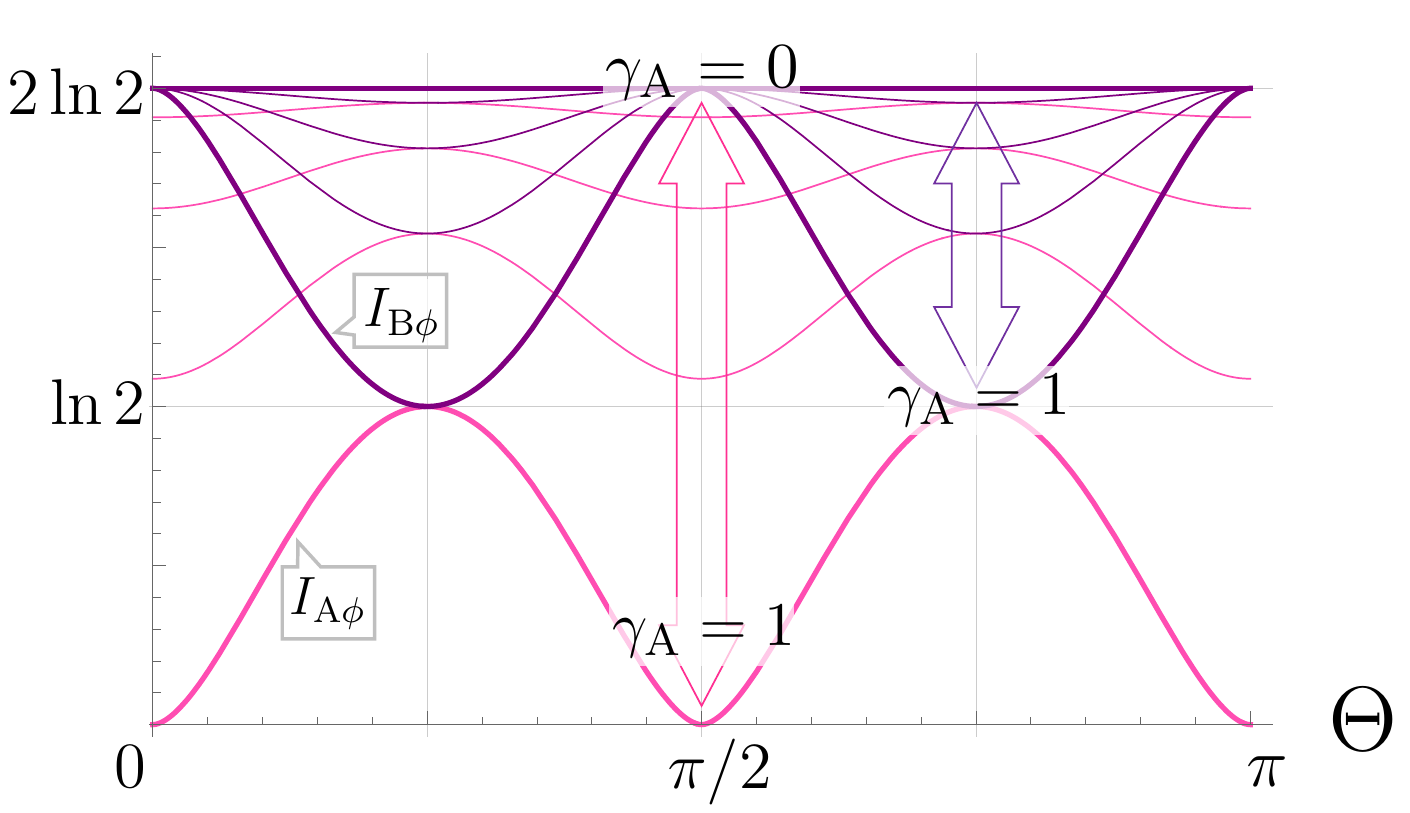}
\caption{The mutual informations between the field and one of the spins, $I_{\tx{A}\phi}$ in pink and $I_{\tx{B}\phi}$ in purple, in the nonadiabatic limit of Bob $\gamma_\tx{B}=0$ as the functions of $\Theta = \mf{G}_\tx{R}^\tx{AB}$ given by (\ref{I_AB-nonadiabatic})  with $\gamma_\tx{A}=0$, $0.25$, $0.5$, $0.75$ and $1$.
In the limit of $\gamma_\tx{A} \to 1$ toward the region (iii) in Fig.~\ref{fig:Adiabaticity}, these correspond to the ones obtained in the previous section with $\gamma_\tx{A} = 1$ and $\gamma_\tx{B} \to 0$, see Fig.~\ref{fig:Isp-adiabatic}. 
Only in this limit, $I_{\tx{A}\phi}$ can vanish with $\Theta = 0$ mod $\pi /2$.
On the other hand, $\gamma_\tx{A} \to 0$ leads to the region (iv) in Fig.~\ref{fig:Adiabaticity} where both take the maximal value $2 \ln 2$.}
\label{fig:Isp-nonadiabatic}
\end{center}
\end{figure}

The amount of carried-out information by the field is quantified by the mutual informations given by (\ref{I_Aphi}) and (\ref{I_Bphi}):
\eqn{
I_{\tx{A}\phi} &=\Sigma \qty(\gamma_\tx{A} \cos ( 2 \Theta ) ) + \Sigma (\gamma_\tx{A})   ~, \\
I_{\tx{B}\phi} &=   2 \ln 2  + \Sigma (\gamma_\tx{A})  - \Sigma \qty(\gamma_\tx{A} \cos ( 2 \Theta ) )   ~,\label{I_A(B)phi-non-adiabatic}
}
depicted in Fig.~\ref{fig:Isp-nonadiabatic}.
Because of the inequality (\ref{conditional entropy}), 
these are bounded from below as $I_{\tx{A}\phi} \geq S(\hat{\rho}_\tx{A})$,  $I_{\tx{B}\phi} \geq S(\hat{\rho}_\tx{B})$.
%%%%%%%%Fig%%%%%%%
%%%%%%%%%%%%%%%%%%%
The fact that $I_{\tx{B}\phi}$ never vanishes,
$I_{\tx{B}\phi} \geq S(\hat{\rho}_\tx{B}) = \ln 2$, implies that there always is some correlation between Bob's spin and the field. 
For instance,
\eqn{
\bra{\Psi_\tx{f}}  \delta \hat{\sigma}^\tx{B}_{z} \delta \hat{\phi} (\vb*{x}) \ket{\Psi_\tx{f}} &=  \int_{t_\tx{i}} ^{t_\tx{f}}\dd t G_\tx{R}(t_\tx{f}-t, \vb*{x}- \vb*{x}_\tx{B})  \lambda_\tx{B}(t) }
does not vanish at $^\forall  \vb*{x}$ on the final time slice.
Speaking of correlations between the field and Alice's spin, for instance,
\eqn{
\bra{\Psi_\tx{f}}  \delta \hat{\sigma}^\tx{A}_{z} \delta \hat{\phi} (\vb*{x}) \ket{\Psi_\tx{f}} &=  \int_{t_\tx{i}} ^{t_\tx{f}}\dd t G_\tx{R}(t_\tx{f}-t, \vb*{x}- \vb*{x}_\tx{A})  \lambda_\tx{A}(t) }
does not vanish with $\gamma_\tx{A} = \exp (-2 \mf{G}_\tx{K}^\tx{AA} ) < 1$.
Another example is 
\eqn{
\bra{\Psi_\tx{f}}  \delta \hat{\sigma}^\tx{A}_{y} \delta \phi(\vb*{x}) \ket{\Psi_\tx{f}} 
=&  -2 \La \hat{\sigma}^\tx{A}_{x}\Ra \int dt G_\tx{K}(t_\tx{f}-t,\vb*{x}-\vb*{x}_\tx{A}) \lambda_\tx{A}(t) \\
& + \La \hat{\sigma}^\tx{A}_{y} \hat{\sigma}^\tx{B}_{z} \Ra \int dt G_\tx{R}(t_\tx{f}-t,\vb*{x}-\vb*{x}_\tx{B}) \lambda_\tx{B}(t) ~,
}
which vanishes at $^\forall  \vb*{x}$ only with $\gamma_\tx{A}  \to 1$ and $\Theta = 0$ mod $\pi/2$ as $I_{\tx{A}\phi}$ in (\ref{I_A(B)phi-non-adiabatic}) implies.

\subsubsection{Analogy with the COW experiment setup}
Since the total system is in a pure state, the spin correlations are all originated from the quantum effects.
However, (\ref{rho_AB-non-adiabatic}) is now in a separable
form of density matrices, that is, Alice's spin and Bob's spin are not quantum entangled;
the only nonzero spin correlation (\ref{connected-part-nonadiabatic}) 
is to be understood as a classical correlation.

The situation is similar to the COW experiment setup \cite{PhysRevLett.34.1472} where the Earth is described as a classical source or a classical statistical ensemble composed of vast numbers of particles. 
The interference pattern of the neutron beam split and recombined is affected by Earth's gravitational potential. However, it does not require entanglement between the neutron and the Earth.
The nonadiabatic limit considered here represents the situation of the COW experiment. 
Ideally, the experiment is to be done without any decoherence of the neutron beam.
Then this corresponds to the limit of $\gamma_\tx{A} \to 1$ in our model, and thus, the region (iii) in Fig.~\ref{fig:Adiabaticity}.
In the $\gamma_\tx{A} \to 1$ limit, the reduced density matrix (\ref{rho_AB-non-adiabatic}) becomes 
\eqn{
\hat{\rho}_\tx{AB} &=   \ket{+\Theta}_\tx{A}\! \bra{+\Theta}  \frac{\ket{+}_\tx{B}\! \bra{+}}{2}  +   \ket{-\Theta}_\tx{A}\! \bra{-\Theta}  \frac{\ket{-}_\tx{B}\! \bra{-}}{2}  ~,
\label{rho_AB-adiabatic-nonadiabatic}}
where $\ket{\pm \Theta}_\tx{A}$ is defined as (\ref{Theta_A(B)}). The state can also be obtained from the density matrix (\ref{rho_AB-adiabatic}) in the region (i) with $\gamma_\tx{B} \to 0$.
The two different eigenvalues of Alice's $z$-spin correspond to two different paths of neutrons at different heights
and the two different eigenvalues of Bob's $z$-spin are understood as two different configurations of particles that compose the Earth in the COW experiment setup.\footnote{
Because of the low dimensionality of the Hilbert space of Bob's variable,
there is no decoherence due to its very presence. 
However,
if one replaces Bob's spin with some continuous variable such as the meter considered in \cite{Hidaka:2022tzk},  the decoherence of Alice's spin is caused even with $\gamma_\tx{A} =1$.
Then, it models the possible decoherence of Alice's spin caused by the gravitational potential sourced by a vast number of massive objects surrounding it.
}

\subsubsection{Nonadiabatic limit of both Alice and Bob}
Furthermore, let us assume that both Alice and Bob turn on and/or turn off their spin-field interactions abruptly.
It leads to the region (iv) in Fig.~\ref{fig:Adiabaticity} since 
\begin{equation}
\mf{G}^\tx{AA}_\tx{K} , ~\mf{G}^\tx{BB}_\tx{K} \to \infty  ~.
\end{equation} 
Note that $\mf{G}^\tx{AB}_\tx{K}$ remains finite in general.
In this case, all the $C$'s vanish, and the entropies are computed as
\eqn{
S(\hat{\rho}_\tx{A}) =  S(\hat{\rho}_\tx{B}) = \ln 2 ~,~~ S(\hat{\rho}_\tx{AB}) = 2 \ln 2 ~. \label{maximally-entangled-with-field}
}
As the result, we get
\eqn{I_\tx{AB} = 0 ~, ~~~I_{\tx{A}\phi} =I_{\tx{B}\phi} = 2 \ln 2 ~,}
that is, Alice and Bob can not gain any information about each other's spins since the quantum coherence of both two spins is totally lost now.

%%%%%%%%%%%%%%%%%%%%%%%%%%%%%%%%%%%%
\subsection{Spacelike separated case \label{sec.Spacelike-separated}}
In the next two sections, we consider two limiting cases in the causal structure of Fig.~\ref{fig:Causality}. 
First, we study the region (IV) where Alice's and Bob's protocols are spacelike separated from each other.
In this case, both of the retarded Green's functions vanish, and hence, we have
\begin{equation}
\mf{G}^\tx{AB}_\tx{R} =  \mf{G}^\tx{BA}_\tx{R} = 0 ~,
\end{equation}
while $\mf{G}^\tx{AB}_\tx{K}$ is finite in general.
As seen below, while spin-spin correlations are induced, 
negativity vanishes and 
there is no quantum entanglement between the two spins generated. 
\subsubsection{Vacuum-induced spin correlation}
In this case, the expectation values in (\ref{Cs}) become
\aln{
C_{xx} &= \La \hat{\sigma}^\tx{A}_x \hat{\sigma}^\tx{B}_x \Ra = \gamma_\tx{A} \gamma_\tx{B} \cosh (4 \mf{G}^\tx{BA}_\tx{K}) ~,~~ &
C_{yy} &= \La \hat{\sigma}^\tx{A}_y \hat{\sigma}^\tx{B}_y \Ra = \gamma_\tx{A}\gamma_\tx{B} \sinh (4 \mf{G}^\tx{BA}_\tx{K})  ~, \notag \\
C_{x0} &= \La \hat{\sigma}^\tx{A}_x \Ra =  \gamma_\tx{A} ~,~~&
C_{0x} &= \La \hat{\sigma}^\tx{B}_x \Ra = \gamma_\tx{B} ~,   \\
C_{yz} &=\La \hat{\sigma}^\tx{A}_y \hat{\sigma}^\tx{B}_z \Ra = 0 ~,~~ &
C_{zy}&=\La \hat{\sigma}^\tx{A}_z \hat{\sigma}^\tx{B}_y \Ra = 0 ~. \notag
}
Then,
two of the spin correlation functions in (\ref{connected-part}) are nonvanishing
\eqn{
&\La \delta \hat{\sigma}_x^\tx{A} \delta \hat{\sigma}_x^\tx{B} \Ra = \gamma_\tx{A} \gamma_\tx{B} \qty[ \cosh (4 \mf{G}^\tx{BA}_\tx{K}) - 1 ]  ~, \\
&\La \delta \hat{\sigma}_y^\tx{A} \delta \hat{\sigma}_y^\tx{B} \Ra =\gamma_\tx{A} \gamma_\tx{B} \sinh (4 \mf{G}^\tx{BA}_\tx{K}) ~,\\
&\La \delta \hat{\sigma}_y^\tx{A} \delta \hat{\sigma}_z^\tx{B} \Ra = \La \delta \hat{\sigma}_z^\tx{A} \delta \hat{\sigma}_y^\tx{B} \Ra =0
\label{connected-part-spacelike}}
which depends only on the Keldysh Green's function  $\mf{G}^\tx{BA}_\tx{K} \ne 0$.
Thus, 
even though Alice's spin and Bob's spin do not have the causally connected direct interactions,
spins get correlated through the correlations of the quantum field in the vacuum.

Since  $C_{yz}$ and $C_{zy}$ vanish, 
both Alice's and Bob's distinguishability vanish $\cl{D}_\tx{A}=\cl{D}_\tx{B}=0$.
Nevertheless, the mutual information becomes nonzero, which will be shown in Fig.~\ref{fig:IAB-spacelike}.
From the discussion below (\ref{negativity-definition})
in Section \ref{sec.Separability condition},   the entanglement negativity vanishes and  the reduced state $\hat{\rho}_\tx{AB}$ is separable.
The two spins are not entangled and their correlations are classical even though they are originated from the entanglement with the field.
In \cite{Matsumura:2021tlu}, it is shown that 
the separability in this case follows from the relation $[\hat{H}_\tx{A}, \hat{H}_\tx{B}] = 0$.

\subsubsection{Spin correlation mediated by created particles}
\begin{figure}
\begin{center}
 \includegraphics[width=8cm]{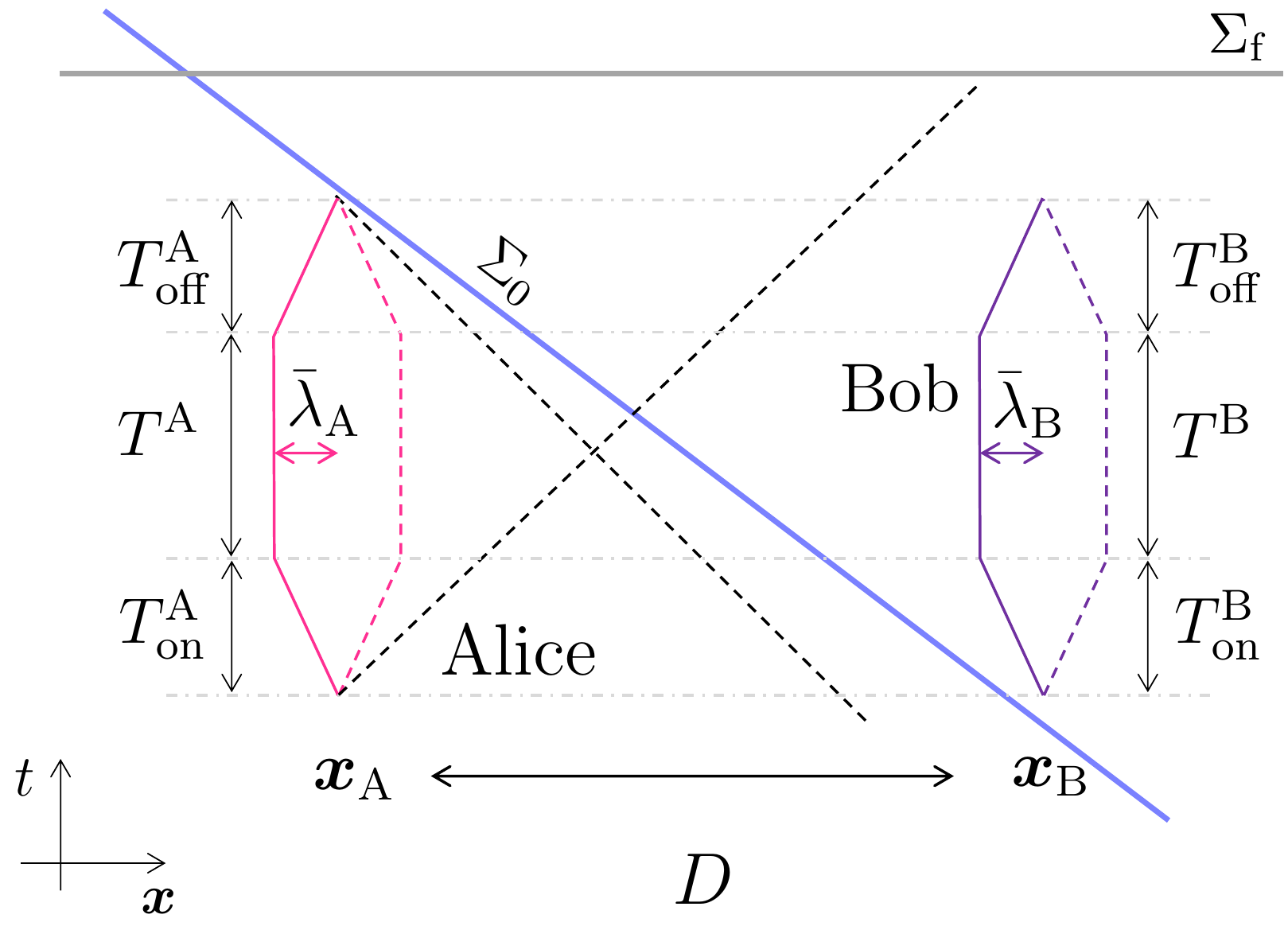}
\caption{
Alice's spin and Bob's spin do not have causal influences on each other.
In order to see that the reduced density matrix of Alice's spin does not care about Bob's spin, one can take a time slice $\Sigma_0$ depicted by the blue line which is mapped to the final time slice $\Sigma_\tx{f}$ by a Lorentz boost.
}
\label{fig:setup-Spacelike}
\end{center}
\end{figure}

As we will see below, particle creation is necessary for the spin correlation to be generated.
The following discussion holds with Alice and Bob interchanged.
Because of relativistic causality, the presence of Bob's spin can not affect Alice's measurement of her own spin,
that is to say,
the reduced density matrix of Alice's spin $\hat{\rho}_\tx{A}$ is independent from whether Bob's spin exists or not.
Then, the entanglement entropy $S(\hat{\rho}_\tx{A})$ of Alice
can be evaluated in the absence of Bob's spin.
That is, $S(\hat{\rho}_\tx{A})$ in the current case measures the amount of entanglement between Alice's spin and the field.
This can be explicitly shown as follows:
Consider a time slice $\Sigma_0$ depicted in Fig.~\ref{fig:setup-Spacelike}.
Since Bob's spin has not yet interacted with the field by this time,
the full density matrix on this slice can be written as a tensor product, $\hat{\rho}_\tx{B,i} \hat{\rho}_0$ where $\hat{\rho}_\tx{B,i}$ 
is the initial state of Bob's spin\footnote{\label{footnote:rho_B}
$\hat{\rho}_\tx{B,i}$ can be any state unless it is entangled with the field and Alice's spin.
For the initial state (\ref{spin-initial-state}), we have $\hat{\rho}_\tx{B,i} = (\ket{+}_\tx{B}+\ket{-}_\tx{B})(\bra{+}_\tx{B}+\bra{-}_\tx{B})/2$.
Note also that, if the spin is dynamical on its own, $\hat{\rho}_\tx{B,i}$ should be replaced with $\hat{U}_\tx{B} \hat{\rho}_\tx{B,i} \hat{U}^\dag_\tx{B}$ where $\hat{U}_\tx{B}$ is the corresponding time evolution operator acting only on the Hilbert space of Bob's spin nontrivially. It does not change the conclusion here.
}
and $\hat{\rho}_0$ is an entangled state of Alice's spin and the field in general.
The density matrix at the slice $\Sigma_0$ is related to the density matrix $\hat{\rho}_\tx{f} := \ket{\Psi_\tx{f}} \bra{\Psi_\tx{f}}$ on the final time slice $\Sigma_\tx{f}$ as $\hat{\rho}_\tx{f} = \hat{L}_{\tx{B}\phi} (\hat{\rho}_0 \hat{\rho}_\tx{B,i}) \hat{L}^\dag_{\tx{B}\phi}$, where $\hat{L}_{\tx{B}\phi}$ is the time evolution operator corresponding to Lorentz boost from $\Sigma_0$ to $\Sigma_\tx{f}$ which is trivial acting on the Hilbert space of Alice's spin.
Therefore, $\hat{\rho}_\tx{A}= \tr_{\tx{B}\phi} \{ \hat{\rho}_\tx{f} \} = \tr_{\tx{B}\phi} \{ \hat{\rho}_0 \hat{\rho}_\tx{B,i} \} = \tr_{\phi} \{ \hat{\rho}_0  \}$, and thus, $S(\hat{\rho}_\tx{A})$ originates from the entanglement between Alice's spin and the field;
if there is no particle creation from Alice's spin, then $S(\hat{\rho}_\tx{A}) = 0$ and the above inequality tells us that $I_\tx{AB} = 0$, which means that there is no correlation between the two spins generated.

Now that we have the explicit forms of $\hat{\rho}_\tx{A}$ and $\hat{\rho}_\tx{B}$ and their eigenvalues given by (\ref{mu_A}) and (\ref{mu_B}), the entanglement entropies (\ref{S_A}) and (\ref{S_B}) are obtained as
\eqn{
S(\hat{\rho}_\tx{A}) = \Sigma (\gamma_\tx{A}) ~, ~~ S(\hat{\rho}_\tx{B}) = \Sigma (\gamma_\tx{B})  ~,
}
which depend only on $\gamma_\tx{A}$ and $\gamma_\tx{B}$, respectively.
Then, in the adiabatic limit where either $\gamma_\tx{A} \to 1$ or $\gamma_\tx{B} \to 1$, the entropies vanish as expected.
Therefore, all of the correlation functions in (\ref{connected-part-spacelike}) vanish in the adiabatic limit with no particle created.

Note also that, once the explicit expressions for the correlation functions in (\ref{connected-part-spacelike}) are obtained, 
the uncertainty relation (\ref{RS-inequality-K}),
\eqn{
(\mf{G}^\tx{BA}_\tx{K})^2 \leq \mf{G}^\tx{AA}_\tx{K} \mf{G}^\tx{BB}_\tx{K}  = \frac{\ln \gamma_\tx{A} \ln \gamma_\tx{B}}{4}   \label{RS-inequality-K-2}
}
guarantees that all the spin correlations vanish in the adiabatic limit.

%%%%%%%%fig%%%%%%%%%%
%%%%%%%%%%%%%%%%%%%%

\subsubsection{Asymptotic behavior of spin correlations}
\begin{figure}
\begin{center}
 \includegraphics[width=9cm]{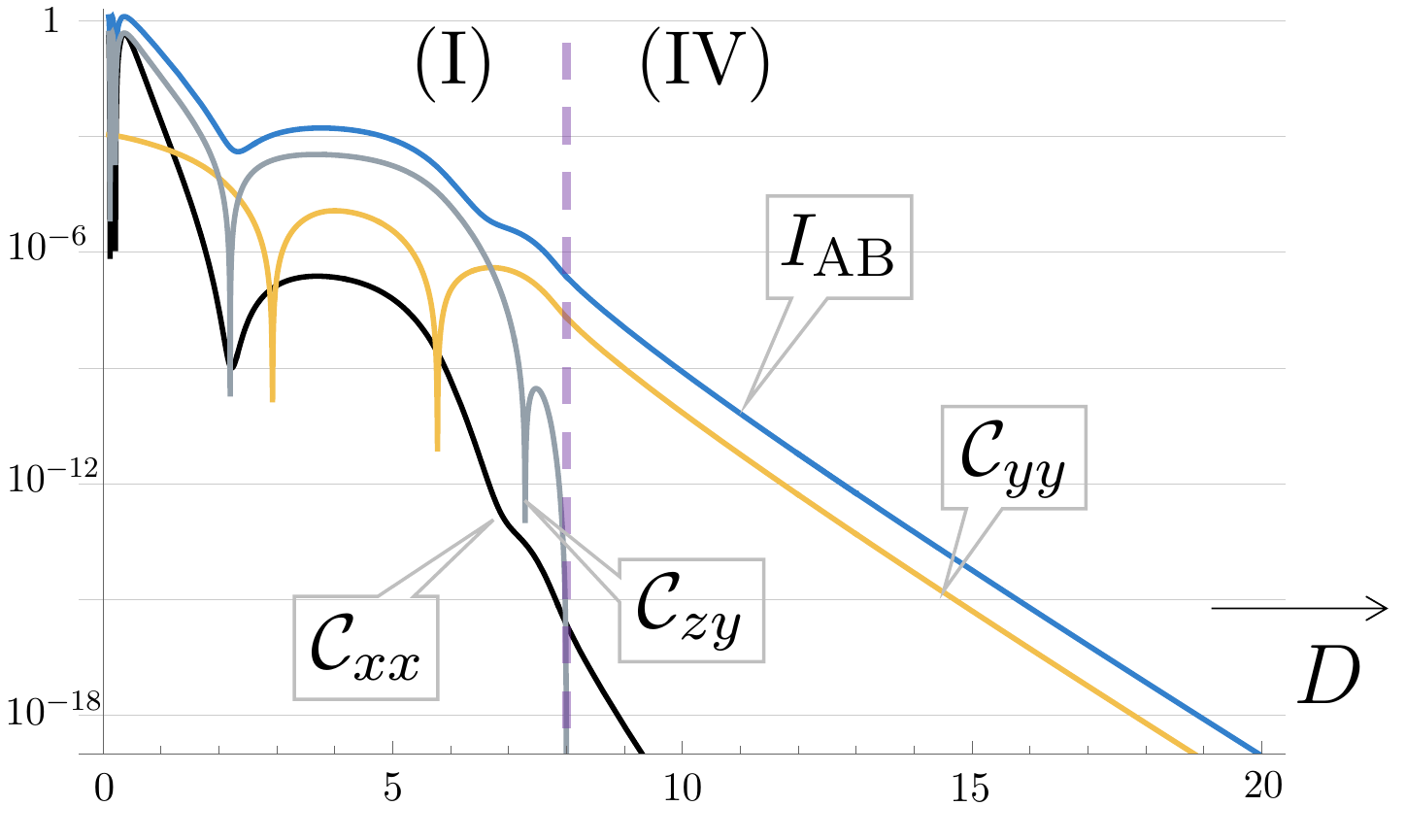}
\caption{
Spin correlation 
$\cl{C}_{w w'} \coloneqq \La \delta \hat{\sigma}_w^\tx{A} \delta \hat{\sigma}_{w'}^\tx{B} \Ra^2 / 2$ with $w, w'=x,y,z$ as a function of $D$ in the spacelike separated case.
With the mass of scalar field $m$ set to be unity,
the parameters are chosen in such a way that the system is symmetric under the swapping of $\lambda_\tx{A}(t)$ and $\lambda_\tx{B}(t)$;
$\bar{\lambda}_\tx{A}=\bar{\lambda}_\tx{B}=1$, $T^\tx{A}_\tx{on}= T^\tx{A}_\tx{off}=  T^\tx{B}_\tx{on}= T^\tx{B}_\tx{off} =2$, $T^\tx{A}=T^\tx{B}=4$ and $t_\tx{on}^\tx{A} = t_\tx{on}^\tx{B}$, see Fig.~\ref{fig:setup-Spacelike}. With these parameters, we get $\mf{G}_\tx{K}^\tx{AA} =\mf{G}_\tx{K}^\tx{BB}\simeq 0.0125$.
With $D\geq 8$, we are in the current limiting case (IV).
The nonzero components, $\cl{C}_{yy}$ and $\cl{C}_{xx}$, are plotted with the yellow and black lines.
Just for comparison, we have included the part with $D< 8$.
Then, the system falls within the region (I) in Fig.~\ref{fig:Causality} where
we have $\cl{C}_{zy}$ and  $\cl{C}_{yz}$ nonvanishing.
Because of the symmetry, $\cl{C}_{zy}=\cl{C}_{yz}$ and it is plotted with the gray line.
The mutual information $I_\tx{AB}$ plotted with the blue line is always larger than $\cl{C}$'s as (\ref{WVHC}) says.}
\label{fig:IAB-spacelike}
\end{center}
\end{figure}

The time integration of the Keldysh function $\mf{G}^\tx{BA}_\tx{K}$ behaves as
\eqn{
\mf{G}^\tx{BA}_\tx{K} \propto \frac{\exp (-m D)}{D^{3/2}} ~, %\to 0 
}
see Fig.~\ref{fig:Gs-spacelike} in Appendix~\ref{app:Numerical evaluation}.
Note that $\mf{G}^\tx{AA}_\tx{K}$ and $\mf{G}^\tx{BB}_\tx{K}$ are independent from $D$.
Therefore, the correlation functions in (\ref{connected-part-spacelike}) behave as $\La \delta \hat{\sigma}_x^\tx{A} \delta \hat{\sigma}_x^\tx{B} \Ra \approx 8 \gamma_\tx{A} \gamma_\tx{B} ( \mf{G}^\tx{BA}_\tx{K})^2$ and $\La \delta \hat{\sigma}_y^\tx{A} \delta \hat{\sigma}_y^\tx{B} \Ra \approx 4 \gamma_\tx{A} \gamma_\tx{B} \mf{G}^\tx{BA}_\tx{K}$, whose squared value is  depicted in Fig.~\ref{fig:IAB-spacelike}.

Let us now explicitly see that the mutual information $I_\tx{AB}$ reflects the behavior of the correlations and satisfies the inequality (\ref{WVHC}).
The eigenvalue (\ref{mu_AB}) of $\hat{\rho}_\tx{AB}$ is given in this case as
\eqn{
\mu_\tx{AB}^{s_1 s_2} =& \frac{1}{4} \biggl\{ 1+s_2 \gamma_\tx{A} \gamma_\tx{B} \cosh (4 \mf{G}^\tx{BA}_\tx{K})  \\
&~~~~~~~~+ s_1 \sqrt{(\gamma_\tx{A}+ s_2 \gamma_\tx{B})^2 + \gamma_\tx{A}^2 \gamma_\tx{B}^2 \sinh^2 (4 \mf{G}^\tx{BA}_\tx{K})   } \biggr\} ~. \label{mu_AB-spacelike}
}
Plugging it into (\ref{S_AB}), we get the entanglement entropy $S(\hat{\rho}_\tx{AB})$.
The mutual information (\ref{I_AB}) is obtained as a function of $\mf{G}^\tx{AA}_\tx{K}$, $\mf{G}^\tx{BB}_\tx{K}$ and $\mf{G}^\tx{BA}_\tx{K}$,
\eqn{
I_\tx{AB} = I_\tx{AB}(\gamma_\tx{A}, \gamma_\tx{B}; \mf{G}^\tx{BA}_\tx{K}) ~. \label{I_AB-spacelike}
}
As seen in Fig.~\ref{fig:IAB-spacelike}, it is always larger than the spin correlation functions entering in to the right-hand side of (\ref{WVHC}) denoted by $\cl{C}_{w w'}$.
For sufficiently large values of $D$, the mutual information depends on $D$ in the same manner as $\cl{C}_{yy}$ does.\footnote{
For $\mf{G}^\tx{BA}_\tx{K} = 0$, we get $S(\hat{\rho}_\tx{AB}) = S(\hat{\rho}_\tx{A}) + S(\hat{\rho}_\tx{B})$ since the eigenvalue (\ref{mu_AB-spacelike}) is simplified as
\[
\mu_\tx{AB}^{s_1 s_2} |_{\mf{G}^\tx{BA}_\tx{K} = 0} = \frac{(1+ s'_1 \gamma_\tx{A}) (1+ s_2 s'_1 \gamma_\tx{B})}{4} ~,
\]
where $s'_1 = s_1$ for $s_2 = +1$ and $s'_1 = s_1 \tx{sign} (\gamma_\tx{A} - \gamma_\tx{B})$ for $s_2 = -1$. Then, $I_\tx{AB} = 0$ follows.
}
Though the inequality (\ref{WVHC}) appears to be a rather weak relation, $I_\tx{AB}$ properly represents the spin-spin correlations.

%%%%%FIG%%%%%%%
%%%%%%%%%%%%

\subsubsection{Spin-field correlations}
In the adiabatic limit of $\gamma_\tx{A} = \exp (-2 \mf{G_\tx{K}^\tx{AA}}) \to 1$, 
the correlations between the field and each spin also must vanish. 
It can be seen from the trade-off relations in (\ref{Information-trade-off}) since the entanglement entropy  in the right-hand side in the relation
$S(\rho_\tx{A})$ in (\ref{S_A}) vanishes  $S(\rho_\tx{A})=\Sigma(1)=0$ and the mutual information behaves as
$I_{\tx{A}\phi} \to 0$.  Similarly we have $I_{\tx{B}\phi} \to 0$.
Thus the correlations between the field and spins must vanish. 
One can explicitly see such behaviors as follows. 
For example,
\eqn{
\bra{\Psi_\tx{f}}  \delta \hat{\sigma}^\tx{A}_{y} \delta \phi(\vb*{x}) \ket{\Psi_\tx{f}}
=& -2  \gamma_\tx{A} \int dt G_\tx{K}(t_\tx{f}-t,\vb*{x}-\vb*{x}_\tx{A}) \lambda_\tx{A}(t) 
}
vanishes in the adiabatic limit.
Note that this is independent of Bob's position $\vb*{x}_\tx{B}$.
Thus, we can take a limit of $\bar{\lambda}_\tx{A} \to 0$ and  $D \to \infty$ 
while keeping Alice and Bob to be separated in the spacelike region. In this limit, the above correlation vanishes. 
Furthermore, with the number operator $\hat{N}$, we get
\eqn{
\bra{\Psi_\tx{f}}  \delta \hat{\sigma}^\tx{A}_{x} \delta \hat{N} \ket{\Psi_\tx{f}} =&  -2 \mf{G}_\tx{K}^\tx{AA} \La \hat{\sigma}_x^\tx{A} \Ra =    \gamma_\tx{A}  \ln  \gamma_\tx{A} ~, \\
\bra{\Psi_\tx{f}}  \delta \hat{\sigma}^\tx{A}_{x} \delta \!\!:\!\!\hat{N}^2 \!\!:\!\! \ket{\Psi_\tx{f}} =& - 4  \qty{ \mf{G}_\tx{K}^\tx{AA} \mf{G}_\tx{K}^\tx{BB}  + (\mf{G}_\tx{K}^\tx{BA})^2 } \La \hat{\sigma}_x^\tx{A} \Ra ~,
}
see (\ref{spin-N-correlation}) and (\ref{spin-N2-correlation}) in Appendix \ref{app.Particle creation and Keldysh function}.
The first one explicitly vanishes with $ \gamma_\tx{A} \to 1$.
The second one also vanishes in the adiabatic limit because of the uncertainty relation (\ref{RS-inequality-K-2}).

%%%%%%%%%%%%%%%%%%%%%%%%%%%%%%%%%%%%
\subsection{A case with one-way causal influence \label{sec.One-way}}
Let us consider the region (II) in  Fig.~\ref{fig:Causality} where the coupling between Alice's spin and the field $\lambda_\tx{A}(t)$ is nonvanishing in the causal past of Bob's protocol, whereas Bob has not turned on the spin-field coupling $\lambda_\tx{B}(t)$ in the causal past of Alice's turning $\lambda_\tx{A}(t)$ off.
Then, we have
\eqn{\mf{G}^\tx{AB} _\tx{R} = 0 ~. \label{G_R^AB=0}}
while $ \mf{G}^\tx{BA} _\tx{R}$ does not vanish.
As seen below, again,  there is no quantum entanglement between the two spins generated in this case.

\subsubsection{Causality vs complementarity}
If one accepted the Newtonian picture in this causally disconnected case and naively used the wave function (\ref{Psi_AB-Newtonian}),
\eqn{
\ket{\Theta}_\tx{AB} 
 =\frac{e^{\ri \Theta}}{2} \qty(\ket{++} + \ket{--}) + \frac{e^{-\ri \Theta}}{2} \qty(\ket{+-} + \ket{-+}) 
}
with $\Theta \propto \bar{\lambda}_\tx{B}$ caused by the Newtonian interaction (\ref{ferromagnetic-interaction}),
it would lead to an apparent paradox as follows.
From  (\ref{expectation-value-Newtonian}), we would have  $\La \hat{\sigma}_x^\tx{A}  \Ra = \cos (2 \Theta)$.  It means that the interference of spin up and down states, $\sigma_z=\pm$, would be reduced 
and vanish at the most entangled case of $\Theta=\pi/4$. On the other hand, due to the causality, Bob cannot affect the result of Alice and the interference must be independent of $\Theta \propto \bar{\lambda}_\tx{B}$. 
The paradox, of course, appeared by the Newtonian approximation where the dynamical effects of the fields are neglected. 
This paradox is first discussed in \cite{Mari:2015qva} in the setup of gravitational interactions as the origin of $\Theta$, and its resolution  is given in \cite{Belenchia:2018szb,Danielson:2021egj} by discussing the vacuum fluctuations of the metric field and the emission of on-shell gravitons.
In particular,  in \cite{Danielson:2021egj}, it is shown in an abstract and sophisticated argument 
that the partial decoherence of Alice's spin, reflected by $\gamma_\tx{A} < 1$ below, is inevitable for Bob to gain the ``which-path'' information. 

Let us look at the trade-off relation between the visibility and distinguishability as the resolution of the paradox. 
With (\ref{G_R^AB=0}), the expectation values in (\ref{Cs}) become
\aln{
C_{xx} &= \La \hat{\sigma}^\tx{A}_x \hat{\sigma}^\tx{B}_x \Ra = \gamma_\tx{A} \gamma_\tx{B}  \cosh (4 \mf{G}^\tx{BA}_\tx{K}) ~,~~ &
C_{yy} &= \La \hat{\sigma}^\tx{A}_y \hat{\sigma}^\tx{B}_y \Ra =  \gamma_\tx{A} \gamma_\tx{B}  \sinh (4 \mf{G}^\tx{BA}_\tx{K}) ~, \notag \\
C_{x0} &= \La \hat{\sigma}^\tx{A}_x \Ra =  \gamma_\tx{A} ~,~~&
C_{0x} &= \La \hat{\sigma}^\tx{B}_x \Ra =  \gamma_\tx{B}  \cos ( 2 \mf{G}^\tx{BA}_\tx{R} ) ~,  \label{C-Wald} \\
C_{yz} &=\La \hat{\sigma}^\tx{A}_y \hat{\sigma}^\tx{B}_z \Ra = 0 ~,~~ &
C_{zy}&=\La \hat{\sigma}^\tx{A}_z \hat{\sigma}^\tx{B}_y \Ra = - \gamma_\tx{B}   \sin ( 2 \mf{G}^\tx{BA}_\tx{R} ) ~. \notag
}
As explicitly seen here, the visibility $\cl{V}_\tx{A}^2=\La \hat{\sigma}^\tx{A}_x \Ra^2+\La \hat{\sigma}^\tx{A}_y \Ra^2=\gamma_\tx{A}^2 $ does not depend on $\bar{\lambda}_\tx{B}$ since the retarded Green's function from Bob to Alice is now irrelevant \cite{Hidaka:2022tzk,Sugiyama:2022wcd}.
The fact that Alice's distinguishability $\cl{D}_\tx{A}=|C_{yz}|=0$  vanishes also indicates that Alice cannot gain the which-path information about Bob's spin.

On the other hand, Bob can distinguish Alice's spin since Bob's spin is causally connected to Alice connected; however, the distinguishability is bounded by $\cl{D}_\tx{B}^2\leq 1- \cl{V}_\tx{A}^2=1-\gamma_\tx{A}^2$ from \eqref{wave-particle-duality},
i.e., the decoherence of Alice's spin or the particle production is inevitable for Bob to gain the which-path information.
The inevitable decoherence of Alice's spin is discussed in a quantitative way in \cite{Sugiyama:2022wcd} with the inequality in (\ref{wave-particle-duality}),
which becomes, in the current case,
\eqn{\gamma_\tx{A}^2 +  \gamma_\tx{B}^2   \sin^2 ( 2 \mf{G}^\tx{BA}_\tx{R} ) \leq 1 ~.
\label{wave-particle-duality-explicitly}}
where $\cl{D}_\tx{B}=\gamma_\tx{B}|\sin(2\mf{G}^\tx{BA}_\tx{R} )|$ is employed.
It is also shown numerically in \cite{Sugiyama:2022wcd} that this inequality is guaranteed by the relation
\eqn{
 \frac{(\mf{G}^\tx{BA}_\tx{R} )^2}{4} \leq  \mf{G}^\tx{AA}_\tx{K} \mf{G}^\tx{BB}_\tx{K} =\frac{\ln \gamma_\tx{A} \ln \gamma_\tx{B}}{4} ~,  \label{Yamamoto-inequality}
}
which can be derived from the Robertson inequality.\footnote{The Robertson inequality $\La \delta \hat{\cl{O}}_1^2 \Ra \La  \delta \hat{\cl{O}}_2^2 \Ra \geq   |\La [\delta \hat{\cl{O}}_1, \delta \hat{\cl{O}}_2] \Ra |^2 /4$ is weaker than the Robertson-Schr\"{o}dinger inequality (\ref{Robertson-Schrodinger}). With the choice of the operators given in (\ref{Operators}), one finds $ \mf{G}^\tx{AA}_\tx{K} \mf{G}^\tx{BB}_\tx{K} \geq (\mf{G}^\tx{BA}_\tx{R} - \mf{G}^\tx{AB}_\tx{R}  )^2 /4$.} 
In the following, we discuss a similar trade-off relation in terms of the mutual informations.
%%%%%%%%%%%%%%%
\subsubsection{Spin correlation mediated by created particles}
\begin{figure}
\begin{center}
 \includegraphics[width=8cm]{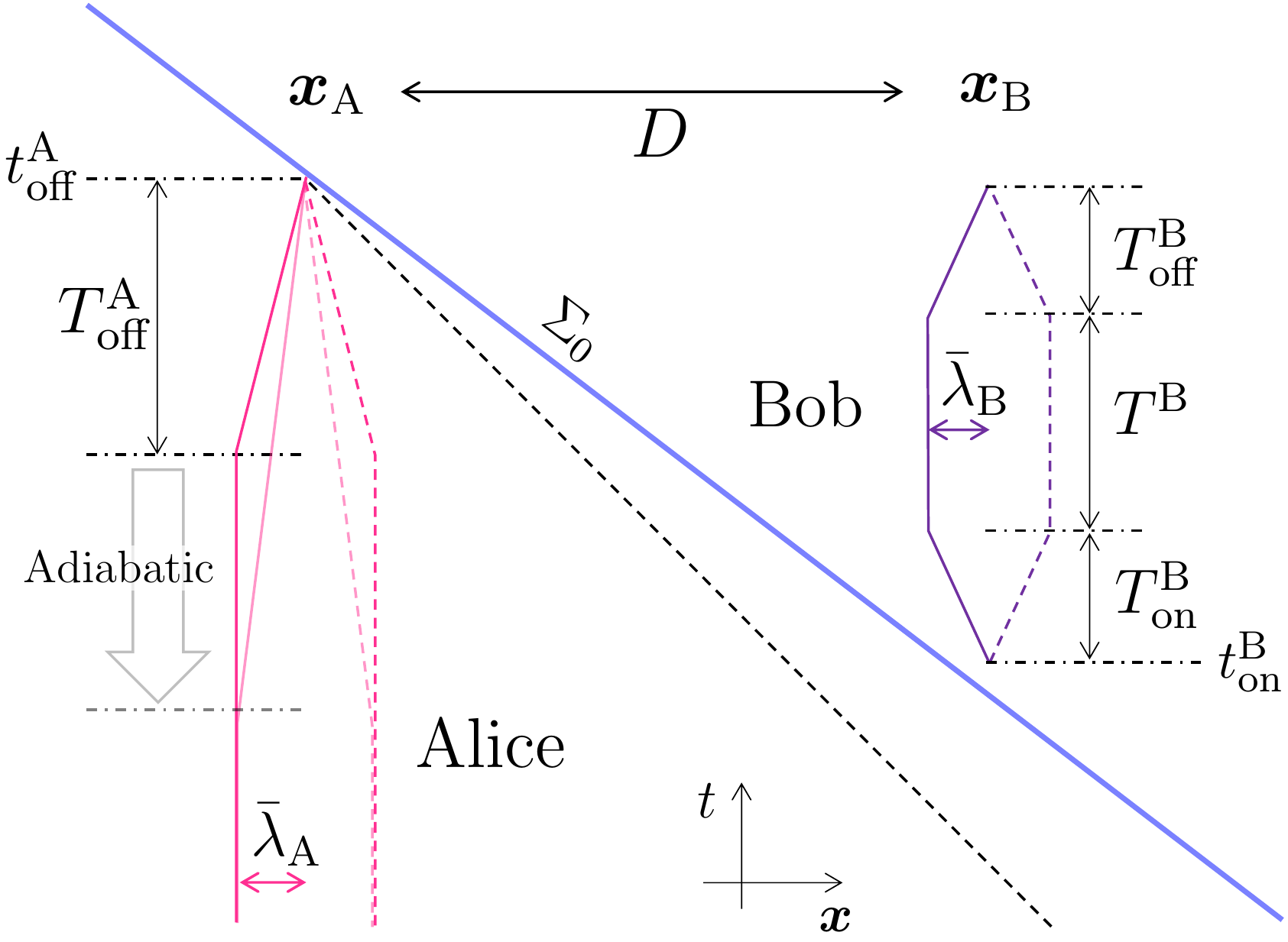}
\caption{
In order to see that the reduced density matrix of Alice's spin does not care about Bob's spin, one can take a time slice $\Sigma_0$ depicted by the blue line.
For taking the adiabatic limit in the case with one-way causal influence,
we take $T^\tx{A}_\tx{on} = \infty$ and change the parameter $T^\tx{A}_\tx{off}$, the length of time that Alice takes to turn off the spin-field interaction,
while keeping all the other parameters unchanged.
In order to restrict ourselves in the region (II) in Fig.~\ref{fig:Causality},
$t_\tx{off}^\tx{A} < t_\tx{on}^\tx{B} + D$ is imposed.
}
\label{fig:Wald-adiabatic}
\end{center}
\end{figure}

At the time slice $\Sigma_0$ depicted in Fig.~\ref{fig:Wald-adiabatic},  Bob's spin is not yet to interact with the field.
Thus Alice's state cannot depend on Bob's action, and for the generation of the spin-spin correlations between Alice and Bob, 
the dynamics of field with particle creation is inevitable as in the case of Sec.~\ref{sec.Spacelike-separated}.
Indeed, 
if there is no particle creation from Alice's spin and $\gamma_\tx{A}=1$, the entanglement entropy $S(\rho_\tx{A})$ in (\ref{S_A})
vanishes since $S(\rho_\tx{A})=\Sigma(1)=0$. Then, from the trade-off relation (\ref{inequalities}), $I_\tx{AB}=0$ and 
no spin correlations are possible.
This can be also seen from the explicit calculations in (\ref{connected-part})  with $\mf{G}_\tx{R}^\tx{AB} =0$. 
If $\mf{G}_\tx{R}^\tx{AB} =0$, the Robertson-Schr\"{o}dinger inequality (\ref{RS-inequality}) becomes 
\eqn{
\frac{(\mf{G}^\tx{BA}_\tx{R})^2}{4}+(\mf{G}^\tx{BA}_\tx{K})^2 \leq \mf{G}^\tx{AA}_\tx{K} \mf{G}^\tx{BB}_\tx{K}  = \frac{\ln \gamma_\tx{A} \ln \gamma_\tx{B}}{4} ~,  \label{RS-inequality-2}
}
and in the adiabatic limit of $\gamma_\tx{A}=1$, we must have $\mf{G}^\tx{BA}_\tx{R}=\mf{G}^\tx{BA}_\tx{K}=0$.\footnote{\label{footnote:Bob's adiabatic limit}
This is not only the case with $\gamma_\tx{A} \to 1$ but also with $\gamma_\tx{B} \to 1$.
The entropy (\ref{S_B}) evaluated with (\ref{mu_B}) in the current case depends not only on $\gamma_\tx{B}$ but also on $\mf{G}_\tx{R}^\tx{BA} $  as
$S(\hat{\rho}_\tx{B}) = \Sigma (\gamma_\tx{B} \cos (2 \mf{G}_\tx{R}^\tx{BA} ))$.
Then, the $\gamma_\tx{B} \to 1$ limit itself does not necessarily make $S(\hat{\rho}_\tx{B})$ vanish.
However, that $\mf{G}_\tx{R}^\tx{BA}$ vanishes with $\gamma_\tx{B} \to 1$ is guaranteed by the inequality (\ref{RS-inequality-2}).
Therefore, the second one of the trade-off relations in (\ref{Information-trade-off}) implies that there are no correlations between the two spins in the adiabatic limit where Bob takes an infinitely long time to turn on and off the spin-field interaction.}
Thus all the correlations in (\ref{connected-part}) 
\eqn{
&\La \delta \hat{\sigma}_x^\tx{A} \delta \hat{\sigma}_x^\tx{B} \Ra =  \gamma_\tx{A}\gamma_\tx{B} \qty[ \cosh (4 \mf{G}_\tx{K}^\tx{BA}) - \cos (2\mf{G}_\tx{R}^\tx{BA})] ~, \\
&\La \delta \hat{\sigma}_y^\tx{A} \delta \hat{\sigma}_y^\tx{B} \Ra = \gamma_\tx{A} \gamma_\tx{B} \sinh (4 \mf{G}_\tx{K}^\tx{BA} )  ~,  \\
&\La \delta \hat{\sigma}_y^\tx{A} \delta \hat{\sigma}_z^\tx{B} \Ra =0  ~,  \\
&\La \delta \hat{\sigma}_z^\tx{A} \delta \hat{\sigma}_y^\tx{B} \Ra = - \gamma_\tx{B}   \sin ( 2 \mf{G}_\tx{R}^\tx{BA} )  ~, \label{connected-part-Wald}
}
must vanish.
When all the correlation functions vanish, it is impossible for Bob to gain any information about Alice's spin; then, the distinguishability $\cl{D}_\tx{B}$ should vanish.
This explicitly follows from 
$\cl{D}_\tx{B}= |\La \delta \hat{\sigma}_z^\tx{A} \delta \hat{\sigma}_y^\tx{B} \Ra|$ in (\ref{D}).

%This type of discussion making use of the degree of freedom to take different time slices was given in \cite{Danielson:2021egj} to argue that the quantum entanglement possibly generated by the Newtonian potential %in the BMV experiment can be strong support for the existence of gravitons.
%Since we take the frame work of the QFT paradigm in this paper, the Newtonian potential is obtained from the retarded potential in the adiabatic limit as explicitly seen in Sec.\ref{sec.Adiabatic limit}.
%In the current case, there is no such retarded potential between Alice's spin and Bob's spin.
%The spin correlation originates solely from the vacuum fluctuations of the field described by $\mf{G}_\tx{K}^\tx{BA}$ in (\ref{connected-part-spacelike}) and it implies the nonzero particle creation for consistency.

%%%%%%%%fig%%%%%%%%%%
%%%%%%%%%%%%%%%%%%%%

\subsubsection{Asymptotic behavior of spin correlations}
\begin{figure}
\begin{center}
 \includegraphics[width=9cm]{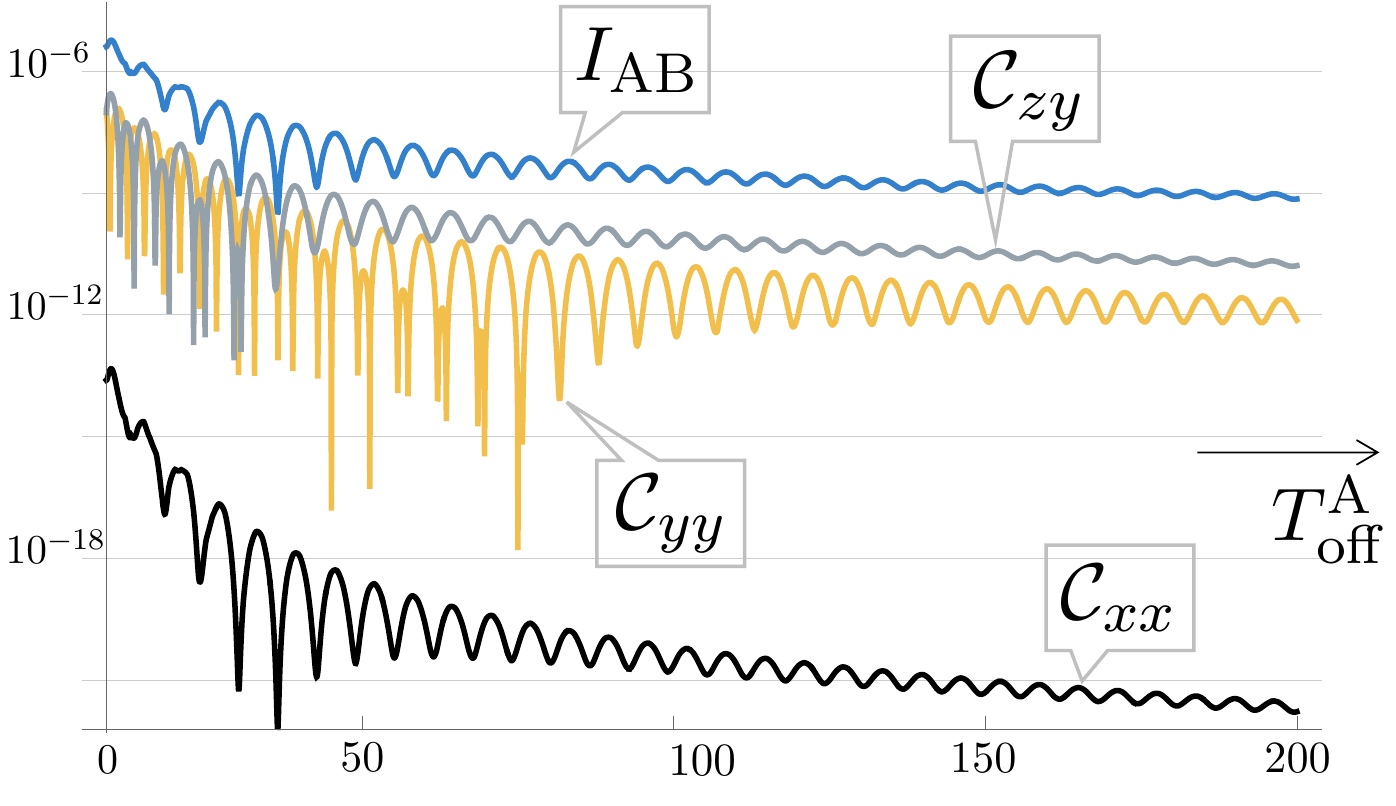}
\caption{
Spin correlation $\cl{C}_{w w'} \coloneqq \La \delta \hat{\sigma}_w^\tx{A} \delta \hat{\sigma}_{w'}^\tx{B} \Ra^2 / 2$ with $w, w'=x,y,z$ as a function of $T^\tx{A}_\tx{off}$ in the case with one-way causal influence.
With the mass of scalar field $m$ set to be unity,
the other parameters are all fixed as follows: $\bar{\lambda}_\tx{A}=\bar{\lambda}_\tx{B}=1$, $T^\tx{A}_\tx{on}= \infty$, $T^\tx{B}_\tx{on}= T^\tx{B}_\tx{off} =1$, $T^\tx{B}=2$, $t_\tx{off}^\tx{A} = t_\tx{on}^\tx{B} + D - 1$ and $D=5$, see Fig.~\ref{fig:Wald-adiabatic}.
we have $\cl{C}_{zy}$, $\cl{C}_{yy}$ and $\cl{C}_{xx}$ nonvanishing in the current case, plotted with the gray, yellow, and black lines, respectively.
The mutual information $I_\tx{AB}$ plotted with the blue line is always larger than $\cl{C}$'s as (\ref{WVHC}) says.
}
\label{fig:IAB-Wald}
\end{center}
\end{figure}

Let us look at an asymptotic behavior of spin correlations near the adiabatic limit $\gamma_\tx{A} = 1$ as sketched in Fig.~\ref{fig:Wald-adiabatic}.
We are interested in the dependence on the adiabatic parameter $T_\tx{off}^\tx{A}$.  
The time integrations of the Keldysh Green's function, $\mf{G}^\tx{BA}_\tx{K}$, $\mf{G}^\tx{AA}_\tx{K}$, and the retarded Green's function, $\mf{G}^\tx{BA}_\tx{R}$, behave as
\eqn{
\mf{G}^\tx{BA}_\tx{K} ~, \mf{G}^\tx{BA}_\tx{R} \propto (T_\tx{off}^\tx{A})^{-1}  ~, ~~ \mf{G}^\tx{AA}_\tx{K} \propto (T_\tx{off}^\tx{A})^{-2} 
}
apart from oscillatory parts,
see Fig.~\ref{fig:Gs-Wald} in Appendix~\ref{app:Numerical evaluation}.
$\mf{G}^\tx{BB}_\tx{K}$ is, of course,  independent of $T_\tx{off}^\tx{A}$.
For small values of  $\mf{G}^\tx{BA}_\tx{K}$ and $\mf{G}^\tx{BA}_\tx{R}$,  
the correlation functions in (\ref{connected-part-Wald}) behave as 
\eqn{
& \La \delta \hat{\sigma}_x^\tx{A} \delta \hat{\sigma}_x^\tx{B} \Ra \approx 8 \gamma_\tx{A}\gamma_\tx{B} [ (\mf{G}^\tx{BA}_\tx{K})^2 + (\mf{G}^\tx{BA}_\tx{R})^2 /4 ], \\
&\La \delta \hat{\sigma}_y^\tx{A} \delta \hat{\sigma}_y^\tx{B} \Ra \approx 4 \gamma_\tx{A}\gamma_\tx{B} \mf{G}^\tx{BA}_\tx{K},  \\
& \La \delta \hat{\sigma}_z^\tx{A} \delta \hat{\sigma}_y^\tx{B} \Ra \approx -2 \gamma_\tx{B} \mf{G}^\tx{BA}_\tx{R} .
}
Their squared values are depicted in Fig.~\ref{fig:IAB-Wald}.

The mutual information $I_\tx{AB}$ reflects the behavior of the correlations and satisfies the inequality (\ref{WVHC}).
In the current case, we have the eigenvalue (\ref{mu_AB}) of $\hat{\rho}_\tx{AB}$ as
\eqn{
\mu_\tx{AB}^{s_1 s_2} =& \frac{1}{4} \biggl\{ 1+s_2 \gamma_\tx{A} \gamma_\tx{B} \cosh (4 \mf{G}^\tx{BA}_\tx{K})  \\
&~~~~~~+ s_1 \sqrt{\gamma_\tx{A}^2 + \gamma_\tx{B}^2 + 2 s_2 \gamma_\tx{A}  \gamma_\tx{B} \cos (2 \mf{G}^\tx{BA}_\tx{R})  + \gamma_\tx{A}^2 \gamma_\tx{B}^2 \sinh^2 (4 \mf{G}^\tx{BA}_\tx{K})  } \biggr\} ~.
\label{mu_AB-Wald}
}
Plugging it into (\ref{S_AB}), we get the entanglement entropy $S(\hat{\rho}_\tx{AB})$, and then 
the mutual information (\ref{I_AB}) is obtained as a function of $\mf{G}^\tx{AA}_\tx{K}$, $\mf{G}^\tx{BB}_\tx{K}$, $\mf{G}^\tx{BA}_\tx{K}$ and $\mf{G}^\tx{BA}_\tx{R}$,
\eqn{
I_\tx{AB} = I_\tx{AB}(\gamma_\tx{A}, \gamma_\tx{B}; \mf{G}^\tx{BA}_\tx{K}; \mf{G}^\tx{BA}_\tx{R}) ~. \label{I_AB-Wald}
}
It is depicted in Fig.~\ref{fig:IAB-Wald}.  
$I_\tx{AB}$ is actually larger than various correlators $\cl{C}_{w w'}$  as the relation (\ref{WVHC}) indicates.

%%%%%FIG%%%%%%%
%%%%%%%%%%%%

\subsubsection{Spin-field correlations}
In the adiabatic limit, the correlations between the field and Alice's spin also  vanish
because we have $S(\hat{\rho}_\tx{A})=0$ 
in the limit $\gamma_\tx{A} = \exp (-2 \mf{G_\tx{K}^\tx{AA}}) \to 1$ and from the trade-off relation in (\ref{Information-trade-off}) 
the mutual information  $I_{\tx{A}\phi}$ also must vanish. 
One can explicitly see this as follows.
For example,
\eqn{
\bra{\Psi_\tx{f}}  \delta \hat{\sigma}^\tx{A}_{y} \delta \phi(\vb*{x}) \ket{\Psi_\tx{f}}
=& -2  \La \hat{\sigma}^\tx{A}_{x}\Ra \int dt G_\tx{K}(t_\tx{f}-t,\vb*{x}-\vb*{x}_\tx{A}) \lambda_\tx{A}(t) 
}
vanishes in the adiabatic limit.
Furthermore, with the number operator $\hat{N}$, we get
\eqn{
\bra{\Psi_\tx{f}}  \delta \hat{\sigma}^\tx{A}_{x} \delta \hat{N} \ket{\Psi_\tx{f}} =&  -2 \mf{G}_\tx{K}^\tx{AA} \La \hat{\sigma}_x^\tx{A} \Ra =    \gamma_\tx{A}  \ln  \gamma_\tx{A} ~, \\
\bra{\Psi_\tx{f}}  \delta \hat{\sigma}^\tx{A}_{x} \delta \!\!:\!\!\hat{N}^2 \!\!:\!\! \ket{\Psi_\tx{f}} =& -4  \qty{ \mf{G}_\tx{K}^\tx{AA} \mf{G}_\tx{K}^\tx{BB} + (\mf{G}_\tx{K}^\tx{BA})^2 + (\mf{G}_\tx{R}^\tx{BA})^2/4 } \La \hat{\sigma}_x^\tx{A} \Ra ~.
}
The first one explicitly vanishes with $ \gamma_\tx{A} \to 1$.
The second one also vanishes in the adiabatic limit because of the uncertainty relation (\ref{RS-inequality-2}).

On the other hand, correlations between the field and Bob's spin do not have to vanish in the limit of $ \gamma_\tx{A} \to 1$.
It is because, when $I_\tx{AB} = 0$, the second one in the trade-off relations in (\ref{Information-trade-off}) becomes $I_{\tx{B}\phi} = 2 S(\hat{\rho}_\tx{B})$ which is nonzero with $\gamma_\tx{B} < 1$.
For instance, the first term of 
\eqn{
\bra{\Psi_\tx{f}}  \delta \hat{\sigma}^\tx{B}_{y} \delta \phi(\vb*{x}) \ket{\Psi_\tx{f}}
=&  -2 \La \hat{\sigma}^\tx{B}_{x}\Ra \int dt G_\tx{K}(t_\tx{f}-t,\vb*{x}-\vb*{x}_\tx{B}) \lambda_\tx{B}(t) \\
& + \La \hat{\sigma}^\tx{A}_{z} \hat{\sigma}^\tx{B}_{y} \Ra \int dt G_\tx{R}(t_\tx{f}-t,\vb*{x}-\vb*{x}_\tx{A}) \lambda_\tx{A}(t) 
}
remains finite  in the limit of $\gamma_\tx{A} \to 1$.
For the correlations with the number operator, we have
\eqn{
\bra{\Psi_\tx{f}}  \delta \hat{\sigma}^\tx{B}_{x} \delta \hat{N} \ket{\Psi_\tx{f}} =&  -2 \mf{G}_\tx{K}^\tx{BB} \La \hat{\sigma}_x^\tx{B} \Ra 
- \mf{G}_\tx{R}^\tx{BA} \La \hat{\sigma}_{z}^\tx{A} \hat{\sigma}_{y}^\tx{B} \Ra ~, \\
\bra{\Psi_\tx{f}}  \delta \hat{\sigma}^\tx{B}_{x} \delta \!\!:\!\!\hat{N}^2 \!\!:\!\! \ket{\Psi_\tx{f}} =& -4  \qty{ \mf{G}_\tx{K}^\tx{AA} \mf{G}_\tx{K}^\tx{BB} + (\mf{G}_\tx{K}^\tx{BA})^2 + (\mf{G}_\tx{R}^\tx{BA})^2/4 } \La \hat{\sigma}_x^\tx{B} \Ra \\
&+ 2  \mf{G}_\tx{R}^\tx{BA} (\mf{G}_\tx{K}^\tx{AA} - \mf{G}_\tx{K}^\tx{BB}) \La \hat{\sigma}^\tx{A}_{z} \hat{\sigma}^\tx{B}_{y} \Ra  ~.
}
While the second one vanishes with $\gamma_\tx{A} \to 1$,
the first one does not and $\gamma_\tx{B}  \ln  \gamma_\tx{B} $ remains.

%%%%%%%%%%%%%%%%%%%%%%%%%%%%%%%%
\subsubsection{Separability}
\label{sec.separaility-proof}
As seen in (\ref{C-Wald}), we have $C_{yz} =0$.
Therefore, the analysis in Section \ref{sec.Separability condition} tells us that the entanglement negativity vanishes, and thus, the reduced state $\hat{\rho}_\tx{AB}$ is separable.
It is consistent with the observation obtained in \cite{Sugiyama:2022ixw} with the electromagnetic field that the entanglement negativity vanishes for $D \gg t^\tx{A}_\tx{off} - t^\tx{A}_\tx{on}$.

%%%%%%%%%%fig%%%%%%%%
\begin{figure}
\begin{center}
 \includegraphics[width=11cm]{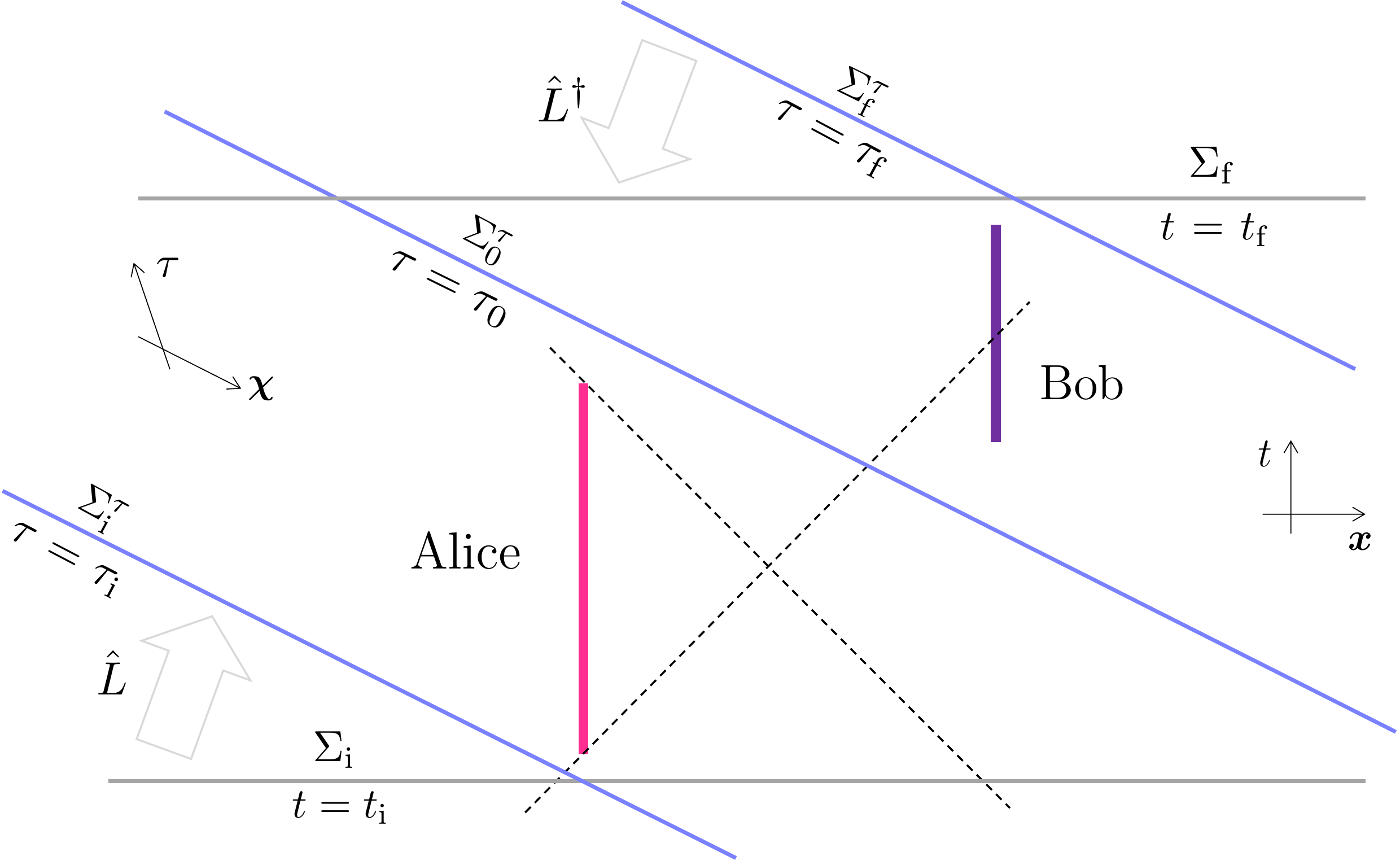}
\caption{Time evolution regarded as a combination of two Lorentz transformation: $\hat{U}(t_\tx{f},t_\tx{i}) = \hat{V}_{\phi \tx{B}} \hat{V}_{\phi \tx{A}}$.
Dashed lines depict light cones. The blue lines depict the time slices with $\tau = \tau_\tx{i}, \tau_{0}$ and $\tau_\tx{f}$. In this example, the retarded Green's function from Alice's spin to Bob's spin and its time integration $\mf{G}_\tx{R}^\tx{BA}$ is nonzero whereas $\mf{G}_\tx{R}^\tx{AB}=0$.}
\label{fig:Lorentz}
\end{center}
\end{figure}
%%%%%%%%%%%%%%%%%%

Here, let us give a simple explanation in Hamiltonian formulation why the entanglement is not generated when one of the retarded Green's functions vanishes because of the causality, applying a no-go theorem for entanglement extraction \cite{Simidzija:2018nku}.
We consider a time evolution by Lorentz boost which brings the initial time slice $\Sigma_\tx{i}$ to $\Sigma'_\tx{i}$, see Fig.~\ref{fig:Lorentz}.
With $\hat{L}$ being the corresponding unitary operator,
the time evolution (\ref{final-state}) is rewritten as
\eqn{
\ket{\Psi_\tx{f}} = \hat{L}^\dag \hat{V}(\tau_\tx{f} ,\tau_\tx{i}) \hat{L} \ket{\Psi_\tx{i}} =  \hat{L}^\dag \hat{V}_{\tx{B}\phi}(\tau_\tx{f} ,\tau_0)  \hat{V}_{\tx{A}\phi}(\tau_0 ,\tau_\tx{i}) \hat{L} \ket{\Psi_\tx{i}}
}
where $\hat{V}(\tau_\tx{f} ,\tau_\tx{i}) : = \hat{L}  \hat{U}(t_\tx{f} ,t_\tx{i}) \hat{L}^\dag $ is decomposed into two parts, the unitary operator $\hat{V}_{\tx{A}\phi}(\tau_0 ,\tau_\tx{i})$ for the time evolution from $\Sigma^\tau_\tx{i}$ to $\Sigma^\tau_0$  and $\hat{V}_{\tx{B}\phi}(\tau_\tx{f} , \tau_0)$ for the evolution from $\Sigma^\tau_0$ to $\Sigma^\tau_\tx{f}$.
Since $\lambda_\tx{B} =0$  for $\tau < \tau_0$ and  $\lambda_\tx{A} =0$ for   $\tau > \tau_0$,
the former one trivially acts on the Hilbert space of Bob's spin, and the latter trivially acts on the Hilbert space of Alice's spin.
Note that $\hat{L}$ does not act on the Hilbert space of the spins since $\lambda_\tx{A} = \lambda_\tx{B} = 0$ during those time evolution by Lorentz boost.
It is convenient to take the interaction picture to describe the evolution from $\Sigma^\tau_0$ to $\Sigma^\tau_\tx{f}$. The state evolves as
\eqn{
\hat{V}_{\phi \tx{B,I}}(\tau_\tx{f} , \tau_0) &= \tx{T} \exp \qty{- \ri \int_{\tau_0 }^{\tau_\tx{f}} \dd \tau \hat{H}^\tau_\tx{B,I}(t') } = e^{\ri \omega_2} \exp \qty{- \ri \int_{\tau_0 }^{\tau_\tx{f}} \dd \tau \hat{H}^\tau_\tx{B,I}(\tau) } ~, \label{V_phiB,I}
}
where
\eqn{
\hat{H}^\tau_\tx{B,I}(\tau) =  - \hat{\sigma}_z^\tx{B} \lambda_\tx{B}(t(\tau))\hat{\phi}(\tau, \vb*{\chi}_\tx{B}(\tau)) ~
}
and
\eqn{\omega_2 &\coloneqq \frac{\ri}{2} \int_{\tau_0 }^{\tau_\tx{f}} \dd \tau \int_{\tau_0 }^\tau \dd \tau' [ \hat{H}^\tau_\tx{B,I}(\tau) , \hat{H}^\tau_\tx{B,I}(\tau') ] = \frac{1}{2} \int \dd t \int \dd t' \lambda_\tx{B}(t) G_\tx{R}(t-t',\vb*{0}) \lambda_\tx{B}(t')~. \label{Omega_2}}
Here, the field operator is defined as $\hat{\phi}(\tau,\vb*{\chi}) = e^{\ri (\tau-\tau_\tx{i}) \hat{H}^\tau_\phi} \hat{L}\hat{\phi}(\vb*{x}) \hat{L}^\dag e^{-\ri (\tau-\tau_\tx{i}) \hat{H}^\tau_\phi }$ with $\hat{H}^\tau_\phi = \hat{L}  \hat{H}_\phi  \hat{L}^\dag$. (See a related discussion in Footnote \ref{source-free}).
In deriving (\ref{V_phiB,I}), 
the time-ordering is evaluated with the Magnus expansion \cite{doi.org/10.1002/cpa.3160070404} as $\hat{V}_{\phi \tx{B,I}} = \exp \qty{ \sum_{k=1}^\infty \hat{\Omega}_k}$ with $\hat{\Omega}_1 = - \ri \int_{\tau_0 }^{\tau_\tx{f}} \dd \tau \hat{H}^\tau_\tx{B,I}(\tau)$, $\hat{\Omega}_2 = \ri \omega_2$ and $\hat{\Omega}_{k \geq 3} = 0$.

In the new coordinate system $(\tau, \vb*{\chi})$,  spin's trajectory is described by $\vb*{\chi}_\tx{B}(\tau)$.
Let us write down the spectral decomposition\footnote{Just for illustrative purposes, we assume $\hat{\Phi}$ has a discreet spectrum. See discussions in \cite{Simidzija:2018nku} for the continuous case.}
of the Hermitian operator $\hat{\Phi} \coloneqq \int_{\tau_0}^{\tau_\tx{f}} \dd \tau  \lambda_z^\tx{B}(t(\tau)) \hat{\phi} (\tau, \vb*{\chi}_\tx{B}(\tau))$  as $\hat{\Phi} = \sum_k \Phi_k \ket{\Phi_k} \bra{\Phi_k}$.
Then, rewriting the evolution operator (\ref{V_phiB,I}) as
\eqn{
\hat{V}_{\phi\tx{B,I}}(\tau_\tx{f} , \tau_0) = e^{\ri \omega_2} \sum_k  \ket{\Phi_k} \bra{\Phi_k} \exp \qty{\ri \hat{\sigma}_z^\tx{B} \Phi_k } ~, \label{simple-generated}
}
we find a reduced density matrix on $\Sigma_\tx{f}$ to be separable:
\eqn{
\hat{\rho}_\tx{AB} = \tr_\tx{\phi} \qty{ \ket{\Psi_\tx{f}} \bra{\Psi_\tx{f}}} = \tr_\tx{\phi} \qty{ \hat{V}_{\phi\tx{B,I}} \hat{\rho}_0 \hat{\rho}_\tx{B,i}  \hat{V}_{\phi\tx{B,I}}^\dag} = \sum_k p_k \hat{\rho}_\tx{A}^{(k)} \hat{\rho}_\tx{B}^{(k)}~,
}
where $\hat{\rho}_\tx{B,i}$ is the initial state of Bob's spin (see Footnote \ref{footnote:rho_B}),
\eqn{ \hat{\rho}_0 \coloneqq \hat{V}_{\tx{A}\phi}(\tau_0 ,\tau_\tx{i}) \hat{L} \ket{\Omega}_\phi 
\frac{1}{2}(\ket{+}_\tx{A}+\ket{-}_\tx{A})(\bra{+}_\tx{A}+\bra{-}_\tx{A})
\bra{\Omega}_\phi\hat{L}^\dag  \hat{V}^\dag_{\tx{A}\phi} (\tau_0 ,\tau_\tx{i} ) }
is the state of the subsystem composed of Alice's spin and the field on $\Sigma_0^\tau$ and
\begin{align}
p_k &\coloneqq \tr_\tx{A} \qty{ \bra{\Phi_k}   \hat{\rho}_0  \ket{\Phi_k} } ~, \\
\hat{\rho}_\tx{A}^{(k)} &\coloneqq \bra{\Phi_k} \hat{\rho}_0   \ket{\Phi_k}/ p_k ~,\\
\hat{\rho}_\tx{B}^{(k)}& \coloneqq e^{\ri \hat{\sigma}_z^\tx{B} \Phi_k } \hat{\rho}_\tx{B,i}  e^{- \ri \hat{\sigma}_z^\tx{B} \Phi_k } ~.
\end{align}
Note that the details of the initial state are irrelevant except for that Bob's spin is not entangled with the rest of the system at the initial time.
It is crucial for this proof of separability that the evolution operator is written in the form of (\ref{simple-generated}) called simple-generated unitary \cite{Simidzija:2018nku}.
There are two reasons here we have this form: Each spin has no energy gap when the coupling to the field is absent, and the commutator
$[\hat{\phi}(x), \hat{\phi}(x')]$ in (\ref{Omega_2}) turns out to be the c-number.
Therefore, when detectors coupling to the field has some dynamics or when the field has nonlinear interactions as is the case with the gravitational field, the quantum entanglement between the detectors can be generated even if one of the retarded Green's functions connecting them vanishes.

%%%%%%%%%%%%%%%%%%%%%%%%%%%%%%%%%%%%%%
\subsection{Short summary of the spin correlations and various Green's functions \label{sec.short summary}}
We give a short summary of the roles played by various Green's functions for the spin correlations and entanglement in various limiting cases. 
After integrating the quantum field, any observables of Alice and Bob are written in terms of the following 5 quantities:
\begin{equation}
    \mf{G}_\tx{R}^\tx{AB}, ~\mf{G}_\tx{R}^\tx{BA}, ~ \mf{G}_\tx{K}^\tx{AA}, ~\mf{G}_\tx{K}^\tx{BB},
~    \mf{G}_\tx{K}^\tx{AB} = \mf{G}_\tx{K}^\tx{BA} ~.
\end{equation}
As we have seen in this section, these quantities are present or absent in different situations and play different roles. 
First $\mf{G}_\tx{K}^\tx{AA}$ and $\mf{G}_\tx{K}^\tx{BB}$ appear as a form of  $\gamma_\tx{A}$ and $\gamma_\tx{B}$, and they only quantify the overall magnitude of the spin correlations. On the other hand, the other three quantities determine whether spin correlations and quantum entanglement appear or disappear.
Thus, we consider the 3 quantities and summarize the situation in the table \ref{table:Gs}. 

%%%%%%%%%Table%%%%%%%
\begin{table}[h]
  \begin{tabular}{c||c|c|c|c} %\hline
     Region & (i)& (ii) & (IV) & (II) \\ \hline 
    $\mf{G}_\tx{R}^\tx{AB}$  &  $\cellcolor{lightgray}=\! \mf{G}_\tx{R}^\tx{BA}$ & $\cellcolor{lightgray}~\ne 0~$ & $~=0~$ & $~=0~$ \\ \hline
    $\mf{G}_\tx{R}^\tx{BA}$  & $\cellcolor{lightgray}=\! \mf{G}_\tx{R}^\tx{AB}$  &$~\ne 0~$ & $~=0~$ &$\cellcolor{lightgray}~\ne 0~$ \\ \hline
    $\mf{G}_\tx{K}^\tx{BA}$  & $~=0~$ &$~\ne 0~$ & \cellcolor{lightgray}$~\ne 0~$ &\cellcolor{lightgray}$~\ne 0~$ \\ \hline
  \end {tabular}
  \caption{Shaded quantities are relevant ones to the spin correlations in each limiting case.}
  \label{table:Gs}
\end {table}
%%%%%%%%%%%%%%%%%%%
In this section, we have studied 4 situations, depicted in Fig.~\ref{fig:Adiabaticity} and Fig.~\ref{fig:Causality}.  
In the region (i) discussed in Sec.~\ref{sec.Adiabatic limit}, 
because of the Robertson-Schr\"{o}dinger inequality (\ref{RS-inequality}),  $\mf{G}_\tx{K}^\tx{BA}$ vanishes and the equality $\mf{G}_\tx{R}^\tx{AB} = \mf{G}_\tx{R}^\tx{BA}$ is required. 
Thus only the quantity $\mf{G}_\tx{R}^\tx{AB} = \mf{G}_\tx{R}^\tx{BA}$ is relevant for the 
absence or presence of correlations. It is the case for the Newtonian approximation. 
In the region (ii) discussed in Sec.~\ref{sec.Non-adiabatic limit},
because the state of Bob's spin is fully decohered by $\gamma_\tx{B}=0$, two quantities
$\mf{G}_\tx{R}^\tx{BA}$ and $\mf{G}_\tx{K}^\tx{BA}$ become irrelevant for determining the spin correlations or entanglement even though they are nonvanishing. 
Thus the only quantity for controlling the correlations is 
the one-way causal influence from Bob to Alice with $\mf{G}_\tx{R}^\tx{AB}$.
In the region (IV) discussed in Sec.~\ref{sec.Spacelike-separated},
both of $\mf{G}_\tx{R}^\tx{AB}$ and $\mf{G}_\tx{R}^\tx{BA}$ vanish due to the causality.
Thus the only quantity $\mf{G}_\tx{K}^\tx{BA}$ representing the vacuum fluctuations induces the spin correlations. 
Finally, in the region (II) discussed in Sec.~\ref{sec.One-way},
$\mf{G}_\tx{R}^\tx{AB}$ vanishes due to the causality.
Then, the one-way causal influence from Alice to Bob with $\mf{G}_\tx{R}^\tx{BA}$ and the vacuum fluctuations with $\mf{G}_\tx{K}^\tx{BA}$ contribute to the spin correlations. 
In this situation, two quantities are relevant. 

The spin correlations do not necessarily indicate the quantum entanglement since spins can be correlated under classical correlations. 
The calculation of the negativity in our setup shows that the quantum entanglement is present only when both of the two retarded Green's functions are nonvanishing. Thus, in regions (IV) and (II), there is no quantum entanglement. In the case of (ii), since Bob's state is completely decohered and decoupled from Alice,
there is no entanglement. Thus, only in the case of (i), entanglement is present.

\section{Summary \label{sec.Summary}}
In this work, we gave a comprehensive study of a field-theoretical toy model for the BMV setup.
The correlation between the two spins is induced  through the local interactions with the scalar field.
We observed that the causal structure and the nonadiabaticity of the setup affect the amount of the spin correlations and the quantum entanglement. 
We first took the Newtonian approximation in Sec.~\ref{sec.Newtonian} to look at how 
entanglement is generated between the spins. It is the  situation considered in the original analysis of the BMV experiment. 
We then introduced a field-theoretical model  in Sec.~\ref{sec.Relativistic}, which can be exactly solved. Introducing various ``tools'' to understand the quantum behavior of the final state (\ref{rho_AB}),  we investigated the model in various limiting situations 
in Sec.~\ref{sec.limiting cases}. We especially focussed on the relativistic causality, the vacuum fluctuations, and particle creations which are absent in the Newtonian approximation.

The quantum entanglement generated by the gravitational potential is the target of the BMV experiment to observe the quantum superposition of spacetime geometries.
In the Newtonian approximation, entanglement is indeed generated between two spins of Alice and Bob
mediated by the Newton potential, 
as  explicitly checked by calculating 
the spin correlations in Fig.~\ref{fig:S_N} or entanglement entropy (\ref{S_N}) in Sec.~\ref{sec.Newtonian}.
In this case, these quantities are sufficient for the quantum entanglement 
since the state of Alice's and Bob's spins is a pure state (\ref{rho_AB-Newtonian}). 

In general, when the quantum field mediating two spins is dynamical, the system composed of Alice's and Bob's spins
is no longer maintained in a pure state. Thus spin correlations or entanglement entropy between spins is not sufficient to describe the behavior of the system.
We thus calculated various  quantities such as the mutual information or negativity. 
The dynamical field plays two important roles, one is the effect of relativistic causality
and the other is the particle creation associated with the nonadiabaticity of Alice's and Bob's protocols. 
These effects are represented by various Green's functions, classified by retarded type and Keldysh type,  summarized in Section \ref{sec.short summary}.

The Keldysh type of Green's functions is related to the nonadiabaticity of the protocols which induce particle creations.  To understand its role, we investigated an adiabatic limit and a nonadiabatic limit. 
The adiabatic limit of Alice's protocol (i) is studied in Sec.~\ref{sec.Adiabatic limit}. In addition to this limit, 
if we take the further adiabatic limit of Bob's protocol, all the decoherence factors disappear and the Newtonian result is reproduced. 
On the other hand,
a nonadiabatic limit of Bob with adiabatic Alice gives an analogous situation to the COW experiment.  The amount of entanglement between Alice's and Bob's spins vanishes as depicted in Fig.~\ref{fig:Negativity}. 

The retarded Green's functions control the causal influence between Alice and Bob.
In Secs.~\ref{sec.Spacelike-separated} and \ref{sec.One-way}, we observed that the entanglement disappears 
unless both of the causal influences from Alice to Bob and from Bob to Alice are present. 
The proof is given in Sec.~\ref{sec.separaility-proof}. 
Due to this property, the nonvanishing spin correlations when spins are spacelike separated 
is not genuine quantum correlation. Indeed the reduced density matrix of the final state
becomes a mixture of product states like (\ref{def-separable}).
Such a separable density matrix can be prepared by LOCC, and hence, 
no quantum field is necessary in principle.
Therefore,
the observation of such spin correlations can not be regarded as evidence of the quantum gravity effect in the BMV experiment.
Nevertheless, the spin correlation attributed to the Keldysh function connecting Alice's spin and Bob's spin is a manifestation of the quantum nature of the field, namely, the vacuum fluctuations.
Thus,  when the causal influence between Alice and Bob is absent,
even though the spin correlations could be generated by LOCC,
they are actually generated by quantum field theoretical interactions. 

While these two types of Green's functions play different roles as above, their effects are constrained by various trade-off relations. 
Our system is composed of Alice's spin, Bob's spin, and the field. 
In the Newtonian limit, the field is decoupled. Then, Alice's visibility of the interference and Bob's distinguishability of Alice's $z$-spin satisfy a relation in (\ref{wave-particle-duality-Newtonian}).
This relation becomes an inequality when we take the dynamical effects of the quantum field into account, given in (\ref{wave-particle-duality}) and called wave particle duality. Note that both quantities of the visibility and the distinguishability are written only in terms of the correlation between Alice's and Bob's spin, and all the information of the field is abandoned. 
In order to take the effects of fields into account, we introduced mutual informations among Alice, Bob and the field in Sec.~\ref{sec.Mutual information} and their trade-off relations similar to the wave-particle duality given in (\ref{tradeof-equality-MI}) in Sec.~\ref{sec.Trade-off relation}. 
From the trade-off relations, it follows that the particle creation is necessary for the spin correlation to be generated when Alice and Bob are spacelike separated as in Secs. \ref{sec.Spacelike-separated} and \ref{sec.One-way}.

\

\begin{acknowledgments}
The work was initiated from discussions at the QUP meetings on quantum sensors and their particle physics applications.
We thank all participants in the meetings for the discussions. 
We also thank S. Kawamoto for useful discussions on entanglement harvesting. 
S.I. is supported in part by the Grant-in-Aid for Scientific research, 18H03708 and 16H06490.
Y.H. is supported in part by the Grant-in-Aid for Scientific research, 21H01084. 
\end{acknowledgments}

\appendix
%%%%%%%%%%%%%%%%%%%%%

%%%%%%%%%%%%%%%%%%%%%
\section{Path-integral formulation on CTP \label{app.Reduced density matrix}}
Here, we apply the path-integral technique to compute the reduced density matrix used in \cite{Hidaka:2022tzk} to the system with the Hamiltonian (\ref{Hamiltonian}), briefly reviewing the Keldysh formalism on the closed time path (CTP).

\subsection{Propagators}
\begin{figure}
\begin{center}
 \includegraphics[width=9cm]{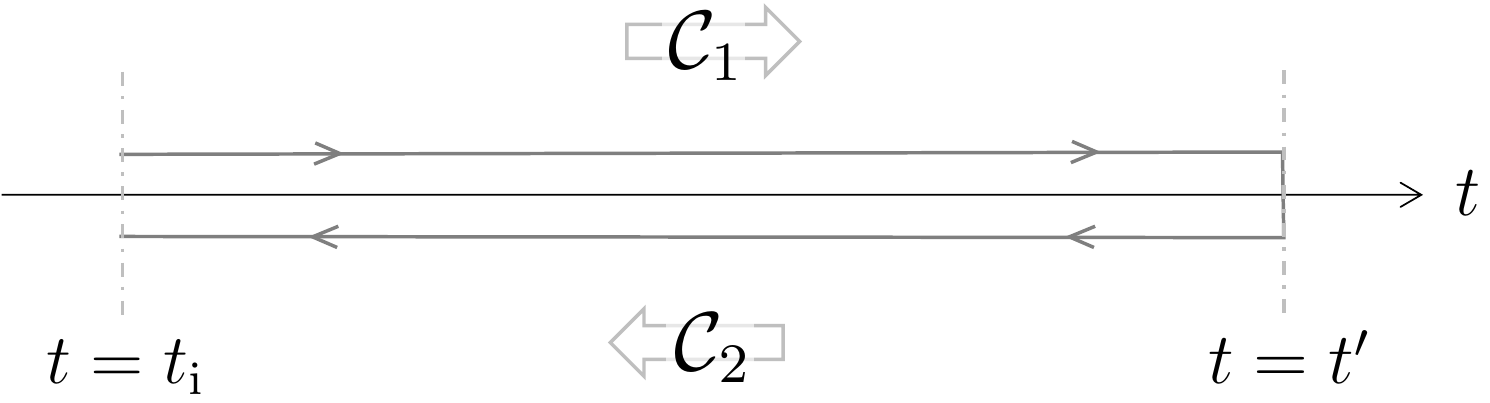}
\caption{Closed time path $\cl{C}= \cl{C}_1 + \cl{C}_2$. Each of the time arguments $x^0$ and $y^0$ of the propagator $G_\cl{C}(x,y)$ is either on the forward path $\cl{C}_1$ or on the backward path $\cl{C}_2$.}
\label{fig:CTP}
\end{center}
\end{figure}

First, we introduce various propagators to be used in the CTP or Keldysh formalism.
For the free, real scalar field described by $H_\phi$ in (\ref{Hamiltonian}) without any source terms,
there are two quantities; the retarded/advanced Green's function (\ref{Retarded}) and the Keldysh function (\ref{Keldysh}) on the vacuum state $\ket{\Omega}_\phi$,
\eqn{
G_\tx{R}(x,y) =G_\tx{A}(y,x) = \ri\theta(x^0-y^0)\bra{\Omega}[\hat{\phi}(x),\hat{\phi}(y)]\ket{\Omega}_\phi ~, \label{app.Retarded}
}
\eqn{
 G_\tx{K}(x,y) &=\frac{1}{2}\bra{\Omega}\{ \hat{\phi}(x),\hat{\phi}(y) \} \ket{\Omega}_\phi  ~, \label{app.Keldysh}
}
which are purely real.
From these two, one can construct the spectral function,
\eqn{
G_\rho (x,y) = G_\tx{R}(x,y) - G_\tx{A}(x,y) = \ri \bra{\Omega}[\hat{\phi}(x),\hat{\phi}(y)]\ket{\Omega}_\phi ~ ~,
}
the Wightman functions,
\eqn{
G_{\lessgtr} (x,y) = G_\tx{K}(x,y) \pm \frac{\ri}{2} G_\rho(x,y) ~, \label{Wightman}
}
the time-ordered propagator,
\eqn{
G_\tx{T} (x,y) = G_\tx{K}(x,y) - \tx{sign} (x^0 -y^0) \frac{\ri}{2} G_\rho(x,y) ~,
}
and the anti-time ordered propagator, $G_{\tilde{\tx{T}}}(x,y)= [G_\tx{T} (x,y)]^*$.
It is also convenient to define the path ordered propagator on the CTP, $\cl{C}= \cl{C}_1 + \cl{C}_2$, sketched in Fig.~\ref{fig:CTP},
\eqn{
G_\cl{C} (x,y) = G_\tx{K}(x,y) - \tx{sign}_\cl{C} (x^0 -y^0) \frac{\ri}{2} G_\rho(x,y) ~, \label{G_C}
}
where $\tx{sign}_\cl{C} (x^0 -y^0)$ is the sign function on the CTP: if $x^0$ is ahead of (behind) $y^0$ in terms of the path $\cal{C}$, it gives $+1$ $(-1)$. 
It can be written with the $2\times 2$ matrix notation as
\eqn{
G_\cl{C} (x,y) &=\pmtx{
G_\tx{T} (x,y) & G_{<} (x,y) \\ G_{>} (x,y)& G_{\tilde{\tx{T}}}(x,y)} \\
&=\pmtx{-\ri /2 & 1\\ \ri /2 & 1} \pmtx{0 &  G_\tx{A} (x,y) \\  G_\tx{R}(x,y)& G_\tx{K}(x,y)} \pmtx{-\ri /2 & \ri /2 \\ 1& 1} ~,
}
where the $(i,j)$ component is for the arguments $x$ and $y$ being on $\cl{C}_i$ and $\cl{C}_j$, respectively.

%%%%%%%%%%fig%%%%%%%%
%%%%%%%%%%%%%%%%%%
%%%%%%%%%%%%%
\subsection{Computing reduced density matrix}
Let us consider an initial state at $t = t_\tx{i}$,
\eqn{
\hat{\rho}_{\tx{AB} \phi}(t_\tx{i}) = \hat{\rho}_\tx{AB} (t_\tx{i}) \otimes \ket{\Omega}_\phi \! \bra{\Omega} ~, \label{total-initial-state}
}
where $\ket{\Omega}_\phi$ is field's ground state with $\lambda_\tx{A} = \lambda_\tx{B}=0$ and $\hat{\rho}_\tx{AB} (t_\tx{i})$ is an arbitrary density matrix of the two spin system at the initial time $t_\tx{i}$.
The density matrix of the total system at some time $t'$ is obtained as
\eqn{\hat{\rho}_{\tx{AB} \phi}(t') = \hat{U}(t',t_\tx{i}) \hat{\rho}_{\tx{AB} \phi}(t_\tx{i})   \hat{U}(t',t_\tx{i})^\dag ~, }
where 
$\hat{U}(t',t_\tx{i}) =  \tx{T} \exp \{ -\ri \int_{t_\tx{i}}^{t'} \dd t \hat{H}(t)  \}$ is the time evolution operator.
Assuming that by this time both Alice and Bob have turned off the spin-field interaction: $t' > \tx{max}\qty{t_\tx{off}^\tx{A}, t_\tx{off}^\tx{B}}$,
we trace out the field $\phi$ to define the reduced density matrix:
\eqn{
\hat{\rho}_{\tx{AB}} &= \tr_{\phi} \qty{ \hat{U}(t',t_\tx{i}) \hat{\rho}_{\tx{AB} \phi}(t_\tx{i})   \hat{U}(t',t_\tx{i})^\dag } ~.  \label{app.rho_AB-without-J^phi}
} 
Note that it does not depend on $t'$ for the assumption $t' > \tx{max}\qty{t_\tx{off}^\tx{A}, t_\tx{off}^\tx{B}}$ since each spin has no dynamics on its own.
For later convenience,  $t'$ is considered to be as large as $+\infty$.

Noting the spins in Hamiltonian are diagonal in $z$-basis, each matrix element can be written in the form of
\eqn{
\bra{\sigma_1^\tx{A} \sigma_1^\tx{B}} \hat{\rho}_\tx{AB} \ket{\sigma_2^\tx{A} \sigma_2^\tx{B}} =f(\sigma^\tx{A}_{1},\sigma^\tx{B}_{1},\sigma^\tx{A}_{2},\sigma^\tx{B}_{2} )  \times Z[J_1,J_2] ~,
}
where
\eqn{ f(\sigma^\tx{A}_{1},\sigma^\tx{B}_{1},\sigma^\tx{A}_{2},\sigma^\tx{B}_{2} )  \coloneqq \bra{\sigma^\tx{A}_{1} \sigma^\tx{B}_{1}} \hat{\rho}_\tx{AB} (t_\tx{i}) \ket{\sigma^\tx{A}_{2} \sigma^\tx{B}_2} }
and
\eqn{ Z[\cl{J}_1, \cl{J}_2] =  \bra{\Omega} \hat{U}_\phi (t' ,t_\tx{i}; J_2)^\dag  \hat{U}_\phi (t' ,t_\tx{i}; J_1) \ket{\Omega}_\phi 
}
with the time evolution operator defined by (\ref{time-evolution-op}) with the ``source'' $J_i$ originated from the interaction between the field and the spins (\ref{source}).
It is written in the path-integral form on the closed time path (CTP) as
\eqn{
 Z[J_1, J_2] =  \! \int \! \dd \phi_{\ri 1} \! \int \! \dd \phi_{\ri 2} \bra{\phi_{\ri 1} } \ket{\Omega}_\phi \! \bra{\Omega}
 \ket{\phi_{\ri 2}}
K_\cl{C}[\phi_{\ri 2}, J_2 ;\phi_{\ri 1}, J_1 ]  \label{app.def-Z}
}
with
\eqn{
K_\cl{C}[\phi_{\ri 2}, J_2 ;\phi_{\ri 1}, J_1 ] :&= \int \! \dd \phi' \bra{\phi_{\ri 2}} \hat{U}_\phi (t' ,t_\tx{i}; J_2)^\dag \ket{\phi'}\bra{\phi'} \hat{U}_\phi (t' ,t_\tx{i}; J_1)   \ket{\phi_{\ri 1}}  \\
&= \int \! \! \dd \phi' \int_{\phi_1 (t_\tx{i}) = \phi_{\ri 1}}^{\phi_1 (t') = \phi'} \hspace{-2em}\cl{D}\phi_1 \int_{\phi_2 (t_\tx{i}) =  \phi_{\ri 2}}^{\phi_2 (t') = \phi' } \hspace{-2em}\cl{D}\phi_2 \exp \qty{ \ri \int_{t_\tx{i}}^{t'} \! \! \! \dd t L[\phi_1, J_1] - \ri \int_{t_\tx{i}}^{t'} \! \! \dd t  L[\phi_2, J_2]} \\
&= \int_{\phi_1 (t_\tx{i}) = \phi_{\ri 1}}^{\phi_2 (t_\tx{i}) = \phi_{\ri 2}} \hspace{-2em} \cl{D}\phi_\cl{C} \exp \qty{ \ri \int_\cl{C} \! \! \dd t  L[\phi_\cl{C}, J_\cl{C}] } ~.
\label{Keldysh-kernel}}
Here, $\cl{C} = \cl{C}_1 + \cl{C}_2$ is the CTP sketched in Fig.~\ref{fig:CTP} and
\eqn{
\phi_\cl{C} = \left\{
\mtx{\phi_1 & \tx{on} & \cl{C}_1 \\ 
     \phi_2 & \tx{on} & \cl{C}_2} \right.~, ~~~  J_\cl{C} = \left\{
\mtx{J_1 & \tx{on} & \cl{C}_1 \\ 
     J_2 & \tx{on} & \cl{C}_2} \right. ~,
}
which appear in the Lagrangian
\eqn{
L[\phi, J] = \int \dd^3 x \qty( -\frac{1}{2} (\partial \phi)^2 -\frac{m^2 }{2} \phi^2 + J \phi ) 
}
in the last line.

Since there is no nonlinear interaction here, the path-integral in (\ref{app.def-Z}) can be exactly evaluated.
The propagator $G_\cl{C}$ defined in (\ref{G_C}) satisfies $(\partial_x^2 -m^2) G_\cl{C}(x,y) =  \ri \delta^{(4)}_\cl{C}(x-y) $
with the vacuum initial boundary condition, where $\delta^{(4)}_\cl{C}(x-y)$ is the delta function on CTP.
With this propagator, we have
\eqn{
\ln Z[J_1, J_2] &= -\frac{1}{2}  \int_\cl{C}\dd^4 x \dd^4 y J_\cl{C}(x) G_\cl{C}(x,y) J_\cl{C}(x) \\
&= +\frac{1}{2}  \int \dd^4 x \dd^4 y \pmtx{ J_\tx{r}(x) & \ri J_\tx{a}(x) } \pmtx{0 &  G_\tx{A} (x,y) \\  G_\tx{R}(x,y)& G_\tx{K}(x,y)}  \pmtx{J_\tx{r}(x) \\ \ri  J_\tx{a}(x)}
} 
with
\eqn{
J_\tx{r}(x) &= \frac{J_\tx{1}(x) + J_\tx{2}(x)}{2} = \sigma_\tx{r}^\tx{A}  \lambda_\tx{A}(x^0)  \delta^{(3)} (\vb*{x} - \vb*{x}_\tx{A}) + \sigma_\tx{r}^\tx{B}  \lambda_\tx{B}(x^0) \delta^{(3)} (\vb*{x} - \vb*{x}_\tx{B}) ~, \\
J_\tx{a}(x) &= J_\tx{1}(x) - J_\tx{2}(x)  = \sigma_\tx{a}^\tx{A}  \lambda_\tx{A}(x^0)  \delta^{(3)} (\vb*{x} - \vb*{x}_\tx{A}) + \sigma_\tx{a}^\tx{B}  \lambda_\tx{B}(x^0) \delta^{(3)} (\vb*{x} - \vb*{x}_\tx{B})  ~,
}
where
$\sigma^\tx{A,B}_\tx{r} = (\sigma^\tx{A,B}_\tx{1} + \sigma^\tx{A,B}_\tx{2})/2$ and $\sigma^\tx{A,B}_\tx{a} = \sigma^\tx{A,B}_\tx{1} - \sigma^\tx{A,B}_\tx{2}$.
Specifying the initial state $\hat{\rho}_\tx{AB}(t_\tx{i})$ in (\ref{total-initial-state}) to be (\ref{spin-initial-state}),
we have $f(\sigma^\tx{A}_{1},\sigma^\tx{B}_{1},\sigma^\tx{A}_{2},\sigma^\tx{B}_{2} ) = 1/4$.
Then, 
the reduced density matrix (\ref{rho_AB-z-basis}) or its Bloch representation (\ref{rho_AB}) is reproduced.

\section{Correlation between field and spin \label{app.Correlation between field and spin}}
In each limiting case in Sec.~\ref{sec.limiting cases}, the correlations between the field and the spin are discussed.
Here, we obtain the generating functional of the correlation functions.
Let us introduce external sources $J^{\phi}_1$ and $J^{\phi}_2$ for the field in the time evolution as
\eqn{\hat{U}(t' ,t_\tx{i}; J_i^\phi ) =  \tx{T} \exp \qty[ -\ri \int_{t_\tx{i}}^{t'} \dd t    \hat{H}(t) + \ri \int_{t_\tx{i}}^{t'} \dd t \int \dd^3 x   J_i^\phi (t,\vb*{x}) \hat{\phi}(\vb*{x}) ] ~,
} 
and define
\eqn{
\hat{\rho}^{J^\phi}_{\tx{AB}} = \tr_{\phi} \qty{\hat{U}(t' ,t_\tx{i}; J_1^\phi ) \hat{\rho}_{\tx{AB} \phi}(t_\tx{i})  \hat{U}(t' ,t_\tx{i}; J_2^\phi )^\dag }  \label{app.rho_AB-with-J^phi}
} 
instead of (\ref{app.rho_AB-without-J^phi}).
Each matrix element is written in the form of
\eqn{
\bra{\sigma_1^\tx{A} \sigma_1^\tx{B}} \hat{\rho}^{J^\phi}_\tx{AB} \ket{\sigma_2^\tx{A} \sigma_2^\tx{B}} =f(\sigma^\tx{A}_{1},\sigma^\tx{B}_{1},\sigma^\tx{A}_{2},\sigma^\tx{B}_{2} )  \times Z[\cl{J}_1, \cl{J}_2] ~,
}
where
\eqn{
\cl{J}_i (x) \coloneqq J_i (x) + J^\phi_i (x) ~. 
}
$Z[\cl{J}_1, \cl{J}_2]$ can be computed in the same manner as  $Z[J_1,J_2]$, and we get
\eqn{
\ln Z[\cl{J}_1, \cl{J}_2] &= -\frac{1}{2}  \int_\cl{C}\dd^4 x \dd^4 y \cl{J}_\cl{C}(x) G_\cl{C}(x,y) \cl{J}_\cl{C}(x) \\
&= +\frac{1}{2}  \int \dd^4 x \dd^4 y \pmtx{ \cl{J}_\tx{r}(x) & \ri  \cl{J}_\tx{a}(x) } \pmtx{0 &  G_\tx{A} (x,y) \\  G_\tx{R}(x,y)& G_\tx{K}(x,y)}  \pmtx{\cl{J}_\tx{r}(y) \\  \ri \cl{J}_\tx{a}(y) }~,
} 
where 
\eqn{
\cl{J}_\tx{r}(x) &= \frac{\cl{J}_\tx{1}(x) + \cl{J}_\tx{2}(x)}{2} = J_\tx{r}(x) +J^\phi_\tx{r}(x) ~, \\
\cl{J}_\tx{a}(x) &= \cl{J}_\tx{1}(x) - \cl{J}_\tx{2}(x)  = J_\tx{a}(x) +J^\phi_\tx{a}(x) ~.
}
The linear combinations of the external sources, $J^\phi_\tx{r}: = (J^\phi_1 + J^\phi_2)/2$ and $J^\phi_\tx{a} \coloneqq J^\phi_1 - J^\phi_2$, come with $\phi_\tx{r}: = (\phi_1 + \phi_2)/2$ and $\phi_\tx{a} \coloneqq \phi_1 - \phi_2$ as
\eqn{
\ri   J^\phi_1 \phi_1 - \ri J^\phi_2  \phi_2 = \ri  J^\phi_\tx{a} \phi_\tx{r} + \ri  J^\phi_\tx{r} \phi_\tx{a} 
}
in $Z$ as seen in (\ref{Keldysh-kernel}).
Therefore, differentiating $Z[\cl{J}_1, \cl{J}_2]$ with respect to $\ri J^\phi_\tx{a/r}$ is equivalent to inserting $\phi_\tx{r/a}$, and thus,
the density matrix (\ref{app.rho_AB-with-J^phi}) with the external sources are the generating functional of the correlation functions of the field and the spins.
For instance, we have
\eqn{
\tr \qty{ \hat{\rho}_\tx{AB} \sigma^\tx{A}_{z}  \hat{\phi}(\vb*{x}) }_\tx{\Sigma_\tx{f}} = \left. \frac{\delta }{\ri \delta J^\phi_\tx{a}(x)} \tr \qty{ \hat{\rho}^{J^\phi}_\tx{AB} \sigma^\tx{A}_{z}  }   \right|_{J^\phi = 0} ~,
}
where $x=(t_\tx{f},\vb*{x})$ is a spacetime point on the time slice $\Sigma_\tx{f}$ before which both Alice and Bob have turned off their spin-field interactions; see the comment below (\ref{app.rho_AB-without-J^phi}).

It is convenient to put it in the Bloch representation as
\eqn{
\hat{\rho}_\tx{AB}^{J^\phi} = \exp \qty{ -\frac{1}{2} J_\tx{a} \cd G_\tx{K} \cd J_\tx{a} +\ri J_\tx{a}\cd G_\tx{R} \cd J_\tx{r}} \frac{1}{4}\sum_{u , v = 0, x, y, z} C_{uv}^J \hat{\sigma}_u^\tx{A} \hat{\sigma}_{v}^\tx{B} ~,
\label{app.rho_AB-with-J^phi-Bloch}}
where ``$\cdot$'' abbreviates the spacetime integration: $J_\tx{a} \cdot G_\tx{K} \cdot J_\tx{a} = \int \dd^4 x \dd^4 y J_\tx{a}(x)  G_\tx{K}(x,y)  J_\tx{a}(y)$, for instance.

In the body of the paper, we are interested in correlations between the field and each spin
with the initial state of the spins given by (\ref{spin-initial-state}).
For Alice's spin, we need the following components in (\ref{app.rho_AB-with-J^phi-Bloch}) to differentiate with respect to $\ri J_\tx{a}^\phi$,
\eqn{
C^{J}_{x0} =& C_{x0} \cos (2 \ri J_\tx{a}^\phi\cd \Phi_\tx{K}^\tx{(A)} ) \cosh (\ri J_\tx{a}^\phi \cd \Phi_\tx{R}^\tx{(B)})
+ C_{yz}  \sin (2 \ri J_\tx{a}^\phi\cd \Phi_\tx{K}^\tx{(A)} ) \sinh (\ri J_\tx{a}^\phi \cd \Phi_\tx{R}^\tx{(B)}) ~, \\
C^{J}_{y0} =& - C_{x0} \sin (2 \ri J_\tx{a}^\phi\cd \Phi_\tx{K}^\tx{(A)} ) \cosh (\ri J_\tx{a}^\phi \cd \Phi_\tx{R}^\tx{(B)})
+ C_{yz} \cos (2 \ri J_\tx{a}^\phi\cd \Phi_\tx{K}^\tx{(A)} ) \sinh (\ri J_\tx{a}^\phi \cd \Phi_\tx{R}^\tx{(B)}) ~, \\
C^{J}_{z0} =& \cosh (\ri J_\tx{a}^\phi \cd \Phi_\tx{R}^\tx{(B)}) \sinh (\ri J_\tx{a}^\phi \cd \Phi_\tx{R}^\tx{(A)})  \label{C^J_i0}
}
with $C_{x0}$ and $C_{yz}$ defined in (\ref{Cs}),
where
\eqn{
\Phi_\tx{K}^\tx{(A)}(t') \coloneqq& \int \dd t  G_\tx{K}(t'-t, \vb*{x}' - \vb*{x}_\tx{A})\lambda_\tx{A}(t) ~, \\
\Phi_\tx{R}^\tx{(A)}(t') \coloneqq& \int \dd t  G_\tx{R}(t'-t, \vb*{x}' - \vb*{x}_\tx{A})\lambda_\tx{A}(t) ~,
}
in addition, $\Phi_\tx{K}^\tx{(B)}(t')$ and $\Phi_\tx{R}^\tx{(B)}(t')$ are defined in the same manner.
Note that why $J^\phi_\tx{r}$ is absent in (\ref{C^J_i0}) is that $J^\phi_i (x)=0$ is assumed for $x^0 < \tx{max}\qty{ t_\tx{off}^\tx{A} ,t_\tx{off}^\tx{B} }$ because, here, we are only interested in correlations between the field and each spin after both Alice and Bob turned off their spin-field interactions.
For correlations between the field and Bob's spin, we need $C^{J}_{0x}$, $C^{J}_{0y}$ and $C^{J}_{0z}$ which are given by the above $C^{J}_{x0}$, $C^{J}_{y0}$ and $C^{J}_{z0}$ with the replacement (A,B) $\to$ (B,A) applied.

%%%%%%%%%%%%%%%%%%%%%%%
\section{Particle creation and propagator \label{app.Particle creation and Keldysh function}}
In the presence of time-dependent external sources, particles are created in general.
Here,  we formulate a computation of the number of particles created and its correlation with the spin operators, based on the Hamiltonian formalism which is more intuitive than the path-integral one.
We first consider a single harmonic oscillator and then extend the discussion to the scalar field case.
Finally, we couple the field to the spin variables s as in the body of the paper and compute correlations between the spin operators and the number of particles created.

\subsection{Harmonic oscillator}
Suppose a system is governed by the Hamiltonian $\hat{H}= \hat{H}_0 + \hat{H}_\tx{s}$ with
\eqn{
\hat{H}_0 = \frac{\hat{p}^2}{2} + \frac{\omega^2 \hat{x}^2}{2} = \omega \qty(\hat{N}  +\frac{1}{2}) ~,
}
\eqn{\hat{H}_\tx{s} = - J(t) \hat{x} = -J(t) \frac{\hat{a} + \hat{a}^\dag}{\sqrt{2\omega}}  ~.}
$\hat{x}$ and $\hat{p}$ denote the position and the momentum operators of a harmonic oscillator with frequency $\omega$. Its mass is set to be unity here.
The number operator $\hat{N} = \hat{a}^\dag \hat{a}$ is constructed from the annihilation operator
$\hat{a} = (\omega \hat{x} + \ri \hat{p})/\sqrt{2 \omega}$ and the creation operator $\hat{a}^\dag = (\omega \hat{x} - \ri \hat{p})/\sqrt{2 \omega}$.
We assume the external source $J(t)$ is zero at the initial time $t=t_\tx{i}$ and the system is in the ground state $\ket{0}$ defined by $\hat{a} \ket{0} = 0$.

It is convenient to take the interaction picture where operators evolve in time with $U_0 (t,t_\tx{i}) = \exp \{-\ri (t-t_\tx{i}) \hat{H}_0 \}$ as $\hat{ \cl{O}}_\tx{I} (t,t_\tx{i}) = \hat{U}_0 (t,t_\tx{i})^\dag \hat{\cl{O}}U_0 (t,t_\tx{i})$
and the state evolves with $\hat{U}_\tx{I} (t,t_\tx{i}) = \tx{T} \exp \qty{ -\ri \int_{t_\tx{i}}^t \dd t'  \hat{H}_\tx{I} (t') }$ where
\eqn{
\hat{ H}_\tx{I} (t) \coloneqq \hat{U}_0 (t,t_\tx{i})^\dag \hat{H}_\tx{s} U_0 (t,t_\tx{i}) = - J(t) x_\tx{I}(t) ~, \label{H_I}
}
\eqn{ x_\tx{I}(t) = \frac{\hat{a} e^{-\ri  \omega (t-t_\tx{i} )} + \hat{a}^\dag e^{+\ri  \omega (t-t_\tx{i})}}{\sqrt{2 \omega}}  ~. \label{x_I}}
Since $\hat{N}_\tx{I}(t) = \hat{U}_0 (t,t_\tx{i})^\dag \hat{N} \hat{U}_0 (t,t_\tx{i}) = \hat{N}$,
we obtain the number of particles at $t=t_\tx{f}$ as
\eqn{
\bra{\Psi_\tx{f}}\hat{N}\ket{\Psi_\tx{f}} = \bra{0} \hat{U}_\tx{I}(t_\tx{f},t_\tx{i})^\dag \hat{N} \hat{U}_\tx{I}(t_\tx{f},t_\tx{i}) \ket{0} = |A|^2 ~, \label{app.N(T)-H.O.}
}
where
\eqn{
A =   \ri \int_{t_\tx{i}}^{t_\tx{f}} \dd t J (t) \frac{e^{+\ri \omega (t-t_\tx{i})}}{\sqrt{2 \omega}} ~. \label{A}
}
This can be proved by noting
\begin{equation}
      [\hat{a} ,  \hat{U}_\tx{I}(t_\tx{f},t_\tx{i}) ] = A \hat{U}_\tx{I}(t_\tx{f},t_\tx{i}) .
\end{equation}
The number of particles (\ref{app.N(T)-H.O.}) can be  rewritten as
\eqn{
\bra{\Psi_\tx{f}}\hat{N}\ket{\Psi_\tx{f}} & = \int_{t_\tx{i}}^{t_\tx{f}} \dd t \int_{t_\tx{i}}^{t_\tx{f}}  \dd t' J(t) G_> (t-t') J(t') \\
&= \int_{t_\tx{i}}^{t_\tx{f}} \dd t \int_{t_\tx{i}}^{t_\tx{f}}  \dd t' J(t) G_\tx{K}(t-t') J(t') ~,
}
where 
\eqn{
G_> (t-t') \coloneqq  \bra{0}  \hat{x}_\tx{I}(t)  \hat{x}_\tx{I}(t')  \ket{0} = \frac{e^{- \ri \omega (t- t') }}{2 \omega}
}
is the Wightman function for a harmonic oscillator and its symmetric part is the Keldysh function,
\eqn{
G_\tx{K}(t-t') \coloneqq \frac{\bra{0} \{ \hat{x}_\tx{I}(t) , \hat{x}_\tx{I}(t')  \} \ket{0}}{2} = \frac{\cos (\omega (t-t'))}{2 \omega} ~.
}

The expectation values of $\hat{N}^n$ with $n \geq 2$ can be computed in the same manner.
For example,
\eqn{
\bra{\Psi_\tx{f}} \hat{N}^2 \ket{\Psi_\tx{f}}=
\bra{\Psi_\tx{f}}  \hat{a}^\dag \hat{a}^\dag\hat{a} \hat{a}
\ket{\Psi_\tx{f}} + \bra{\Psi_\tx{f}} \hat{N} \ket{\Psi_\tx{f}} ~,
}
where the first term is given by
\eqn{
\bra{0} \hat{U}_\tx{I}(t_\tx{f},t_\tx{i})^\dag  \hat{a}^\dag \hat{a}^\dag\hat{a} \hat{a} \hat{U}(t_\tx{f},t_\tx{i}) \ket{0} = (  |A|^2 )^2 = \bra{\Psi_\tx{f}} \hat{N} \ket{\Psi_\tx{f}}^2  ~.
}

\subsection{Free scalar field}
The free scalar field is nothing but the sum of infinitely many harmonic oscillators:
\eqn{
\hat{\phi} (\vb*{x}) = \int \frac{\dd^3 k}{(2\pi)^3} \hat{\phi} (\vb*{k}) e^{+\ri \vb*{k} \cdot \vb*{x} }   ~,
}
\eqn{ \hat{\phi} (\vb*{k}) =  \frac{ \hat{a}_{\vb*{k}} + \hat{a}_{-\vb*{k}}^\dag }{\sqrt{2 \omega_k }} ~,}
where $\omega_k = \sqrt{m^2 +  |\vb*{k}|^2 }$ with $m$ being field's mass.
Now, we suppose the system is governed by the Hamiltonian $\hat{H}= \hat{H}_0 + \hat{H}_\tx{s}$ with
\eqn{
\hat{H}_0  = \int \frac{\dd^3 k}{(2\pi)^3} \omega_k \qty( \hat{N}_{\vb*{k}}  +\frac{1}{2}) ~,
}
\eqn{\hat{H}_\tx{s} = - \int \frac{\dd^3 k}{(2\pi)^3} \tilde{J}(t, -\vb*{k})  \hat{\phi} (\vb*{k})   ~. \label{H_s}}
Here, $\hat{H}_0$ is the same as $\hat{H}_\phi$ in (\ref{Hamiltonian}) and $\hat{N}_{\vb*{k}}  = \hat{a}_{\vb*{k}}^\dag  \hat{a}_{\vb*{k}}$ is the number density operator of particle with momentum $\vb*{k}$.
On the other hand, $\hat{H}_\tx{s}$ is the external source term and  $\tilde{J}(t, \vb*{k})$ is a Fourier component of the external source $J(t, \vb*{x})$:
\eqn{
J(t, \vb*{x}) = \int \frac{\dd^3 k}{(2\pi)^3} \tilde{J} (t,\vb*{k}) e^{+\ri \vb*{k} \cdot \vb*{x} } ~.
}
Note that $\tilde{J}^* (t, -\vb*{k}) = \tilde{J}(t, \vb*{k})$ for $J(t, \vb*{x})$ to be real.

In the interaction picture, the field operator becomes
\eqn{ \hat{\phi}_\tx{I} (t,\vb*{k}) =  \frac{\hat{a}_{\vb*{k}} e^{-\ri  \omega_k (t-t_\tx{i})} + \hat{a}_{-\vb*{k}}^\dag e^{+\ri  \omega_k (t-t_\tx{i})}}{\sqrt{2 \omega_k}}  ~,}
and the interaction Hamiltonian is given by
\eqn{
\hat{ H}_\tx{I} (t) =  - \int \frac{\dd^3 k}{(2\pi)^3} \tilde{J}(t, -\vb*{k})  \hat{\phi}_\tx{I} (t,\vb*{k}) ~.
}
Then, we get
\eqn{
A_{\vb*{k}} = \ri \int_{t_\tx{i}}^{t_\tx{f}} \dd t \tilde{J} (t,\vb*{k}) \frac{e^{+\ri \omega_k (t-t_\tx{i})}}{\sqrt{2 \omega_k}} ~. \label{A_k(T)}
}
As in the harmonic oscillator case,
the operator measuring the number of particles created
\eqn{
\hat{N} = \int \frac{\dd^3 k}{(2\pi)^3} 
\hat{a}^\dag_{\vb*{k}}
\hat{a}_{\vb*{k}}
}
itself does not evolve in the interaction picture: $\hat{N}_\tx{I}(t) = \hat{N}$.
Therefore, starting from the vacuum initial state $\ket{\Omega}_\phi$,
we find
\eqn{
\bra{\Psi_\tx{f}}\hat{N}\ket{\Psi_\tx{f}} &= \bra{\Omega} \hat{U}_\tx{I}(t_\tx{f},t_\tx{i})^\dag \hat{N} \hat{U}_\tx{I}(t_\tx{f},t_\tx{i}) \ket{\Omega}_\phi = \int \frac{\dd^3 k}{(2\pi)^3}   |A_{\vb*{k}}|^2 \\
&=  \int_{t_\tx{i}}^{t_\tx{f}} \dd t \int_{t_\tx{i}}^{t_\tx{f}}  \dd t' \int \frac{\dd^3 k}{(2\pi)^3} \tilde{J}(t, -\vb*{k}) \frac{e^{- \ri \omega_k (t- t') } }{2 \omega_k}  \tilde{J}(t', \vb*{k}) \\
&= \int_{t_\tx{i}}^{t_\tx{f}} \dd t  \int_{t_\tx{i}}^{t_\tx{f}}  \dd t' \int \dd^3 x \int \dd^3 x' J(t, \vb*{x}) G_> (t-t',\vb*{x}-\vb*{x}') J(t', \vb*{x}')  \\
&= \int_{t_\tx{i}}^{t_\tx{f}} \dd t  \int_{t_\tx{i}}^{t_\tx{f}}  \dd t' \int \dd^3 x \int \dd^3 x' J(t, \vb*{x}) G_\tx{K} (t-t',\vb*{x}-\vb*{x}') J(t', \vb*{x}') ~,
}
where
\eqn{
G_> (t-t',\vb*{x}-\vb*{x}') \coloneqq& \bra{\Omega}  \hat{\phi}_\tx{I}(t,\vb*{x}),  \hat{\phi}_\tx{I}(t',\vb*{x'}) \ket{\Omega}_\phi  \\
=& \int \frac{\dd^3 k}{(2 \pi)^3} \frac{  e^{- \ri \omega_k (t- t') } }{2 \omega_k} e^{+\ri \vb*{k}\cdot (\vb*{x}-\vb*{x}')}
}
is the Wightman function  and
\eqn{
G_\tx{K} (t-t',\vb*{x}-\vb*{x}') \coloneqq& \frac{\bra{\Omega} \{ \hat{\phi}_\tx{I}(t,\vb*{x}),  \hat{\phi}_\tx{I}(t',\vb*{x'}) \}\ket{\Omega}_\phi }{2} \\
=& \int \frac{\dd^3 k}{(2 \pi)^3} \frac{\cos \qty( \omega_k (t-t') )}{2 \omega_k} e^{+\ri \vb*{k}\cdot (\vb*{x}-\vb*{x}')}
}
is the Keldysh Green's function for the free field,
\eqn{
\hat{\phi}_\tx{I} (t,\vb*{x}) \coloneqq \int \frac{\dd^3 k}{(2\pi)^3} \frac{\hat{a}_{\vb*{k}} e^{-\ri  \omega_k (t - t_\tx{i}) +\ri \vb*{k}\cdot \vb*{x}} + \hat{a}_{\vb*{k}}^\dag e^{+\ri \omega_k (t - t_\tx{i}) - \ri \vb*{k}\cdot \vb*{x}}}{\sqrt{2 \omega_k}}   ~.
}
In the body of paper, the subscript ``I'' of $\hat{\phi}_\tx{I} (t,\vb*{x})$ is dropped, see Footnote \ref{source-free} in Sec.~\ref{sec.Reduced density matrix}.

The expectation values of $\hat{N}^n$ with $n \geq 2$ can be computed similarly.
For instance,
\eqn{
\bra{\Psi_\tx{f}} \hat{N}^2 \ket{\Psi_\tx{f}}= (\bra{\Psi_\tx{f}} \hat{N} \ket{\Psi_\tx{f}} )^2 
+ \bra{\Psi_\tx{f}} \hat{N} \ket{\Psi_\tx{f}} ~,
}
where the first term is calculated as
\eqn{
& \int \frac{\dd^3 k}{(2\pi)^3} \int \frac{\dd^3 k'}{(2\pi)^3} \bra{\Omega} \hat{U}_\tx{I}(t_\tx{f},t_\tx{i})^\dag  \hat{a}_{\vb*{k}}^\dag \hat{a}_{\vb*{k}'}^\dag\hat{a}_{\vb*{k}'} \hat{a}_{\vb*{k}}\hat{U}(t_\tx{f},t_\tx{i}) \ket{\Omega}_\phi \\
&= \qty[ \int \frac{\dd^3 k}{(2\pi)^3}   |A_{\vb*{k}}|^2 ]^2 = (\bra{\Psi_\tx{f}} \hat{N} \ket{\Psi_\tx{f}} )^2  ~.
}

\subsection{Free scalar field coupled with spin}
Now, let us identify $\hat{H}_\tx{s}$ in (\ref{H_s}) as the spin-field interaction $\hat{H}_\tx{A} + \hat{H}_\tx{B}$ in (\ref{Hamiltonian}).
Then the current becomes operators acting on the spin variables,
\eqn{
J(t, \vb*{x}) = \hat{\sigma}_{z}^\tx{A}  \lambda_\tx{A}(t)  \delta^{(3)} (\vb*{x} - \vb*{x}_\tx{A}) + \hat{\sigma}_{z}^\tx{B}  \lambda_\tx{B}(t) \delta^{(3)} (\vb*{x} - \vb*{x}_\tx{B})  ~,
}
which is an operator acting on spins' Hilbert space.
Note that $\hat{\sigma}^\tx{A}_z$ and $\hat{\sigma}^\tx{B}_z$ commute with the free-field Hamiltonian $\hat{H}_0 = \hat{H}_\phi$, and the operator in the interaction picture is the same as in the Schr\"{o}dinger picture; $\hat{\sigma}^\tx{A,B}_{z \tx{I}} = \hat{\sigma}^\tx{A,B}_z$. It is also true for other components of the spin. 
Therefore, (\ref{A_k(T)}) is replaced by an operator acting on the spin variables,
\eqn{
A_{\vb*{k}}  = \ri \int_{t_\tx{i}}^{t_\tx{f}} \dd t \frac{e^{+\ri \omega_k (t-t_\tx{i})}}{\sqrt{2 \omega_k}} \qty( \hat{\sigma}_{z}^\tx{A}  \lambda_\tx{A}(t)  e^{-\ri \vb*{k}\cdot \vb*{x}_\tx{A}} + \hat{\sigma}_{z}^\tx{B}  \lambda_\tx{B}(t) e^{-\ri \vb*{k}\cdot \vb*{x}_\tx{B}} ) ~.
}
With the initial state  $\ket{\Psi_\tx{i}}$ given in (\ref{initial-state}),
the number of particles created due to the nonadiabaticity is computed as
\eqn{
\bra{\Psi_\tx{f}}\hat{N}\ket{\Psi_\tx{f}} &=
\bra{\Psi_\tx{i}} \hat{U}_\tx{I}(t_\tx{f},t_\tx{i})^\dag \hat{N} \hat{U}_\tx{I}(t_\tx{f},t_\tx{i}) \ket{\Psi_\tx{i}} \\
 &= \int \frac{\dd^3 k}{(2\pi)^3}  \bra{\Psi_\tx{i}} \hat{U}_\tx{I}(t_\tx{f},t_\tx{i})^\dag A_{\vb*{k}}^\dag  A_{\vb*{k}} \hat{U}_\tx{I}(t_\tx{f}) \ket{\Psi_\tx{i}} \\
&= \sum_{\tx{X},\tx{X}'= \tx{A,B}}  \int_{t_\tx{i}}^{t_\tx{f}} \dd t \int_{t_\tx{i}}^{t_\tx{f}}  \dd t'  \La \hat{\sigma}_{z}^\tx{X} \hat{\sigma}_{z}^{\tx{X}'}  \Ra \lambda_\tx{X}(t) G_> (t-t',\vb*{x}_\tx{X}-\vb*{x}_{\tx{X}'}) \lambda_{\tx{X}'}(t')  \\
&= \sum_{\tx{X}= \tx{A,B}}  \int_{t_\tx{i}}^{t_\tx{f}} \dd t  \int_{t_\tx{i}}^{t_\tx{f}}  \dd t'  \lambda_\tx{X}(t) G_\tx{K} (t-t',\vb*{0}) \lambda_\tx{X}(t') \\
&= \mf{G}_\tx{K}^\tx{AA} + \mf{G}_\tx{K}^\tx{BB} ~.
}
On the third equality, we have used $(\hat{\sigma}_z^\tx{A})^2=(\hat{\sigma}_z^\tx{B})^2=1$ 
and $\La  \hat{\sigma}_{z}^\tx{A} \hat{\sigma}_{z}^\tx{B}  \Ra = \tr \{ \hat{\rho}_\tx{AB}  \hat{\sigma}_{z}^\tx{A} \hat{\sigma}_{z}^\tx{B}  \}    = 0$ with the initial state of the spins given by (\ref{spin-initial-state}) as seen in (\ref{Cs}).
In the same manner, we find
\begin{equation}
    \bra{\Psi_\tx{f}} \hat{N}^2 \ket{\Psi_\tx{f}}=
  \bra{\Psi_\tx{f}}  \! \! :\! \! \hat{N}^2 \! \! :\!  \ket{\Psi_\tx{f}}
+ \bra{\Psi_\tx{f}} \hat{N} \ket{\Psi_\tx{f}} ~,
\end{equation}
where the first term is given by
\eqn{
&\bra{\Psi_\tx{f}}  \! \! :\! \! \hat{N}^2 \! \! :\!  \ket{\Psi_\tx{f}}
 = \int \frac{\dd^3 k}{(2\pi)^3} \int \frac{\dd^3 k'}{(2\pi)^3}  \bra{\Psi_\tx{i}} \hat{U}_\tx{I}(t_\tx{f},t_\tx{i})^\dag A_{\vb*{k}}^\dag   A_{\vb*{k}'}^\dag  A_{\vb*{k}'} A_{\vb*{k}} \hat{U}_\tx{I}(t_\tx{f},t_\tx{i}) \ket{\Psi_\tx{i}} \\
&= \sum_{\tx{X},\tx{X}',\tx{Y},\tx{Y}'= \tx{A,B}}  \int_{t_\tx{i}}^{t_\tx{f}} \dd u \int_{t_\tx{i}}^{t_\tx{f}}  \dd u' \int_{t_\tx{i}}^{t_\tx{f}} \dd v \int_{t_\tx{i}}^{t_\tx{f}}  \dd v' \La \hat{\sigma}_{z}^\tx{Y} \hat{\sigma}_{z}^\tx{X} \hat{\sigma}_{z}^{\tx{X}'} \hat{\sigma}_{z}^{\tx{Y}'} \Ra   \\
& \hspace{5em}\times \lambda_\tx{X}(u) G_> (u-u',\vb*{x}_\tx{X}-\vb*{x}_{\tx{X}'}) \lambda_{\tx{X}'}(u') \lambda_\tx{Y}(v) G_> (v-v',\vb*{x}_\tx{Y}-\vb*{x}_{\tx{Y}'}) \lambda_{\tx{Y}'}(v') \\
& = (\mf{G}_\tx{K}^\tx{AA} + \mf{G}_\tx{K}^\tx{BB})^2 + 4 (\mf{G}_\tx{K}^\tx{BA})^2 ~.
}
It can be easily shown by noting that,  
due to $\La \hat{\sigma}_{z}^\tx{A} \hat{\sigma}_{z}^\tx{B}  \Ra = 0$, four spin correlations vanish unless two of them are A and the other two are B, or all four of them are A or B. 
The statement is expressed by the following equality,
\eqn{
\hat{\sigma}_{z}^\tx{Y} \hat{\sigma}_{z}^\tx{X} \hat{\sigma}_{z}^{\tx{X}'} \hat{\sigma}_{z}^{\tx{Y}'} =& \delta_{\tx{X}\tx{X}'}\delta_{\tx{Y}\tx{Y}'} + 4 \delta_{\tx{X}(\tx{A}}\delta_{\tx{B})\tx{X}'} \delta_{\tx{Y}(\tx{A}}\delta_{\tx{B})\tx{Y}'} \\
& + 2 (\delta_{\tx{X}\tx{X}'} \delta_{\tx{Y}(\tx{A}}\delta_{\tx{B})\tx{Y}'} + \delta_{\tx{Y}\tx{Y}'} \delta_{\tx{X}(\tx{A}}\delta_{\tx{B})\tx{X}'}) \hat{\sigma}_{z}^\tx{A} \hat{\sigma}_{z}^\tx{B}   ~.
}
Here, $\delta_{\tx{X}(\tx{A}}\delta_{\tx{B})\tx{X}'}\coloneqq
(\delta_{\tx{X}\tx{A}}\delta_{\tx{B}\tx{X}'}
+\delta_{\tx{X}\tx{B}}\delta_{\tx{A}\tx{X}'})/2
$.

\subsection{Correlation between spin and number of particles}
One can compute the correlation functions between the spin and the number of particles in the same way as above.
For instance, inserting $\hat{\sigma}^\tx{A}_{w \tx{I}} = \hat{\sigma}^\tx{A}_w$, we get 
\eqn{
\bra{\Psi_\tx{f}} \hat{\sigma}^\tx{A}_w \hat{N}\ket{\Psi_\tx{f}} &= 
 \int \frac{\dd^3 k}{(2\pi)^3}  \bra{\Psi_\tx{i}} \hat{U}_\tx{I}(t_\tx{f},t_\tx{i})^\dag A_{\vb*{k}}^\dag  \hat{\sigma}^\tx{A}_w A_{\vb*{k}} \hat{U}_\tx{I}(t_\tx{f}) \ket{\Psi_\tx{i}} \\
&= \sum_{\tx{X},\tx{X}'= \tx{A,B}}  \int_{t_\tx{i}}^{t_\tx{f}} \dd t \int_{t_\tx{i}}^{t_\tx{f}}  \dd t'  \La \hat{\sigma}_{z}^\tx{X} \hat{\sigma}^\tx{A}_w \hat{\sigma}_{z}^{\tx{X}'} \Ra  \lambda_\tx{X}(t) G_> (t-t',\vb*{x}_\tx{X}-\vb*{x}_{\tx{X}'}) \lambda_{\tx{X}'}(t')  ~.
}
Most of the terms in the expectation value of
\eqn{
\hat{\sigma}_{z}^\tx{X} \hat{\sigma}_{w}^\tx{A}  \hat{\sigma}_{z}^{\tx{X}'} =
& 2 \delta_{wz}  \delta_\tx{XA}\delta_{\tx{X}'\tx{A}} \hat{\sigma}_{z}^\tx{A}    +2  \delta_{wz}   \delta_\tx{X(A}\delta_{\tx{B})\tx{X}'} \hat{\sigma}_{z}^\tx{B}\\
&~ + (\delta_\tx{XB}\delta_{\tx{X}'\tx{B}} - \delta_\tx{XA}\delta_{\tx{X}'\tx{A}} )  \hat{\sigma}_{w}^\tx{A}   +  2 \ri \epsilon_{wvz} \delta_\tx{X[A}\delta_{\tx{B}]\tx{X}'}  \hat{\sigma}_{v}^\tx{A} \hat{\sigma}_{z}^\tx{B}  
}
vanish because of the initial state (\ref{spin-initial-state}) assumed here, see (\ref{Cs}).
Here, $\delta_\tx{X[A}\delta_{\tx{B}]\tx{X}'} =
(\delta_\tx{XA}\delta_{\tx{B}\tx{X}'} 
-\delta_\tx{XB}\delta_{\tx{A}\tx{X}'})/2
$, and $\epsilon_{wvu}$ denotes the totally anti-symmetric tensor with $\epsilon_{xyz}=+1$.
Therefore,
\eqn{
&\bra{\Psi_\tx{f}}\hat{\sigma}^\tx{A}_x \hat{N} \ket{\Psi_\tx{f}}
= (\mf{G}_\tx{K}^\tx{BB} - \mf{G}_\tx{K}^\tx{AA} ) \La \hat{\sigma}_{x}^\tx{A} \Ra + (\mf{G}_\tx{R}^\tx{AB} - \mf{G}_\tx{R}^\tx{BA}) \La \hat{\sigma}_{y}^\tx{A} \hat{\sigma}_{z}^\tx{B} \Ra   ~, \\
&\bra{\Psi_\tx{f}} \hat{\sigma}^\tx{A}_y \hat{N}  \ket{\Psi_\tx{f}} =0 ~, ~~~
\bra{\Psi_\tx{f}} \hat{\sigma}^\tx{A}_z \hat{N} \ket{\Psi_\tx{f}} =0 ~. \label{spin-N-correlation}
}
For the correlations with Bob's spin, one can simply make the replacement (A,B) $\to$ (B,A).

Similarly, we find
\eqn{
\bra{\Psi_\tx{f}} \hat{\sigma}_{w}^\tx{A}   \! :\! \! \hat{N}^2 \! \! :\!   \ket{\Psi_\tx{f}}
& = \int \frac{\dd^3 k}{(2\pi)^3} \int \frac{\dd^3 k'}{(2\pi)^3}  \bra{\Psi_\tx{i}} \hat{U}_\tx{I}(t_\tx{f},t_\tx{i})^\dag A_{\vb*{k}}^\dag   A_{\vb*{k}'}^\dag   \hat{\sigma}_{w}^\tx{A}  A_{\vb*{k}'}(t_\tx{f}) A_{\vb*{k}}   \hat{U}_\tx{I}(t_\tx{f},t_\tx{i})  \ket{\Psi_\tx{i}} \\
&= \sum_{\tx{X},\tx{X}',\tx{Y},\tx{Y}'= \tx{A,B}}  \int_{t_\tx{i}}^{t_\tx{f}} \dd u \int_{t_\tx{i}}^{t_\tx{f}}  \dd u' \int_{t_\tx{i}}^{t_\tx{f}} \dd v \int_{t_\tx{i}}^{t_\tx{f}}  \dd v'  \La \hat{\sigma}_{z}^\tx{Y} \hat{\sigma}_{z}^\tx{X}
 \hat{\sigma}^\tx{A}_w
 \hat{\sigma}_{z}^{\tx{X}'} \hat{\sigma}_{z}^{\tx{Y}'} \Ra    \\
&~~~~~~~ \times \lambda_\tx{X}(u) G_> (u-u',\vb*{x}_\tx{X}-\vb*{x}_{\tx{X}'}) \lambda_{\tx{X}'}(u') 
 \lambda_\tx{Y}(v) G_> (v-v',\vb*{x}_\tx{Y}-\vb*{x}_{\tx{Y}'}) \lambda_{\tx{Y}'}(v') ~,
}
where most of the terms in the expectation value of
\eqn{
 \hat{\sigma}_{z}^\tx{Y} \hat{\sigma}_{z}^\tx{X} \hat{\sigma}_{w}^\tx{A}  \hat{\sigma}_{z}^{\tx{X}'} \hat{\sigma}_{z}^{\tx{Y}'}  
=&  8\delta_{zw} \delta_\tx{X(A}\delta_{\tx{B})\tx{Y}} \delta_\tx{X'(A}\delta_\tx{B)Y'}  \hat{\sigma}_{z}^\tx{A}     \\
&+ 2 \delta_{zw} ( \delta_\tx{X'Y'} \delta_\tx{X(A}\delta_\tx{B)Y}  +\delta_\tx{XY} \delta_\tx{X'(A}\delta_\tx{B)Y'} )   \hat{\sigma}_{z}^\tx{B}   \\
&+(\delta_\tx{XY}\delta_{\tx{X}'\tx{Y}'} - 4 \delta_\tx{X(A}\delta_{\tx{B})\tx{Y}} \delta_\tx{X'(A}\delta_\tx{B)Y'})   \hat{\sigma}_{w}^\tx{A}   \\
&+ 2 \ri \epsilon_{wvz} ( \delta_\tx{X'Y'} \delta_\tx{X(A}\delta_\tx{B)Y}  -\delta_\tx{XY} \delta_\tx{X'(A}\delta_\tx{B)Y'} )   \hat{\sigma}_{v}^\tx{A}  \hat{\sigma}_{z}^\tx{B}    
}
vanish with the initial state (\ref{spin-initial-state}) assumed.
Then,
\eqn{
\La \hat{\sigma}_{x}^\tx{A}   \! :\! \! \hat{N}^2 \! \! :  \Ra  =& ( (\mf{G}_\tx{K}^\tx{BB}- \mf{G}_\tx{K}^\tx{AA})^2 - (\mf{G}_\tx{R}^\tx{AB} - \mf{G}_\tx{R}^\tx{BA})^2 ) \La \hat{\sigma}_{x}^\tx{A}  \Ra \\
& + 2 ( \mf{G}_\tx{R}^\tx{AB} - \mf{G}_\tx{R}^\tx{BA} ) ( \mf{G}_\tx{K}^\tx{BB}- \mf{G}_\tx{K}^\tx{AA} ) \La  \hat{\sigma}_{y}^\tx{A} \hat{\sigma}_{z}^\tx{B} \Ra ~, \\
\La \hat{\sigma}_{y}^\tx{A}   \! :\! \! \hat{N}^2 \! \! :  \Ra  =& 0 ~, ~~~\La \hat{\sigma}_{z}^\tx{A}   \! :\! \! \hat{N}^2 \! \! :  \Ra  =0 ~.
\label{spin-N2-correlation}}
Again, one can make the replacement (A,B) $\to$ (B,A) to obtain the correlations with Bob's spin.

%%%%%%%%%%%%%%%%%%%%%%%

\section{Time integration of propagator \label{app:Numerical evaluation}}
The reduced density matrix (\ref{rho_AB}) and various quantities obtained from it are all given as functions of $\mf{G}$'s defined in (\ref{mathfrak-G_R}) and (\ref{mathfrak-G_K}).
Here, we present explicit forms of  $\mf{G}$'s and their numerical values used to draw Figs.~\ref{fig:IAB-spacelike} and \ref{fig:IAB-Wald}.
As a function of $s\coloneqq  (t-t')^2 - |\vb*{x}-\vb*{x}'|^2$,
the retarded Green's function of the source-free field defined in (\ref{Retarded}) is written as
\eqn{
G_\tx{R}(x,x')  =   \frac{- m}{4 \pi  \sqrt{s}}   J_1(m \sqrt{s})  + \frac{\delta (s)}{2 \pi} ~~\tx{for} ~~ s \geq 0 ~, \label{app.Retarded_explicit}
}
and it vanishes for $s < 0$ reflecting the relativistic causality.
Here, $J_\alpha$ is the Bessel function of the first kind.
The Keldysh function (\ref{Keldysh}) can be written as
\eqn{
G_\tx{K}(x,x') =  \frac{1}{2} \frac{m}{2 \pi^2}  \times \left\{ \mtx{
s^{-1/2} (\pi /2) Y_1(m \sqrt{s})    &  \tx{for} ~ s \geq 0 \\
(-s)^{-1/2} K_1(m \sqrt{-s})& \tx{for} ~ s< 0
}\right. ~,
}
where $Y_\alpha$ is the Bessel function of the second kind and $K_\alpha$ is the modified Bessel function of the second kind.
Although $G_\tx{K}$ itself is divergent at $s=0$, an integration over $s=0$ gives a finite number since $(\pi /2) Y_1(x)/K_1(x) \to -1$ for $x\searrow 0$.
Then, it is convenient to express it in terms of Fourier modes as
\eqn{
G_\tx{K}(x,x') =& \int \frac{\dd^3 k}{(2 \pi)^3} \frac{\cos \qty( \omega_k (t-t') )}{2 \omega_k} e^{\ri \vb*{k}\cdot (\vb*{x}-\vb*{x}')} \\
=&\frac{2^{-1}}{2 \pi^2} \int_0^\infty \dd k \frac{k^2}{\omega_k}  \cos \qty( \omega_k (t-t') ) \frac{\sin (k |\vb*{x}-\vb*{x}'|)}{k |\vb*{x}-\vb*{x}'|} ~. \label{app.Keldysh_explicit}
}

For the time dependence of the couplings, we assume a simple form
\eqn{
\lambda_\tx{A}(t) =& \bar{\lambda}_\tx{A} \biggl\{ \theta(t^\tx{A}_\tx{off}-t) \theta(t- \tilde{t}^\tx{A}_\tx{off}) \frac{t^\tx{A}_\tx{off}- t}{T^\tx{A}_\tx{off}}  \\
&~~~~~~~~+ \theta(\tilde{t}^\tx{A}_\tx{off} -t) \theta(t - \tilde{t}^\tx{A}_\tx{on} )  + \theta( \tilde{t}^\tx{A}_\tx{on}  -t )  \theta( t - t^\tx{A}_\tx{on}  ) \frac{t - t^\tx{A}_\tx{on} }{T^\tx{A}_\tx{on}} \biggr\} ~,\label{time-dependence}
}
where
$T^\tx{A}_\tx{off} = t^\tx{A}_\tx{off} - \tilde{t}^\tx{A}_\tx{off} \geq 0$, $T^\tx{A}_\tx{on} =  \tilde{t}^\tx{A}_\tx{on}  - t^\tx{A}_\tx{on} \geq 0$ and $T^\tx{A}= \tilde{t}^\tx{A}_\tx{off} - \tilde{t}^\tx{A}_\tx{on} \geq 0$,
as sketched in Fig.~\ref{fig:Causality}.
Bob's one $\lambda_\tx{B}(t)$ is defined by replacing ``A'' with ``B''.
Then, $\mf{G}_\tx{R}$'s are numerically evaluated simply by plugging (\ref{app.Retarded_explicit}) and (\ref{time-dependence}) into the definitions in (\ref{mathfrak-G_R}) and implementing the time integration.
On the other hand, for $\mf{G}_\tx{K}$'s,
the time integration is analytically done with the expression (\ref{app.Keldysh_explicit}) plugged in the definition (\ref{mathfrak-G_K}):
\eqn{
\mf{G}_\tx{K}^\tx{BA} = \frac{\bar{\lambda}_\tx{A}\bar{\lambda}_\tx{B}}{4  \pi^2} \int \dd k \frac{k^2}{\omega_k^3}  \frac{\sin (k D)}{k D} \cl{I}_{k}(T_\tx{off}^\tx{A},T_\tx{on}^\tx{A},t_\tx{off}^\tx{A},t_\tx{on}^\tx{A}; T_\tx{off}^\tx{B},T_\tx{on}^\tx{B},t_\tx{off}^\tx{B},t_\tx{on}^\tx{B}) 
}
with
\eqn{
&\hspace{-1em}\cl{I}_{k}(T_\tx{off}^\tx{A},T_\tx{on}^\tx{A},t_\tx{off}^\tx{A},t_\tx{on}^\tx{A}; T_\tx{off}^\tx{B},T_\tx{on}^\tx{B},t_\tx{off}^\tx{B},t_\tx{on}^\tx{B}) \\
:&=\frac{4 \omega_k^{-2}}{T_\tx{off}^\tx{A}T_\tx{off}^\tx{B}} \sin \qty(\frac{\omega_kT_\tx{off}^\tx{A}}{2}) \sin \qty(\frac{\omega_kT_\tx{off}^\tx{B}}{2}) \cos \qty(\omega_k (t_\tx{off}^\tx{A} - t_\tx{off}^\tx{B})- \omega_k\frac{T_\tx{off}^\tx{A} - T_\tx{off}^\tx{B}}{2}) \\
&+\frac{4 \omega_k^{-2}}{T_\tx{on}^\tx{A}T_\tx{on}^\tx{B}} \sin \qty(\frac{\omega_k T_\tx{on}^\tx{A}}{2}) \sin \qty(\frac{\omega_k T_\tx{on}^\tx{B}}{2}) \cos \qty(\omega_k(t_\tx{on}^\tx{A} - t_\tx{on}^\tx{B})+\omega_k\frac{T_\tx{on}^\tx{A} - T_\tx{on}^\tx{B}}{2}) \\
&-\frac{4 \omega_k^{-2}}{T_\tx{off}^\tx{A}T_\tx{on}^\tx{B}} \sin \qty(\frac{\omega_k T_\tx{off}^\tx{A}}{2}) \sin \qty(\frac{\omega_k T_\tx{on}^\tx{B}}{2}) \cos \qty(\omega_k(t_\tx{off}^\tx{A} - t_\tx{on}^\tx{B})-\omega_k\frac{T_\tx{off}^\tx{A} + T_\tx{on}^\tx{B}}{2}) \\
&-\frac{4  \omega_k^{-2}}{T_\tx{on}^\tx{A}T_\tx{off}^\tx{B}} \sin \qty(\frac{\omega_k T_\tx{on}^\tx{A}}{2}) \sin \qty(\frac{\omega_k T_\tx{off}^\tx{B}}{2}) \cos \qty(\omega_k(t_\tx{on}^\tx{A} - t_\tx{off}^\tx{B})+\omega_k\frac{T_\tx{on}^\tx{A} + T_\tx{off}^\tx{B}}{2}) ~,
}
and
\eqn{
\mf{G}_\tx{K}^\tx{AA} = \frac{\bar{\lambda}_\tx{A}\bar{\lambda}_\tx{B}}{4  \pi^2} \int \dd k \frac{k^2}{\omega_k^3}  \bar{\cl{I}}_{k}(T_\tx{off}^\tx{A},T_\tx{on}^\tx{A},t_\tx{off}^\tx{A},t_\tx{on}^\tx{A})
}
with
\eqn{
\bar{\cl{I}}_k(T_\tx{off}^\tx{A},T_\tx{on}^\tx{A},t_\tx{off}^\tx{A},t_\tx{on}^\tx{A}) :&=\cl{I}_{k}(T_\tx{off}^\tx{A},T_\tx{on}^\tx{A},t_\tx{off}^\tx{A},t_\tx{on}^\tx{A}; T_\tx{off}^\tx{A},T_\tx{on}^\tx{A},t_\tx{off}^\tx{A},t_\tx{on}^\tx{A} ) \\
&=\frac{4 \omega_k^{-2}}{(T_\tx{off}^\tx{A})^2} \sin^2 \qty(\frac{\omega_k T_\tx{off}^\tx{A}}{2}) +\frac{4 \omega_k^{-2}}{(T_\tx{on}^\tx{A})^2} \sin^2 \qty(\frac{\omega_k T_\tx{on}^\tx{A}}{2})  \\
&-\frac{8 \omega_k^{-2}}{T_\tx{off}^\tx{A}T_\tx{on}^\tx{A}} \sin \qty(\frac{\omega_k T_\tx{off}^\tx{A}}{2}) \sin \qty(\frac{\omega_k T_\tx{on}^\tx{A}}{2}) \cos \qty(\omega_k \bar{T}^\tx{A}) ~,
}
where $\bar{T}^\tx{A} \coloneqq T^\tx{A} + \qty( T^\tx{A}_\tx{off}+T^\tx{A}_\tx{on})/2$.
Also, we get a similar form for $\mf{G}_\tx{K}^\tx{BB}$.
Then, the momentum integration is numerically implemented.

%%%%%%%%FIG%%%%
\begin{figure}
\begin{center}
 \includegraphics[width=9cm]{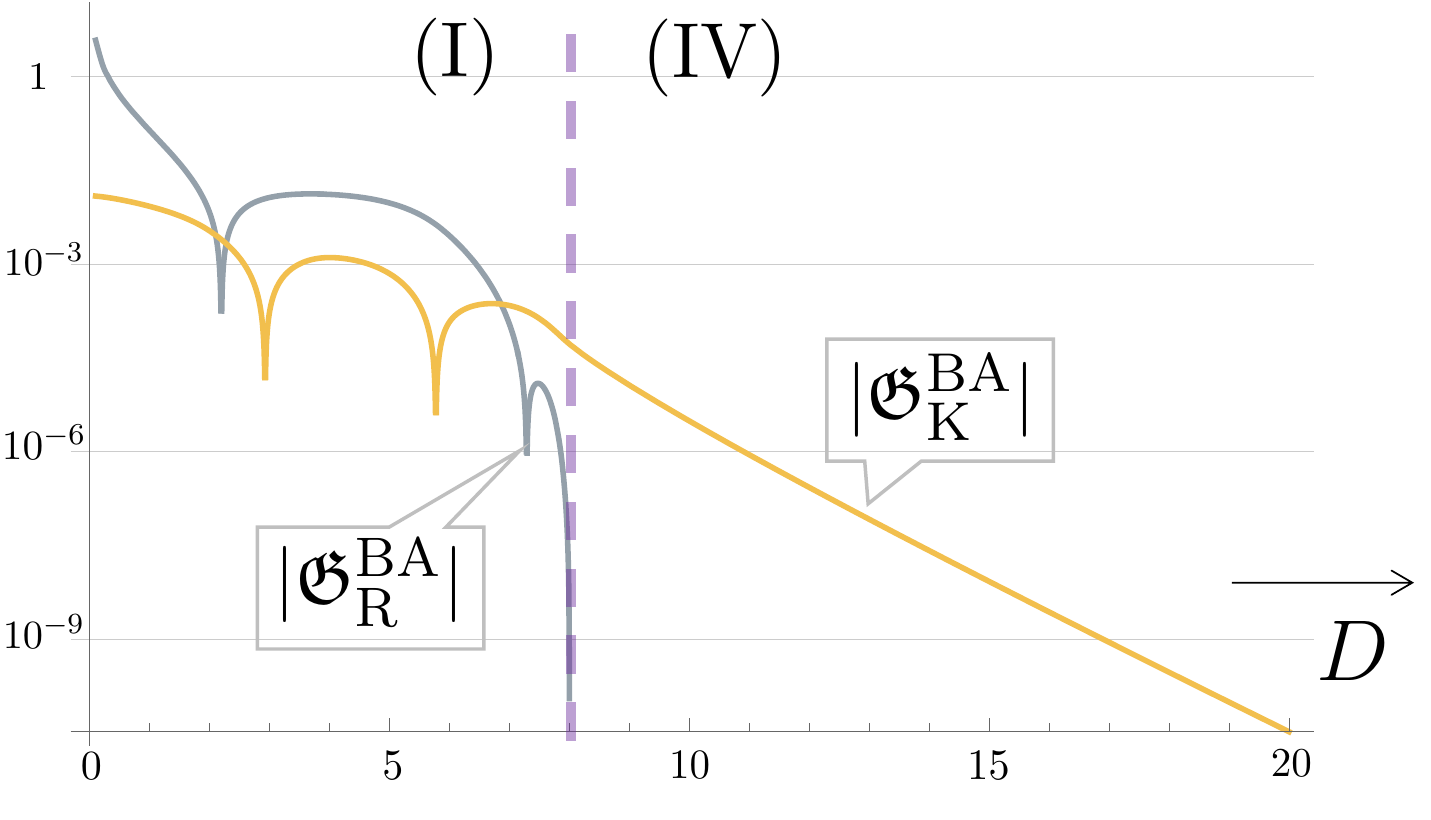}
\caption{The absolute value of $\mf{G}^\tx{BA}_\tx{K}$ is plotted as a function of the spatial distance $D$ between Alice and Bob with the yellow line.
We have included the part
with $D< 8$ where the system falls within the region (I) in Fig.~\ref{fig:Causality}; the retarded Green's functions connecting Alice's spin and Bob's spin are nonvanishing. Because of the symmetry, we have $\mf{G}_\tx{R}^\tx{AB} = \mf{G}_\tx{R}^\tx{BA}$. Its absolute value is plotted with the gray line.
}
\label{fig:Gs-spacelike}
\end{center}
\end{figure}
%%%%%%%%%%%%%
For Fig.~\ref{fig:IAB-spacelike} in Sec.\ref{sec.Spacelike-separated}, with the mass of scalar field $m$ set to be unity, 
parameters are chosen in such a way that the system is symmetric under the swapping of $\lambda_\tx{A}(t)$ and $\lambda_\tx{B}(t)$;
$\bar{\lambda}_\tx{A}=\bar{\lambda}_\tx{B}=1$, $T^\tx{A}_\tx{on}= T^\tx{A}_\tx{off}=  T^\tx{B}_\tx{on}= T^\tx{B}_\tx{off} =2$, $T^\tx{A}=T^\tx{B}=4$ and $t_\tx{on}^\tx{A} = t_\tx{on}^\tx{B}$.
With these parameters, we get $\mf{G}_\tx{K}^\tx{AA} =\mf{G}_\tx{K}^\tx{BB}\simeq 0.0125$.
The absolute values of $\mf{G}^\tx{BA}_\tx{K}$ and $\mf{G}_\tx{R}^\tx{BA}$ are plotted in Fig.~\ref{fig:Gs-spacelike}.
Note that,
because of the symmetry, we have $\mf{G}_\tx{R}^\tx{AB} = \mf{G}_\tx{R}^\tx{BA}$.
With $D=0$, we get $\mf{G}^\tx{BA}_\tx{K} = \mf{G}^\tx{AA}_\tx{K} = \mf{G}^\tx{BB}_\tx{K}$, and then, the Robertson-Schr\"{o}dinger relation (\ref{RS-inequality}) saturates.

%%%%%%%%FIG%%%%
\begin{figure}
\begin{center}
 \includegraphics[width=9cm]{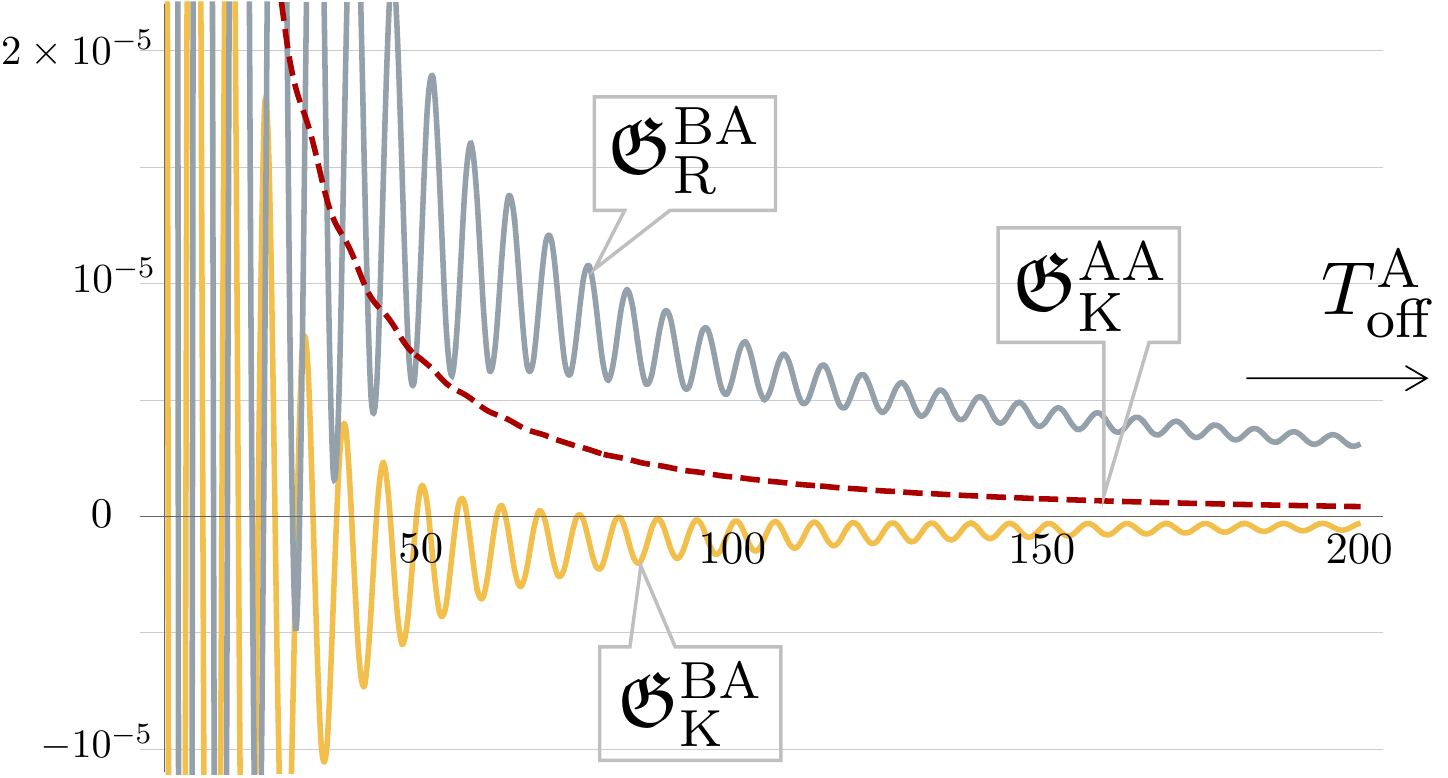}
\caption{
$\mf{G}_\tx{R}^\tx{BA}$ plotted with the gray line and  $\mf{G}_\tx{K}^\tx{BA}$ plotted with the yellow line decreases as $\sim (T^\tx{A}_\tx{off})^{-1}$, where $T^\tx{A}_\tx{off}$ is the time that Alice takes to turn off the spin-field interaction.
On the other hand, $\mf{G}_\tx{K}^\tx{AA}$ plotted with the dashed line decreases as $\sim (T^\tx{A}_\tx{off})^{-2}$.
}
\label{fig:Gs-Wald}
\end{center}
\end{figure}

%%%%%%%%%%%%%%
For Fig.~\ref{fig:IAB-Wald} in Sec.~\ref{sec.One-way}, 
parameters are chosen as follows: $\bar{\lambda}_\tx{A}=\bar{\lambda}_\tx{B}=1$, $T^\tx{A}_\tx{on}= \infty$, $T^\tx{B}_\tx{on}= T^\tx{B}_\tx{off} =1$, $T^\tx{B}=2$, $t_\tx{off}^\tx{A} = t_\tx{on}^\tx{B} + D - 1$ and $D=5$, see Fig.~\ref{fig:Wald-adiabatic}. With these parameters, we get $\mf{G}_\tx{K}^\tx{BB}\simeq 0.037$.
$\mf{G}_\tx{R}^\tx{BA}$, $\mf{G}_\tx{K}^\tx{BA}$ and $\mf{G}_\tx{K}^\tx{AA}$ are plotted in Fig.~\ref{fig:Gs-Wald}.
It is checked that the ratio of the left-hand side to the right-hand side of the inequality (\ref{RS-inequality-2}) is always less than unity; for large $T^\tx{A}_\tx{off}$, it becomes $\sim 2 \times 10^{-4}$ and oscillates around it.
%%%%%%%%%%%%%%%%%%%%

%%%%%%%%%%%%%%%%%%%%%%%%%%
\section{Consistency condition from nonnegativity \label{app.Consistency condition}}
Since the total system is governed by the well-defined theory with the Hamiltonian  (\ref{Hamiltonian}),
the reduced density matrices are necessarily positive semidefinite.
Here, let us observe that the nonnegativity of the density matrix (\ref{rho_AB}) for the initial state given by (\ref{initial-state}) requires an inequality
similar to the one (\ref{RS-inequality}) derived from the Robertson-Schr\"{o}dinger relation.

The eigenvalues of (\ref{rho_AB}) are computed as (\ref{mu_AB}).
For $s_1 = +1$, the nonnegativity is trivially satisfied.
On the other hand, for $s_1 = -1$, 
it turns out that 
\eqn{
-s_2 \qty[\cosh \qty(4 \mf{G}^\tx{BA}_\tx{K}) -\cos \qty(2 (\mf{G}^\tx{BA}_\tx{R}-\mf{G}^\tx{AB}_\tx{R}))] &\leq \qty( \gamma_\tx{A}^{-1} - \gamma_\tx{A}) \qty( \gamma_\tx{B}^{-1} - \gamma_\tx{B})/2 \\
& =2 \sinh (2 \mf{G}^\tx{AA}_\tx{K}) \sinh (2 \mf{G}^\tx{BB}_\tx{K} )  ~.
}
This is trivially satisfied for $s_2 = +1$. However, for $s_2 =-1$, it becomes 
\eqn{
\cos \qty(2 (\mf{G}^\tx{BA}_\tx{R}-\mf{G}^\tx{AB}_\tx{R})) 
\geq  1- 2 \qty[ \sinh (2 \mf{G}^\tx{AA}_\tx{K}) \sinh (2 \mf{G}^\tx{BB}_\tx{K} ) - \sinh^2 (2 \mf{G}^\tx{BA}_\tx{K})] ~. \label{positivity-condition}
}
Note that, this is merely a consistency condition, and thus, it can be confirmed by explicitly evaluating both sides with the definitions (\ref{mathfrak-G_R}) and (\ref{mathfrak-G_K}). 

Especially, for the case with $\mf{G}^\tx{AA}_\tx{K}, \mf{G}^\tx{BB}_\tx{K}, \mf{G}^\tx{BA}_\tx{K} \ll 1$,
the consistency condition (\ref{positivity-condition}) reads
\eqn{
1- \cos \qty(2 (\mf{G}^\tx{BA}_\tx{R}-\mf{G}^\tx{AB}_\tx{R}))  
\leq 8 \qty( \mf{G}^\tx{AA}_\tx{K} \mf{G}^\tx{BB}_\tx{K} - (\mf{G}^\tx{BA}_\tx{K})^2 ) ~, \label{positivity-condition-adiabatic-limit}
}
which is guaranteed by the Robertson-Schr\"{o}dinger inequality (\ref{RS-inequality}).

\bibliographystyle{apsrev4-1}
\bibliography{Relativistic_Quantum_Information}
\end{document}